\documentclass[10pt]{article}

\usepackage{graphicx} % Required for inserting images
\usepackage{amsfonts}
\usepackage{amsmath}
\usepackage{mathrsfs}
\usepackage{amssymb}
\usepackage{url}
\usepackage{fancyhdr}
\usepackage{indentfirst}
\usepackage{enumerate}
\usepackage{cases}
\usepackage{dsfont,footmisc}
\usepackage{bm}

\usepackage{subcaption}
\usepackage{amsthm}
\usepackage{color}
\usepackage{natbib}
\usepackage[british]{babel}
\usepackage[colorlinks=true,citecolor=blue]{hyperref}
\usepackage{tikz}
\usetikzlibrary{arrows.meta}
\usetikzlibrary{decorations.pathreplacing}

\usepackage[right,pagewise,mathlines]{lineno}

\usepackage[font=scriptsize]{caption} 
\usepackage{soul,xcolor}
\setstcolor{red}

\usepackage[title]{appendix}
\def\d{\mathrm{d}}
\def\laweq{\buildrel \d \over =}

\newcommand{\VaR}{\mathrm{VaR}}

\newcommand{\RVaR}{\mathrm{RVaR}}

\newcommand{\ES}{\mathrm{ES}}

\newcommand{\E}{\mathbb{E}}
\newcommand{\R}{\mathbb{R}}

\newcommand{\p}{\mathbb{P}}

\newcommand{\id}{\mathds{1}}

\newcommand{\X}{\mathcal X}

\newcommand{\mA}{\mathcal{A}}
\newcommand{\esssup}{\mathrm{ess\mbox{-}sup}}
\newcommand{\essinf}{\mathrm{ess\mbox{-}inf}}
\renewcommand{\ge}{\geqslant}
\renewcommand{\le}{\leqslant}
\renewcommand{\geq}{\geqslant}
\renewcommand{\leq}{\leqslant}
\renewcommand{\epsilon}{\varepsilon}

\theoremstyle{plain}
\newtheorem{theorem}{Theorem}%[section]
\newtheorem{corollary}{Corollary}%[section]
\newtheorem{lemma}{Lemma}%[section]
\newtheorem{proposition}{Proposition}%[section]

\theoremstyle{definition}
\newtheorem{definition}{Definition}%[section]
\newtheorem{example}{Example}%[section]
\theoremstyle{remark}
\newtheorem{remark}{Remark}%[section]

\theoremstyle{definition}

\renewcommand{\cite}{\citet}
\renewcommand{\cdots}{\dots}

\usepackage[onehalfspacing]{setspace}

\usepackage{geometry}

\usepackage{float}[H]

\usepackage{booktabs}
\hypersetup{allcolors=blue}

\begin{document}
\title{Optimal transport and risk measurement for order statistics}
\author{
Liyuan Lin\thanks%
 {Department of Econometrics and Business Statistics, 
 Monash University,
 Clayton, VIC, Australia.
 E-mail: \href{mailto:liyuan.lin@monash.edu}{liyuan.lin@monash.edu}.} \and
 Liulei Sun\thanks%
 {Department of Statistics and Actuarial Science,
 University of Waterloo,
 Waterloo, Ontario, Canada.
 E-mail: \href{mailto:l229sun@uwaterloo.ca}{l229sun@uwaterloo.ca}.} \and
 Ruodu Wang\thanks%
 {Department of Statistics and Actuarial Science,
 University of Waterloo,
 Waterloo, Ontario, Canada.
 E-mail: \href{mailto:wang@uwaterloo.ca}{wang@uwaterloo.ca}.}}
\date{}
\maketitle

\begin{abstract}

We study the  optimal transport (OT) problem for a function of an order statistic of a risk portfolio, as well as the corresponding   risk assessment problem under dependence uncertainty. 
They are special cases of multi-marginal OT and distorted OT problems.
For the maximum,
a tail counter-monotonic coupling and the comonotonic
coupling give the largest and smallest laws in stochastic order,
respectively. These couplings attain the distribution bounds at all
thresholds and thus provide sharp upper and lower bounds for monotone OT costs and
law-invariant and monotone risk measures, such as Value-at-Risk,
Expected Shortfall, and the mean.  
The corresponding problems for general order statistics are more complicated. For arbitrary marginals, we derive sharp
distribution bounds at each threshold and characterize when a single
coupling attains that sharp bounds at every threshold simultaneously.
These pointwise distribution bounds yield sharp Value-at-Risk bounds. For marginals with finite first moments, we further derive bounds for Expected Shortfall and the mean and characterize when the bounds are sharp. Finally, we highlight our theoretical results with a reserve capacity problem using flight disruption records from eight US airports.

\textbf{Keywords}: {multi-marginal optimal transport, order statistics, tail dependence, Expected Shortfall.} 
\end{abstract}

\clearpage

\section{Introduction}\label{sec:introduction}

Optimal transport  (OT) theory, initially studied by Monge and Kantorovich,  is widely used in operations research, statistics,
machine learning, quantitative finance and economics.   \citet{bib:ambrosio-2003}, \citet{bib:villani-2003,bib:villani-2009}, \citet{bib:galichon-2016} and \citet{bib:santambrogio-2015} provide  the theory and   applications of OT in various fields. 
We study a particular type of OT problem in which the cost is a function of an order statistic of random variables with given marginals. 
When the objective is the expected cost, this is a multi-marginal OT
problem \citep{bib:ruschendorf-2013, P15, bib:altschuler-boix-adsera-2023} on order statistics;
when the objective is a general risk measure, this is a  specific problem of risk aggregation with dependence uncertainty \citep{EPR13, EWW15} 
or distorted OT \citep{bib:liu-et-al-2026}, applied to order statistics.

Optimization problems with objectives that depend on order statistics arise widely in operations research and economics.
As a celebrated masterpiece of economic theory, in the
auction model of \citet{bib:vickrey-1961}, the winner pays the second-highest bid.
In discrete choice, \citet{bib:natarajan-song-teo-2009} maximize the expected
utility of the best alternative over joint laws with given marginals. In hypothesis testing, rank-based combination rules use the $k$-th smallest p-value   \citep{bib:ruger-1978,bib:vovk-wang-2021}.  In
delay propagation in airline networks
\citep{bib:vranas-et-al-1994,bib:arikan-et-al-2013}, the $k$-th
largest daily disruption   identifies days when at least $k$ airports exceed
a common threshold. For general ranks, moment information yields bounds on
the probability that at least $k$ events occur \citep{bib:prekopa-1990} and
on the expected $k$-th order statistic \citep{bib:bertsimas-natarajan-teo-2006}. 
Our paper studies the combination of  order statistics and OT.

For  $n\ge 2$ and a random vector $\mathbf{X}=(X_1,\ldots,X_n)$, let $X_{(k)}$ denote its $k$-th smallest order statistic. Given marginal distribution functions $F_1,\dots,F_n$, let $\mathcal F(F_1,\dots,F_n)$ be the set of all random vectors $\mathbf{X}$ with $X_i\sim F_i$ for $i\in \{1,\dots,n\}$. We consider two types of optimal transport problems for order statistics. The first objective evaluates the expected cost of an order statistic, and the second evaluates its risk under a law-invariant risk measure. To be more specific, we consider 
\begin{equation}\label{eq:intro-cost}
\inf_{\mathbf{X}\in \mathcal F(F_1,\dots,F_n)} \E\big[c(X_{(k)})\big]
\quad\text{and}\quad
\sup_{\mathbf{X}\in \mathcal F(F_1,\dots,F_n)} \E\big[c(X_{(k)})\big],
\end{equation}
where $c:\R\to \R$ is a measurable cost function, and
\begin{equation}\label{eq:intro-rm}
\inf_{\mathbf{X}\in \mathcal F(F_1,\dots,F_n)} \rho(c(X_{(k)}))
\quad\text{and}\quad
\sup_{\mathbf{X}\in \mathcal F(F_1,\dots,F_n)} \rho(c(X_{(k)})),
\end{equation}
where $\rho$ is a law-invariant risk measure, that is, it maps a random variable to a real number that  depends only on the distribution of the random variable. When $\rho$ is Value-at-Risk (VaR) or Expected Shortfall
(ES), a distorted expectation replaces the linear expectation, and
\eqref{eq:intro-rm} is a distorted OT problem \citep{bib:liu-et-al-2026}.

We first analyze the maximum and the minimum, the two extreme order
statistics. For this, we introduce  a new concept of $\mathbf{p}$-tail counter-monotonicity, where the large values of different risks never occur together. 
Theorem~\ref{thm:char} characterizes (multivariate) counter-monotonicity with $\mathbf{p}$-tail counter-monotonicity for all $\mathbf{p}$,  which complements the characterization of comonotonicity through common tail events in \citet[Theorem 4]{bib:wang-zitikis-2021}. Different from counter-monotonicity,   $\mathbf{p}$-tail counter-monotonicity (for a given $\mathbf p$) is compatible with any set of  marginals (Proposition~\ref{prop:exist}). Comonotonicity and $\mathbf{p}$-tail counter-monotonicity provide the lower and upper bounds on the maximum in stochastic order (Theorem~\ref{thm:max}), respectively.
These couplings
solve \eqref{eq:intro-cost} for every increasing $c$ and
\eqref{eq:intro-rm} for every monotone law-invariant risk measure $\rho$ with increasing $c$.

Next, we move to a general order statistic $X_{(k)}$. Since $X_{(k)}\le t$ if and only if at least $k$ of the events $\{X_i\le t\}$ occur, it holds that
\begin{equation}\label{eq:intro-pooled}
    \p(X_{(k)}\le t)
    \le 1\wedge\frac{1}{k}\sum_{i=1}^{n}F_i(t),
    \qquad t\in\R.
\end{equation}
For identical marginals, \eqref{eq:intro-pooled} is sharp
\citep{bib:caraux-gascuel-1992}. Otherwise,
\citet{bib:rychlik-1995} gives necessary and sufficient conditions under which one coupling attains \eqref{eq:intro-pooled} at every threshold. The difficulty is visible
even for the maximum of two risks. At any $t$ where $F_1(t)\ne F_2(t)$,
the sharp upper bound is $\min\{F_1(t),F_2(t)\}$, strictly below the
average in \eqref{eq:intro-pooled}. For an intermediate rank, a further
difficulty arises because the coupling attaining the sharp bound can
change with $t$.  Three questions then arise.
\begin{enumerate}[(a)]
\item What are the sharp bounds on the distribution function of $X_{(k)}$ at a fixed threshold?
\item When does one dependence structure attain this bound at all thresholds simultaneously?
\item If no such dependence structure exists, how can \eqref{eq:intro-cost} and \eqref{eq:intro-rm} be solved?
\end{enumerate}
To answer questions (a)--(c), we obtain several explicit results on the OT problems for order statistics, extending those in the literature. 
First, building on \citet{bib:ruger-1978} and \citet{bib:salako-2025}, we use layered arrangements to construct couplings with given marginals that attain the upper and lower bounds on $\p(X_{(k)}\le t)$ at
each fixed threshold $t$
(Proposition~\ref{prop:order}). These distribution bounds yield exact VaR
bounds for general marginals (Theorem~\ref{thm:var-attain}).

Second, for each of the two distribution bounds, we give necessary and
sufficient conditions on the given marginals for a single coupling to attain the bound at every threshold simultaneously (Theorem~\ref{thm:universal-attainment}
and Proposition~\ref{prop:upper}). Therefore, such a coupling attains the corresponding optimum of $X_{(k)}$ for every increasing cost and every monotone law-invariant risk measure.

Third, for marginals with finite first moments, we give bounds on ES and the
mean and characterize exactly when they are sharp (Theorems~\ref{thm:mean-bounds}--\ref{thm:es-bounds}). We also characterize all attainable laws
of $X_{(k)}$ for general marginals and show how to compute the
expected cost and ES bounds even when no coupling attains the distribution
bound at all thresholds simultaneously (Appendix~\ref{app:attainable-distributions}).
 
Answering these questions leads to many applications, and we present one
in the context of airport reserve capacity. Specifically, using flight
disruption records from eight US airports, we apply the sharp
distribution bound to determine the minimum shared reserve capacity
needed to meet a service target under dependence uncertainty. The application illustrates how assuming independence or comonotonicity
without justification can lead to insufficient reserve capacity.

The rest of the paper is organized as follows.
Section~\ref{sec:tail-counter-monotonicity} proposes
$\mathbf p$-tail counter-monotonicity as a new dependence concept and studies its properties. 
Section~\ref{sec:maximum}
 solves the OT problems for the two extreme order statistics. Section~\ref{sec:layered-arrangements} provides a technical step by constructing $k$-layered arrangements
of events with given probabilities that maximize the probability that at
least $k$ of them occur.
Section~\ref{sec:distribution-bounds} uses these arrangements to obtain
sharp upper and lower bounds on $\p(X_{(k)}\le t)$ at each fixed threshold $t$.
For each bound, Section~\ref{sec:stochastic-order} characterizes the
marginals for which a single coupling attains it at every threshold.  Section~\ref{sec:risk-measure-bounds} provides the corresponding risk measure bounds. Section~\ref{sec:airport-reserve-capacity} presents the empirical
application to airport reserve capacity, and
Section~\ref{sec:conclusion} concludes.
Appendix~\ref{app:attainable-distributions} characterizes the attainable laws and gives computational methods.

%Section \ref{sec:tail-counter-monotonicity} introduces $\bp$-tail counter-monotonicity. Section \ref{sec:maximum} gives the smallest and largest maximum in stochastic order, as well as sharp bounds for law-invariant and monotone risk measures. Section \ref{sec:layered-arrangements} introduces $k$-layered arrangements of events and the corresponding $k$-layered $\bp$-tail optimal dependence structures. Section \ref{sec:distribution-bounds} uses these structures to obtain sharp bounds on the distribution function of the k-th order statistic at each fixed threshold. Section \ref{sec:stochastic-order} gives necessary and sufficient conditions on the marginals under which a single coupling attains these bounds at all thresholds simultaneously. Section \ref{sec:risk-measure-bounds} derives VaR, ES, and mean bounds for general order statistics and characterizes when they are attained. Section \ref{sec:airport-reserve-capacity} applies our results to a reserve capacity problem using flight disruption records from eight US airports, and Section \ref{sec:conclusion} concludes. Appendix \ref{app:attainable-distributions} characterizes the attainable distributions of the order statistic and shows how to compute the bounds on the expected cost and ES for general marginals.

\paragraph{Related literature.} 
In the risk management literature, replacing $X_{(k)}$ in \eqref{eq:intro-cost} and \eqref{eq:intro-rm}
by the sum $X_1+\dots+X_n$ gives the classical risk aggregation problem
under dependence uncertainty; see
\citet[Section~8.4.4]{bib:mcneil-et-al-2015} and
\citet{bib:ruschendorf-2013}. 
Many results were developed for this problem. 
For the sum, comonotonic maximizes convex order,  a counter-monotonic coupling \citep{bib:cheung-lo-2014,bib:lauzier-et-al-2023}  or a joint mix \citep{bib:wang-wang-2016} minimizes  convex order, and VaR maximization or minimization involves a specific dependence structure \citep{bib:wang-et-al-2013, bib:blanchet-et-al-2025}.

In multi-marginal optimal transport,
\citet{bib:altschuler-boix-adsera-2023} develop polynomial-time
algorithms for classes of structured costs and apply their framework
to extremal network reliability. For nonlinear objectives,
\citet{bib:liu-et-al-2026} study how distorted expectations affect
optimal couplings in the two-marginal setting.
\citet{bib:ghossoub-hall-saunders-2023} establish a transport duality
for worst-case spectral risk with two given factor marginals.
\citet{bib:ennaji-et-al-2024} reformulate spectral risk maximization
with several given factor marginals as an ordinary multi-marginal optimal
transport problem with one additional marginal.

We consider two important risk measures, VaR and ES, which serve as popular objectives in robust portfolio optimization.
\citet{EOO03} study worst-case VaR with partial information on returns,
\citet{NPS08} incorporate asymmetric distributional information, and
\citet{ZF09} optimize worst-case conditional VaR. 
For overlapping discrete marginals satisfying the running intersection
property, \citet{bib:doan-li-natarajan-2015} give a linear programming
formulation for minimizing worst-case conditional VaR.
\citet{bib:mao-wang-wu-2026} relate worst-case VaR and ES to suprema
in first-order stochastic dominance and increasing convex order,
respectively.

% The applications encompass several distinct perspectives. In multiple testing, \citet{bib:vovk-et-al-2022} study the admissibility
% of merging rules for p-values under arbitrary dependence, and
% \citet{bib:gasparin-wang-ramdas-2025} develop exchangeable and randomized
% variants of R{\"u}ger's rank-based rule. In airline operations,
% \citet{bib:sohoni-et-al-2004} study reserve crew staffing for trips left
% uncovered by disruptions. At the network level, \citet{bib:lee-et-al-2020} use forecasts of
% airport congestion to guide recovery decisions in anticipation of
% further disruptions.

\section{Tail counter-monotonicity}\label{sec:tail-counter-monotonicity}

Fix an atomless probability space $(\Omega,\mathcal A,\p)$. We treat almost surely (a.s.)~equal random variables as identical; this means that 
all equalities and inequalities for random variables hold in the a.s.~sense, and we omit ``a.s." in all our statements.
For $p\in (0,\infty)$, denote by $L^p=L^p(\Omega,\mathcal A,\p)$ the set of all random variables $X$ with $\E[|X|^p]<\infty$ where $\E$ is the expectation under $\p$. Furthermore, $L^\infty=L^\infty(\Omega,\mathcal A,\p)$ is the space of all essentially bounded random variables, and $L^0=L^0(\Omega,\mathcal A,\p)$ is the space of all random variables. 
The set $\X$ is the set of all random variables on this space, thus, $\X=L^0$.
Terms like ``increasing" are in the non-strict sense. 
 Let $n$ be a positive integer  and $[n]=\{1,\dots,n\}$. Throughout,  
we consider $n\ge 2$.  Let $F^{-1}$ be the left inverse function of a distribution function $F: \R \to [0,1]$, defined as 
$F^{-1}(y)=\inf\{x \in \R: F(x)\ge y\}$.  We say $X$ and $Y$ are \emph{comonotonic} if there exist increasing functions $f$ and $g$ and a random variable $Z$ such that $X=f(Z)$ and $Y=g(Z)$. We say $X$ and $Y$ are \emph{counter-monotonic} if  $X$ and $-Y$ {are} comonotonic. Throughout this paper, we write a random vector as $\mathbf{X}=(X_1,\ldots,X_n)$. A random vector is comonotonic (resp. counter-monotonic) if each pair of its components is comonotonic (resp. counter-monotonic).

 A \emph{$p$-tail event} of a random variable $X$, introduced by \cite{bib:wang-zitikis-2021}, is an event $A\in \mathcal{A}$ such that $X(\omega)\geq X(\omega')$ holds for {$\p\otimes\p$-almost every $(\omega,\omega')\in A\times A^{c}$} and  $\p(A)=p$.  
Denote $$\Delta_n=\left\{(\theta_1,\dots,\theta_n)\in [0,1]^n:  \sum_{i=1}^n \theta_i=1\right\} 
~~\mbox{and}~~\Delta_n^{0}=\mathrm{conv}(\Delta_n,\mathbf{0})=\left\{(\theta_1, \dots, \theta_n) \in [0,1]^n: \sum_{i=1}^n \theta_i\le 1\right\}.$$
We first introduce a new dependence structure called \emph{$\mathbf{p}$-tail counter-monotonicity} for some $\mathbf{p}  \in \Delta_n^{0} $.
\begin{definition}[$\mathbf{p}$-tail counter-monotonicity]
A random vector $\mathbf{X}$ is said to be \emph{$\mathbf{p}$-tail counter-monotonic} for some $\mathbf{p}=(p_1, \dots, p_n) \in \Delta_n^{0}$ if there exist mutually exclusive events $A_1, \dots, A_n$ such that 
 $A_i$ is a $p_i$-tail event of $X_i$ for all $i \in [n]$. 
\end{definition}
%Let $\mathbf e_{1},\dots, \mathbf e_{n}$ be the column vectors of the $n\times n$ identity matrix. 
For $i \in [n]$, let $\mathbf{p}$ be a vector such that $p_i\in [0,1]$ and $p_j=0$ for $j \in [n]$ and $j \neq i$.  For this $\mathbf{p}$, every random vector is $\mathbf{p}$-tail counter-monotonic. Below, we present a nesting property of $\mathbf{p}$-tail counter-monotonicity. For two vectors $\mathbf{x} $ and $\mathbf{y} $, the inequality $\mathbf{x}\le \mathbf{y}$ is componentwise.

\begin{proposition}\label{prop:tail-nesting}
For $\mathbf{p},\mathbf{q} \in \Delta_n^{0}$  with  $\mathbf{q}\le \mathbf{p}$, $\mathbf{p}$-tail counter-monotonicity implies $\mathbf{q}$-tail counter-monotonicity.
\end{proposition}
\begin{proof}
Let $\mathbf{q}=(q_1, \dots, q_n), \mathbf{p}=(p_1, \dots, p_n)\in \Delta_n^{0}$ and  $\mathbf{X}$ be $\mathbf{p}$-tail counter-monotonic. There exist disjoint events $A_1, \dots, A_n$ such that
 $A_i$ is a $p_i$-tail event of $X_i$ for all $i \in [n]$. Since $A_i$ is a $p_i$-tail event of $X_i$,  there exists $B_i \subseteq A_i$ such that $B_i$ is a $q_i$-tail event for $X_i$. %{\color{blue}To obtain $B_i$, take the upper tail within $A_i$ and split any ties at its cutoff using atomlessness.} 
 Since $A_1, \dots, A_n$ are disjoint, $B_1, \dots, B_n$ are disjoint. Hence $\mathbf{X}$ is $\mathbf{q}$-tail counter-monotonic.
\end{proof}

\cite{bib:lauzier-et-al-2023} provide a stochastic representation for counter-monotonic random vectors with $n \ge 3$ based on mutual exclusivity. In the following proposition, we provide a similar stochastic representation for the $\mathbf{p}$-tail counter-monotonic random vector.
 Denote by $\mathcal X$ the set of all random variables.
% \begin{proposition}
% Let $\bp=(p_1, \dots, p_n) \in \Delta_n^{0}$. The following are equivalent.
% \begin{enumerate}[(i)]
% \item  $(X_1, \dots, X_n)$  is $\bp$-tail counter-monotonic.
% \item There exist disjoint events $A_1, \dots, A_n$ with $\p(A_i)=p_i$, $S_i, K_i \in \X$ with $S_i\ge K_i$, $i\in [n]$ such that
% $$X_i=S_i\id_{A_i}+K_i\id_{{\color{blue} A_i^c}},~~~\mbox{for all} ~~~i \in [n],$$
% {\color{blue}and for every $i$ with $0<p_i<1$,
% \[
%  \essinf_{A_i} S_i \ge \esssup_{A_i^c} K_i.
% \]}
% \end{enumerate}
% \end{proposition}
% \begin{proof}
% From (i) to (ii), by definition, we have disjoint events $A_1, \dots, A_n$ such that $A_i$ is a $p_i$-tail event of $X_i$.
% {\color{blue}If $0<p_i<1$, choose $c_i\in\R$ such that $X_i\ge c_i$ on $A_i$ and $X_i\le c_i$ on $A_i^c$, almost surely. Put $S_i=X_i\vee c_i$ and $K_i=X_i\wedge c_i$. Then $S_i\ge K_i$, the displayed representation holds, and the restricted essential bounds are separated by $c_i$. If $p_i$ is $0$ or $1$, take $S_i=K_i=X_i$.}
% From (ii) to (i), {\color{blue}the separation of the essential bounds ensures that $A_i$ is a $p_i$-tail event of $X_i$ when $0<p_i<1$. For $p_i\in\{0,1\}$, the tail condition is automatic.} By definition, $(X_1, \dots, X_n)$ is $\bp$-tail counter-monotonic.
% \end{proof}
\begin{proposition}
Let $\mathbf{p}=(p_1, \dots, p_n) \in \Delta_n^{0}$. The following are equivalent.
\begin{enumerate}[(i)]
\item  $\mathbf{X}$  is $\mathbf{p}$-tail counter-monotonic.
\item There exist disjoint events $A_1, \dots, A_n$ with $\p(A_i)=p_i$, $S_i, K_i \in \X$ with $\essinf_{A_i} S_i \ge \esssup_{A_i^c} K_i$, $i\in [n]$, such that 
$$X_i=S_i\id_{A_i}+K_i\id_{A_i^c},~~~\mbox{for all} ~~~i \in [n].$$
\end{enumerate}
\end{proposition}
\begin{proof}
From (i) to (ii), by definition, we have disjoint events $A_1, \dots, A_n$ such that  $A_i$ is a $p_i$-tail event of $X_i$. Let  $S_i=X_i\id_{A_i}$ and $K_i=X_i\id_{A_i^c}$. It is clear that $\essinf_{A_i} S_i \ge \esssup_{A_i^c} K_i$ and 
$S_i\id_{A_i}+K_i\id_{A_i^c}=X_i$. 
From (ii) to (i), it is clear that $A_i$ is a $p_i$-tail event of $X_i$ since $\essinf_{A_i} S_i \ge \esssup_{A_i^c} K_i$. By definition, we have $\mathbf{X}$ is $\mathbf{p}$-tail counter-monotonic.
\end{proof}

In the special case that $K_i$ are constants, $\mathbf{X}$ is counter-monotonic in the classic sense; see \cite{bib:cheung-lo-2014} and \cite{bib:lauzier-et-al-2023}. %The $\bp$-tail counter-monotonicity requires the tail events of $X_1, \dots, X_n$ to be mutually exclusive.
 In the next theorem, we further characterize counter-monotonicity with $\mathbf{p}$-tail counter-monotonicity.
For $\mathbf{p} \in \Delta_n$, the choice of $A_1, \dots, A_n$ is in the set of all $n$-compositions of $\Omega$, that is, $$\Pi_n=\left\{(A_1,\dots,A_n)\in \mathcal A^n: \bigcup_{i\in [n]} A_i=\Omega \mbox{~and~$A_1,\dots,A_n$ are disjoint}\right\}.$$
In other words, a composition of $\Omega$ is a partition of $\Omega$ with order.

\begin{theorem}\label{thm:char}
For an $n$-dimensional random vector $\mathbf{X}$, the following statements are equivalent.
\begin{enumerate}[(i)]
\item $\mathbf{X}$ is  counter-monotonic. 
\item $\mathbf{X}$ is $\mathbf{p}$-tail counter-monotonic for all $\mathbf{p}\in \Delta_n$.
\item $\mathbf{X}$ is $\mathbf{p}$-tail counter-monotonic for all $\mathbf{p}\in \Delta_n^{0}$.
\end{enumerate}
\end{theorem}
{\begin{proof}
The direction from (iii) to (ii) is trivial as $\Delta_n\subseteq \Delta_n^{0}$. 
 For any $\mathbf{p}_1 \in \Delta_n^{0}$, we can find $\mathbf{p}_2 \in \Delta_n$ such that $\mathbf{p}_1\le \mathbf{p}_2$. By Proposition~\ref{prop:tail-nesting}, we have (ii) to (iii).
 
We first show (i) to (ii). For $n=2$, assume $(X_1, X_2)$ is counter-monotonic. There exist a random variable $Z$, an increasing function $f$, and a decreasing function $g$ such that $X_1=f(Z)$ and $X_2=g(Z)$. For any $p \in [0,1]$, we can take $A$ as the $p$-tail event of $Z$. Since $f$ is increasing and $g$ is decreasing, we have $A$ is the $p$-tail event of $X_1$ and $A^c$ is the $(1-p)$-tail event of $X_2$, which implies $(X_1, X_2)$ is $(p, 1-p)$-tail counter-monotonic. For $n\ge 3$, assume $\mathbf{X}$ is counter-monotonic and $\mathbf{p}=(p_1, \dots, p_n) \in \Delta_n$. We consider the following two cases.
\begin{enumerate}
\item[(1)] Assume $X_i$, $i\in [n]\setminus[2]$ are constant. By the $n=2$ case, we have $(X_1, X_2)$ is $(p,1-p)$-tail counter-monotonic for all $p\in [0,1]$. By (ii) to (iii), we have $(X_1, X_2)$ is $(p_1,p_2)$-tail counter-monotonic for all $p_1, p_2\in [0,1]$ such that $p_1+p_2\le 1$. Since $X_i$ are constant for $i\in [n]\setminus[2]$, it is clear that $\mathbf{X}$ is $\mathbf{p}$-tail counter-monotonic. 
% $X_1=f(X)$ and $X_2=g(Z)$ for some random variable $Z$, increasing function $f$, and  decreasing function $g$. Let $A_1$ be the $p_1$-tail event of $Z$ and $A^c_2$ be the $(1-p_2)$-tail event of  $Z$. Since $p_1+p_2\le 1$, we can take $A_1 \cap A_2 =\varnothing$. Let $C=\Omega\setminus(A_1 \cup A_2)$ and $A_i \subseteq C$ for $i \in [n]\setminus C$ such that $A_i\cap A_j=\varnothing$ for all $i,j \in [n]\setminus[2]$ and $\p(A_i)=p_i$ for all $i \in [n]\setminus[2]$. Since $X_3, \dots, X_n$ are constant, it is clear that $A_i$ is $p_i$-tail event of $X_i$ for $i \in [n]\setminus[2]$.
% It is clear that $(A_1, \dots, A_n) \in \Pi_n$ and $A_i$ is the $p_i$-tail event of $X_i$. Therefore, $(X_1, \dots, X_n)$ is $\bp$-tail counter-monotonic. 
\item[(2)]
Assume at least three components of $\mathbf{X}$ are non-degenerate. By \citet[Theorem 1]{bib:lauzier-et-al-2023}, there exist $m_1, \dots, m_n \in \R$, $(A_1, \dots, A_n) \in \Pi_n$ and $Z \in \X_\pm$ such that $X_i=Z\id_{A_i}+m_i$ for all $i \in [n]$. First, we consider $Z \ge 0$. Without loss of generality, assume $p_i\le \p(A_i)$ for $i \in [k]$ and $p_i> \p(A_i)$ for $i \in [n]\setminus[k]$ for some $k \in [n]$.  For $i \in [k]$, let $B_i \subseteq A_i$ such that $B_i$ is a $p_i$-tail event of $Z\id_{A_i}$. Let $C=\bigcup_{i\in [k]} (A_i \setminus B_i)$ and 
$C_i \subseteq C$ such that $\p(C_i)=p_i-\p(A_i)$ for all $i \in [n]\setminus [k]$ and $C_i \cap C_j=\varnothing$ for all $i, j \in [n]\setminus[k]$. For $i\in [n]\setminus [k]$, let $B_i=A_i \cup C_i$. It is clear that $B_i$ is a $p_i$-tail event of $X_i$ for each $i \in [n]$ and $(B_1, \dots, B_n) \in \Pi_n$. 
Therefore, $\mathbf{X}$ is $\mathbf{p}$-tail counter-monotonic. 

Next, suppose $Z\leqslant 0$. If $p_i>\p(A_i^c)$ for some $i\in[n]$, then such $i$ is unique. Indeed, if  $p_i>1-\p(A_i)$ and $p_j>1-\p(A_j)$ for some $i\neq j$, then $p_i+p_j>2-\p(A_i)-\p(A_j)\ge 1$, which contradicts $\mathbf{p}\in \Delta_n$. We further consider the following two cases.

(a)   Suppose $p_n>\p(A^c_n)$ and $p_i\le \p(A^c_i)$ for all $i \in [n-1]$.
Let $D\subseteq A_n$ with $\p(D)=p_n-\p(A_n^c)$ such that $Z(\omega)\ge  Z(\omega')$ for  $\omega \in D$ and $\omega'\in A_n\setminus D$. Let $B_n=A_n^c\cup D$. As $X_n=m_n$ on $A_n^c$,  it is clear that $B_n$ is  $p_n$-tail event of $X_n$. Note that $\p(A_n\setminus D)=\p(A_n)-p_n+\p(A_n^c)=1-p_n=\sum_{i\in [n-1]} p_i$. We can  partition $A_n\setminus D$ into $B_1,\dots,B_{n-1}$ such that $\p(B_i)=p_i$, $i\in[n-1]$. Since $B_i\subseteq A_n\subseteq A_i^c$, the event $B_i$ is a $p_i$-tail event of $X_i$ for $i\in[n-1]$, which implies $\mathbf{X}$ is $\mathbf{p}$-tail counter-monotonic.

(b) Suppose that $p_i\le \p(A_i^c)$ for all $i\in[n]$. For each $j\in[n]$ we  partition $A_j$ into disjoint events $E_{ij}$, $i\in[n]$, with $\p(E_{ij})=x_{ij}$ such that $x_{ii}=0$, $\sum_{i\in [n]} x_{ij}=\p(A_j)$ and $\sum_{j \in [n]} x_{ij}=p_i$. Let $B_i=\bigcup_{j\in [n]} E_{ij}$. Then $(B_1,\dots,B_n)\in\Pi_n$, $\mathbb P(B_i)=p_i$, and $B_i\subseteq\bigcup_{j\neq i}A_j=A_i^c$ as $E_{ii}=\emptyset$. Hence, $B_i$ is a $p_i$-tail event of $X_i$ for all $i\in[n]$ and $B_1, \dots, B_n$ are disjoint, which implies $\mathbf{X}$ is $\mathbf{p}$-tail counter-monotonic.

We find such $\{x_{ij}\}_{i,j\in [n]}$ by
considering a supply-demand problem in which supplier $j\in[n]$ has supply $\p(A_j)$, customer $i\in[n]$ has demand $p_i$, and supplier $j$ ships  $x_{ij}$ to  customer $i$ with the condition that supplier $i$ cannot serve customer $i$. By the supply-demand theorem in \cite{bib:gale-1957}, a feasible shipment $\{x_{ij}\}_{i, j \in [n]}$ exists if and only if Hall's condition
\[
\sum_{i\in S}p_i\leqslant \sum_{j\in N(S)}\p(A_j)
\]
holds for all $S \subseteq [n]$ where $N(S)$ is the set of all suppliers that can serve at least one customer in $S$. If $S=\{i\}$, we have $N(S)=[n]\setminus\{i\}$ and  Hall's condition holds as  $p_i\le \p(A_i^c)=\sum _{j \in N(\{i\})}\p(A_j)$. If  $|S|\geqslant 2$, we have $N(S)=[n]$ and   Hall's condition holds as $\sum_{i\in S} p_i\le 1$. 
\end{enumerate}
To prove that (ii) implies (i), fix $i\ne j$ and cutoffs $x,y$. Put
$u=\p(X_i>x)$ and $v=\p(X_j>y)$. 
By (ii), we have $(X_i, X_j)$ is $(u, 1-u)$-tail counter-monotonic. Since a $u$-tail event of
$X_i$ coincides with $A=\{X_i>x\}$  a.s. Consequently,
$A^c$ is a $(1-u)$-tail event of $X_j$. The event $\{X_j>y\}$ is
contained in $A^c$ when $v\le 1-u$ and contains $A^c$ when $v\ge 1-u$.
Hence,
$
 \p(X_i>x,X_j>y)=(u+v-1)_+,
$
which implies $X_i$ and $X_j$ are counter-monotonic. Therefore, every pair of components of $\mathbf{X}$ is counter-monotonic.
\end{proof}}

\begin{remark}
One can compare Theorem~\ref{thm:char} with \citet[Theorem 4]{bib:wang-zitikis-2021}, in which comonotonicity is characterized by $p$-concentration; that is, the components of the comonotonic random vector $\mathbf{X}$ share a common tail event of probability $p$ for all $p\in (0,1)$.  {Theorem~\ref{thm:char} shows that a counter-monotonic random vector admits a choice of mutually exclusive tail events for every $\mathbf{p}\in\Delta_n^{0}$.}
\end{remark}

Different from counter-monotonicity, tail counter-monotonicity does not guarantee negative dependence in general. 
For a fixed $\mathbf{p}\in\Delta_n^{0}$, $\mathbf{p}$-tail counter-monotonicity does not   guarantee negative dependence in general; for instance, negative association, negative orthant dependence, and negative supermodular dependence do not necessarily hold. The $\mathbf{p}$-tail counter-monotonicity only guarantees mutual exclusivity of the specified tail events.

In the following proposition, we show that the tail counter-monotonicity is invariant under strictly increasing transformations of the marginal distributions. Such an invariance property holds for many concepts of dependence, such as comonotonicity, $p$-concentration, counter-monotonicity, positive association, negative association, and others.

\begin{proposition}\label{prop:increase}
Let $\mathbf{p} \in \Delta_n^{0}$. If $\mathbf{X}$ is $\mathbf{p}$-tail counter-monotonic, then $(f_1(X_1), \dots, f_n(X_n))$ is  $\mathbf{p}$-tail counter-monotonic for all increasing functions $f_1, \dots, f_n$.
\end{proposition}
\begin{proof}
By definition, we can find disjoint events $A_1, \dots, A_n$ such that  $A_i$ is a $p_i$-tail event of $X_i$. Fix $i \in [n]$. Since $f_i$ is increasing, for {almost all pairs} $\omega \in A_i$ and $\omega' \in A_i^c$, we have $f_i(X_i(\omega))\ge f_i(X_i(\omega'))$, which implies $A_i$ is a $p_i$-tail event of {$f_i(X_i)$}. By definition, {$(f_1(X_1), \dots, f_n(X_n))$} is $\mathbf{p}$-tail counter-monotonic.
\end{proof}

It is well known that the counter-monotonic copula does not exist for $n \ge 3$. 
As a consequence, 
not all marginal distributions allow for a counter-monotonic random vector; see \cite{bib:puccetti-wang-2015} for a review of this fact. In contrast, $\mathbf{p}$-tail counter-monotonic random vectors always exist for any $\mathbf{p}\in \Delta_n^{0}$ and any marginal distributions.

\begin{proposition}
\label{prop:exist}
For any $\mathbf{p}\in \Delta_n^{0}$ and any given marginal distributions $F_1,\dots,F_n$, there exists a $\mathbf{p}$-tail counter-monotonic random vector with the specified marginal distributions.
\end{proposition}

\begin{proof} 
Fix $\mathbf{p}=(p_1,\dots,p_n)\in \Delta_n^{0}$. 
Let
\begin{equation}\label{eq:copula}
U_i=\left(1-p_i+p_iV_i\right)\id_{A_i}+(1-p_i)V_i\id_{A_i^c}, \qquad i \in [n],
\end{equation}
where $ V_1, \dots, V_n \sim \mathrm{U}[0,1]$ and $A_1,\dots,A_n$ are disjoint events such that  $V_1, \dots, V_n$ are independent of $(A_1, \dots, A_n)$ and $\mathbb{P}(A_i)=p_i$ for all $i\in [n]$. It is clear that $A_i$ is the $p_i$-tail event of $U_i$. Hence, the distribution of $(U_1, \dots, U_n)$ is a $\mathbf{p}$-tail counter-monotonic copula.
By taking $X_i^{\rm tc}=F_i^{-1}(U_i)$ for $i \in [n]$ and applying Proposition~\ref{prop:increase}, we can get a $(p_1, \dots, p_n)$-tail counter-monotonic $(X_1^{\rm tc},\dots, X_n^{\rm tc}) $ with the given marginals.
\end{proof}

\section{Maximum of dependent random variables}\label{sec:maximum}

For random variables $X$ and $Y$, $X$ is smaller than $Y$ in \emph{stochastic order}, denoted by $X\le_{\rm st}Y$, if $\p(X\le t)\ge \p(Y\le t)$ for all $t\in\R$. 
We refer to \cite{bib:muller-stoyan-2002} and  \cite{bib:shaked-shanthikumar-2007} for properties of stochastic order. 
In this section, we show that tail counter-monotonicity provides an upper bound on the maximum of a portfolio of risks in the stochastic order.

Let $\mathcal M$ be the set of cdfs on $\R$. For  $F_1, \dots, F_n\in \mathcal{M}$, define the set
$$
\mathcal F (F_1, \dots, F_n) =\{\mathbf{X}=(X_1,\ldots,X_n): X_i\sim F_i, ~i \in [n]\}.
$$ 
Denote $(x)_+=\max\{0, x\}$. By the Fr\'echet inequalities, it is well known that the distributions for $\mathbf{X} \in \mathcal{F}(F_1, \dots, F_n)$ satisfy
\begin{equation}\label{eq:frechet}
\left(\sum_{i=1}^n F_i(x_i)-(n-1)\right)_+  \le \p(X_1\le x_1, \dots, X_n \le x_n) \le \min\left\{F_1(x_1), \dots, F_n(x_n)\right\}.
\end{equation}
It is well known that the upper bound is the {joint distribution function} for comonotonic $(X_1^{\rm co}, \dots, X_n^{\rm co}) \in \mathcal{F}(F_1, \dots, F_n)$, but the lower bound is not necessarily a distribution when $ n \ge 3$. 
We now introduce an intermediate quantity $\tau\in\R$ that is useful in the identification of extremal dependence scenarios. 
\begin{proposition}
For any $F_1,\dots,F_n\in\mathcal M$, there exist $\tau\in\mathbb R$
and $\mathbf{p}=(p_1,\dots,p_n)\in\Delta_n$ such that
\begin{equation}\label{eq:exist-tau}
    \sum_{i=1}^n F_i(\tau-)\le n-1\le \sum_{i=1}^n F_i(\tau)
    ~~~\mbox{and}~~~
    F_i(\tau-)\le 1-p_i\le F_i(\tau)
    \quad\text{for all }i\in[n].
\end{equation}
\end{proposition}

\begin{proof}
Let $\tau=\inf\{t\in\mathbb R:\sum_{i=1}^n F_i(t)\ge n-1\}$.
By monotonicity and right-continuity of cdfs, $\sum_{i=1}^n F_i(\tau-)\le n-1\le \sum_{i=1}^n F_i(\tau)$ holds.
Thus we can choose $a_i\in[F_i(\tau-),F_i(\tau)]$ for $i\in[n]$
such that $\sum_{i=1}^n a_i=n-1$. Set $p_i=1-a_i$, $i\in[n]$. This proves
$\mathbf{p}\in\Delta_n$ and \eqref{eq:exist-tau}.
\end{proof}

In the next result, we find a $\mathbf{p}$-tail counter-monotonic vector $(X^{\rm tc}_1, \dots, X^{\rm tc}_n)$ that attains the lower bound in \eqref{eq:frechet} when $x_1=\dots=x_n=\tau$, {which is referred to as ``maximally dependent'' by \cite{bib:lai-robbins-1976}.}
\begin{theorem}\label{thm:max}
Suppose $F_1, \dots, F_n \in \mathcal M$,   $\mathbf{p}\in \Delta_n$,  
and $\tau\in \mathbb{R}$ satisfy \eqref{eq:exist-tau}. 
Let $(X_1^{\rm tc},\dots, X_n^{\rm tc})\in \mathcal F(F_1, \dots, F_n)$ be $\mathbf{p}$-tail counter-monotonic and $(X_1^{\rm co},\dots, X_n^{\rm co})\in \mathcal F(F_1, \dots, F_n)$ be comonotonic. We have 
$$\max\{X_1^{\rm co}, \dots, X_n^{\rm co}\} \le_{\rm st} \max\{X_1, \dots, X_n\}  \le_{\rm st} \max\{X_1^{\rm tc}, \dots, X_n^{\rm tc}\}$$
for all $\mathbf{X} \in\mathcal F(F_1,\dots,F_n)$.
\end{theorem}

\begin{proof}
For any $t \in \R$, by \eqref{eq:frechet}, we have
\begin{align*}\p\left(\max\{X_1^{\rm co}, \dots, X_n^{\rm co}\}\le t\right)& =\p\left(X_1^{\rm co}\le t, \dots, X_n^{\rm co}\le t\right)= \min\left\{F_1(t), \dots, F_n(t)\right\} \\
& \ge \p\left(X_1\le t, \dots, X_n\le t\right)=\p\left(\max\{X_1, \dots, X_n\}\le t\right).
\end{align*}
Hence, we have the first stochastic order.

Since $\mathbf{p} \in \Delta_n$, there exists $(A_1, \dots, A_n) \in \Pi_n$ such that $A_i$ is a $p_i$-tail event of ${X_i^{\rm tc}}$ for all $i \in [n]$. {Since $1-F_i(\tau)\le p_i\le 1-F_i(\tau-)$, these tail events satisfy $\{X_i^{\rm tc}>\tau\}\subseteq A_i\subseteq\{X_i^{\rm tc}\ge\tau\}$ up to null sets. For $t<\tau$, it follows that $\{X_i^{\rm tc}\le t\}\subseteq A_i^c$.} Hence, 
$$\p\left(X_1^{\rm tc}\le t, \dots X_n^{\rm tc}\le t\right)\le \p\left(\bigcap_{i=1}^n A_i^c\right)=1-\p\left(\left(\bigcap_{i=1}^n A_i^c\right)^c\right)=1-\p\left(\bigcup_{i=1}^n A_i\right)=0.$$
Moreover, we have $F_i(t) \le F_i(\tau-)\le 1-p_i$ for all $i \in [n]$. That is, 
$$\p\left(X_1^{\rm tc}\le t, \dots X_n^{\rm tc}\le t\right)=0=\left(\sum_{i=1}^n F_i(t)-(n-1)\right)_+.$$
For $t \ge \tau$, since the tail events of $X_1, \dots, X_n$ are mutually exclusive, we have $A_i \subseteq \{X_j^{\rm tc} \le \tau\}\subseteq \{X_j^{\rm tc} \le t\}$ for all $i \in [n]$, $j \in [n]$ and $j \neq i$. Let $B=\bigcap_{i=1}^n \{ X_i^{\rm tc}\le t\}$. As a result, we have $B \cap A_i=\{ X_i^{\rm tc}\le t\} \cap A_i$ for all $i \in [n]$.  Hence,
$$\p\left(X_1^{\rm tc}\le t, \dots X_n^{\rm tc}\le t\right)=\sum_{i=1}^n\p\left(B\cap A_i\right)=\sum_{i=1}^n \p\left(\{ X_i^{\rm tc}\le t\} \cap A_i\right).$$
It is clear that $A^c_i \cap \{X_i^{\rm tc}>t\}=\varnothing$. Hence,
$$\p\left(\{ X_i^{\rm tc}\le t\} \cap A_i\right)=1-\p\left(\{ X_i^{\rm tc}> t\} \cup A^c_i\right)=1-(1-F_i(t))-(1-p_i)=F_i(t)+p_i-1.$$
Therefore, 
$$\p\left(X_1^{\rm tc}\le t, \dots X_n^{\rm tc}\le t\right)=\sum_{i=1}^n\left(F_i(t)+p_i-1\right)=\left(\sum_{i=1}^n F_i(t)-(n-1)\right)_+.$$
By \eqref{eq:frechet}, we have 
\begin{align*}
\p\left(\max\{X_1^{\rm tc}, \dots, X_n^{\rm tc}\}\le t \right)&=\p\left(X_1^{\rm tc}\le t, \dots, X_n^{\rm tc}\le t\right)
=\left(\sum_{i=1}^n F_i(t)-(n-1)\right)_+ \\
&\le \p(X_1\le t, \dots, X_n \le t)=\p\left(\max\{X_1, \dots, X_n\}\le t \right),
\end{align*}
which gives the second stochastic order.
\end{proof}

\begin{remark}\label{rem:distribution}
By the proof of Theorem~\ref{thm:max}, we have the universal upper bound and lower bound for the distribution of  $\max\{X_1, \dots, X_n\}$ for $\mathbf{X} \in \mathcal{F}(F_1, \dots, F_n)$. That is, for all $t\in \R$,
$$ \left(\sum_{i=1}^n F_i(t)-(n-1)\right)_+\le \p\left(\max\{X_1, \dots, X_n\}\le t\right)  \le \min\{F_1(t), \dots, F_n(t)\},$$
where the lower bound is attained by $(X_1^{\rm tc}, \dots, X_n^{\rm tc})$, and the upper bound is attained by $(X_1^{\rm co}, \dots, X_n^{\rm co})$.
\end{remark}

A risk measure $\rho$ is a mapping  from $\X_\rho$ to $\R$, where $\X_\rho$ is  a convex cone of random variables  on  $(\Omega,\mathcal A,\p)$ representing losses faced by a financial institution or an investor, and $\X_\rho$ is assumed to include all constants (i.e., degenerate random variables). 
For a risk measure $\rho$, we assume that its domain $\X_\rho\subseteq L^0$ is closed under equality in distribution, that is,
$X\in\X_\rho$, $Y\in L^0$, and $X\laweq Y$ imply $Y\in\X_\rho$.
A risk measure is \emph{monotone} if $\rho(X)\le \rho (Y)$ for all $X\le Y$. A risk measure is \emph{law-invariant} if $\rho(X)= \rho (Y)$ whenever $X\laweq Y$.

Consider the dependence uncertainty bounds for the maximum of risks, where the marginal distributions are known, but the dependence structure is not, measured by $\rho$. 
Theorem~\ref{thm:max} provides the upper bound and lower bound for the dependence uncertainty problem when $\rho$ is law-invariant and monotone. 
\begin{corollary}\label{cor:maximum-risk-bounds}
Suppose that $\rho$ is law-invariant and monotone. For $F_1,\dots,F_n\in \mathcal M$, we have
\begin{align*}
\sup_{\mathbf X\in \mathcal F(F_1, \dots,F_n)}\rho\left(\max\{X_1, \dots, X_n\}\right)&=\rho(\max\{X_1^{\rm tc}, \dots, X_n^{\rm tc}\}); \\  \inf_{\mathbf X\in \mathcal F(F_1, \dots,F_n)}\rho\left(\max\{X_1, \dots, X_n\}\right)&=\rho(\max\{X_1^{\rm co}, \dots, X_n^{\rm co}\})
\end{align*}
for $(X_1^{\rm tc},\dots, X_n^{\rm tc})$ and $(X_1^{\rm co},\dots, X_n^{\rm co})$ in Theorem~\ref{thm:max}.
\end{corollary}
\begin{proof}
Law-invariance and monotonicity of $\rho$ imply $\rho(X)\le \rho(Y)$ if $X \le_{\rm st } Y$. By Theorem~\ref{thm:max}, we have 
$$\rho\left(\max\{X_1^{\rm co}, \dots, X_n^{\rm co}\}\right) \le \rho\left( \max\{X_1, \dots, X_n\}\right)  \le \rho \left( \max\{X_1^{\rm tc}, \dots, X_n^{\rm tc}\}\right)$$
for all $\mathbf{X} \in\mathcal F(F_1,\dots,F_n)$, which completes the proof.
\end{proof}
The two popular classes of risk measures in banking and insurance practice are VaR and ES, which are monotone and law-invariant.  The VaR at level $\alpha \in [0,1)$ is defined as the left $(1-\alpha)$-quantile, that is, 
$$
\VaR_\alpha(X)=\inf\{x\in \R: \p(X\le x) \ge 1-\alpha\},~~~X\in L^0.
$$
For $\alpha \in (0,1]$, we also write $\VaR^+_\alpha$ for the right $(1-\alpha)$-quantile, that is,
$$
\VaR^+_\alpha(X)=\inf\{x\in \R: \p(X\le x) > 1-\alpha\},~~~X\in L^0.
$$
The ES (also called CVaR, TVaR or AVaR) at level $\alpha \in (0,1)$ is defined as
$$
\ES_{\alpha}(X) = \frac 1 \alpha \int_{0}^\alpha \VaR_\beta(X) \d \beta,~~~X\in L^1,
$$
and  $\ES_0(X)=\esssup(X)=\VaR_0(X)$, which may be $\infty$. Note that $\ES_1(X)=\E[X]$ holds for $X\in L^1$.
Range Value-at-Risk (RVaR) is a formal generalization and extension of ES. For $0\le p<q\le 1$, The RVaR over the interval $[p,q]$ is defined as
$$\RVaR_{p,q}(X)=\frac{1}{q-p}\int_p^q \VaR_\beta(X) \d\beta,~~~X\in L^1.$$
We calculate the bounds for VaR, ES, and the mean when the marginal distributions are identical.

\begin{proposition}\label{prop:bounds}
Let $F \in \mathcal{M}$ and $X \sim F$. For each $\alpha\in(0,1)$, we have
$$\sup_{\mathbf{X}\in\mathcal F(F,\ldots,F)}\VaR_\alpha(\max\{X_1, \dots, X_n\})=\VaR_{\alpha/n}(X),~~~\inf_{\mathbf{X}\in\mathcal F(F,\ldots,F)}\VaR_\alpha(\max\{X_1, \dots, X_n\})=\VaR_{\alpha}(X).$$
Moreover, if $X\in L^1$, we have
$$\sup_{\mathbf{X}\in\mathcal F(F,\ldots,F)}\ES_\alpha(\max\{X_1, \dots, X_n\})
=\ES_{\alpha/n}(X),~~~\inf_{\mathbf{X}\in\mathcal F(F,\ldots,F)}\ES_\alpha(\max\{X_1, \dots, X_n\})=\ES_{\alpha}(X),$$
$$\sup_{\mathbf{X}\in\mathcal F(F,\ldots,F)}\E(\max\{X_1, \dots, X_n\})=\ES_{1/n}(X),~~~\mbox{and}~~~\inf_{\mathbf{X}\in\mathcal F(F,\ldots,F)}\E(\max\{X_1, \dots, X_n\})=\E[X].$$
\end{proposition}

\begin{proof}
Since $\VaR$, $\ES$ and $\E$ are law-invariant and monotone, by
Corollary~\ref{cor:maximum-risk-bounds}, the upper bound is attained by $(X_1^{\rm tc}, \dots, X_n^{\rm tc})$ and  the lower bound is attained by $(X_1^{\rm co}, \dots, X_n^{\rm co})$.
By Remark~\ref{rem:distribution}, we have 
$$\p(\max\{X_1^{\rm tc}, \dots, X_n^{\rm tc}\}\le t)=(nF(t)-(n-1))_+~~~\mbox{and} ~~~\p(\max\{X_1^{\rm co}, \dots, X_n^{\rm co}\}\le t)= F(t).$$
Hence, the three lower bounds are straightforward. For the upper bound, we further have
\begin{align*}
\VaR_\alpha(\max\{X_1^{\rm tc}, \dots, X_n^{\rm tc}\}) 
=\inf\{x\in\R:(nF(x)-n+1)_+\ge 1-\alpha\}=\inf\left\{x\in\R:F(x)\ge 1-\frac{\alpha}{n}\right\}.
\end{align*}
Hence, $\VaR_\alpha(\max\{X_1^{\rm tc}, \dots, X_n^{\rm tc}\}) =\VaR_{\alpha/n}(X)$, which implies $$\sup_{\mathbf{X}\in\mathcal F(F,\ldots,F)}\VaR_\alpha(\max\{X_1, \dots, X_n\})=\VaR_\alpha(\max\{X_1^{\rm tc}, \dots, X_n^{\rm tc}\})=\VaR_{\alpha/n}(X),$$ 
and
\begin{align*}
\sup_{\mathbf{X}\in\mathcal F(F,\ldots,F)}\ES_\alpha(\max\{X_1, \dots, X_n\})
=\ES_\alpha(\max\{X_1^{\rm tc}, \dots, X_n^{\rm tc}\}) 
=\frac1\alpha \int_0^\alpha \VaR_{\beta/n}(X)\,\d\beta
=\ES_{\alpha/n}(X).
\end{align*}
The result for the mean directly holds because $\E=\ES_1$.
\end{proof}
\begingroup
If the cost function  $c$ is increasing, then the comonotonic $(X_1^{\rm co},\dots, X_n^{\rm co})$ and $\mathbf{p}$-tail counter-monotonic $(X_1^{\rm tc},\dots, X_n^{\rm tc})$ give the lower bound and upper bound for the expected cost $\E[c(\max\{X_1, \dots, X_n\})]$, respectively, whenever the expectations exist. For a distortion $g:[0,1]\to[0,1]$ that is increasing with $g(0)=0$ and $g(1)=1$, the same conclusion applies to the distorted expectation
\[
 \rho_g(Y)=\int_0^\infty g(\p(Y>t))\,\d t+
 \int_{-\infty}^0\bigl(g(\p(Y>t))-1\bigr)\,\d t
\]
 whenever  $\rho_g$ is finite. 
%In Section \ref{sec:3}, we will see that the universal optimizers might not exists for  general order statistics $X_{(k)}$.
% Section \ref{sec:4}, we distinguish when the universal optimizers can be extended for general order statistics $X_{(k)}$.
% {\color{red}I am not sure whether the above two sentences are necessary or not.}

\endgroup

\begin{remark}\label{rem:min}
It is easy to get the upper bound and lower bound for $\min\{X_1, \dots, X_n\}$ in stochastic order by taking $\min\{X_1, \dots, X_n\}=-\max\{-X_1, \dots, -X_n\}$. Let $(-\widetilde X_1^{\rm tc}, \dots,-\widetilde X_n^{\rm tc}) \in \mathcal{F}(\widehat F_1, \dots, \widehat F_n)$ be a $\mathbf{p}$-tail counter-monotonic vector as constructed in Theorem~\ref{thm:max}  where $ \widehat F_i$ is the distribution of $-X_i$, $i \in [n]$. Then, 
$$\min\{{\widetilde X_1^{\rm tc}}, \dots, {\widetilde X_n^{\rm tc}}\} \le_{\rm st} \min\{X_1, \dots, X_n\} \le_{\rm st}  \min\{X_1^{\rm co}, \dots, X_n^{\rm co}\}.$$
Hence, for a law-invariant and monotone risk measure $\rho$, we have 
$$\rho\left(\min\{{\widetilde X_1^{\rm tc}}, \dots, {\widetilde X_n^{\rm tc}}\} \right)\le  \rho\left(\min\{X_1, \dots, X_n\}\right) \le \rho\left( \min\{X_1^{\rm co}, \dots, X_n^{\rm co}\}\right).$$
For the case that $ X, X_1, \dots, X_n \sim F$, using Proposition~\ref{prop:bounds}, we have
$$\VaR_{(n-1+\alpha)/n}(X)\le \VaR_{\alpha}\left(\min\{X_1, \dots, X_n\}\right)\le \VaR_\alpha(X)$$
for all $\mathbf{X} \in \mathcal{F}(F, \dots, F)$.
Moreover, if $X\in L^1$, we have
$$\RVaR_{(n-1)/n,(\alpha+n-1)/n}(X)\le \ES_{\alpha}\left(\min\{X_1, \dots, X_n\}\right)\le \ES_\alpha(X)$$
and 
$$\RVaR_{(n-1)/n,1}(X)\le \E\left(\min\{X_1, \dots, X_n\}\right)\le \E[X]$$
for all $\mathbf{X} \in \mathcal{F}(F, \dots, F)$.
\end{remark}

\section{Technical tool: Layered arrangements}\label{sec:layered-arrangements}

To study general order statistics, we next introduce the
arrangements of $n$ events with given probabilities that
maximize the probability that at least $k$ occur. We construct such
arrangements by folding intervals into layers and use them in
Section~\ref{sec:distribution-bounds} to obtain sharp bounds on the
distribution of order statistics.

For $n$ events $A_1, \dots, A_n$ with $p_i=\p(A_i)$ for $i \in [n]$ such that $p_1\le \dots \le p_n$, the bounds for the probability that at least $k$ of the events $A_1, \dots, A_n$  occur have been provided in \cite{bib:ruger-1978} and \cite{bib:salako-2025},
\begin{equation}\label{eq:event-count-bounds}
\left(1-\min_{0\leq r\leq n-k}\frac{\sum_{j=r+1}^{n}(1-p_{j})}{n-k+1-r}\right)_+
\le \p\left(\sum_{i=1}^n \id_{A_i}\ge k\right)
\le 1\wedge \left(\min_{0\leq r\leq k-1}\frac{\sum_{j=1}^{n-r}p_{j}}{k-r}\right).
\end{equation}
Let $k\in[n]$, $\mathbf{p}=(p_1,\ldots,p_n)\in[0,1]^n$ and $p_{(1)}\le\cdots\le p_{(n)}$ be the order statistics of $\mathbf{p}$. Define
\begin{equation}\label{eq:bk-def}
    b_k(\mathbf{p})=\sup\left\{b\in[0,1]:\sum_{i=1}^n(p_i\wedge b)\ge kb\right\}.
\end{equation}
We first show that the quantity $b_k(\mathbf{p})$ is the sharp upper bound in \eqref{eq:event-count-bounds}.
\begin{lemma}\label{lem:bk-closed}
For $k\in[n]$ and $\mathbf{p}\in[0,1]^n$, we have 
$$b_k(\mathbf{p})=1\wedge \left(\min_{0\le r\le k-1}\frac{\sum_{j=1}^{n-r}p_{(j)}}{k-r}\right).$$
Moreover, $b_k(\mathbf{p})=1$ if and only if $\sum_{i=1}^n p_i\ge k$.
\end{lemma}

\begin{proof}
Let $a=1\wedge\min_{0\le r\le k-1}\bigl(\sum_{j=1}^{n-r}p_{(j)}\bigr)/(k-r)$. For any  $b\in[0,1]$ such that  $\sum_{i=1}^n (p_i\wedge b)\ge kb$, we have 
\begin{equation*}
    kb\le\sum_{i=1}^n(p_i\wedge b)\le \sum_{j=1}^{n-r}p_{(j)}+rb
\end{equation*}
for all $r\in\{0,\ldots,k-1\}$.
Hence, $b\le \bigl(\sum_{j=1}^{n-r}p_{(j)}\bigr)/(k-r)$ for all $0\le r\le k-1$. It is clear that $b_k(\mathbf{p})\le 1$. Thus, $b_k(\mathbf{p})\le a$.

%It remains to show that $a$ is feasible. 
Let $r_a=\vert\{i:p_i>a\}\vert$. If $r_a\ge k$, then $\sum_{i=1}^n (p_i\wedge a)\ge r_a a\ge ka$. If $r_a\le k-1$, then $p_{(i)}\le a$ for $i \in {[n-r_a]}$. By definition of $a$, we have $(k-r_a)a\le \sum_{j=1}^{n-r_a} p_{({j})}$. Therefore,
\begin{equation*}
    \sum_{i=1}^n(p_i\wedge a)=\sum_{j=1}^{n-r_a}p_{(j)}+r_a a\ge(k-r_a)a+r_a a=ka.
\end{equation*}
Thus, $b_k(\mathbf{p})\ge a$. In conclusion, we have  $b_k(\mathbf{p})= a$.

Finally, if $b_k(\mathbf{p})=1$, $\sum_{i=1}^n p_i/k\ge 1$ follows from setting $r=0$. Hence, $\sum_{i=1}^n p_i\ge k$, which implies $\sum_{i=1}^n (p_i/b \wedge 1)\ge \sum_{i=1}^n p_i\ge k$ for all $b \in (0,1]$. Thus, $\sum_{i=1}^n (p_i\wedge b)\ge kb$ for all $b \in [0,1]$ and $b_k(\mathbf{p})=1$.
\end{proof}

Next, we define the arrangement of $A_1, \dots, A_n$ that achieves the sharp upper bound $b_k(\mathbf{p})$.
\begin{definition}\label{def:k-layered-arrangement}
Let $\mathbf{p}\in [0,1]^n$ and $A_1,\ldots,A_n\in\mA$ such that 
 $\p(A_i)=p_i$. Denote $N_A=\sum_{i=1}^n \id_{A_i}$. We say that $A_1,\ldots,A_n$ form a \emph{$k$-layered $\mathbf{p}$-arrangement} if the following conditions hold.

\begin{enumerate}[(i)]
\item Non-saturated case. If  $b_k(\mathbf{p})<1$, then there exists an event $B_{\mathbf{p}}\in\mA$ with $\mathbb P(B_{\mathbf{p}})=b_k(\mathbf{p})$ such that $N_A=k$ on $B_{\mathbf{p}}$ and $N_A<k$ on $B_{\mathbf{p}}^c$. %$A_i\subseteq B_{\bq}$ whenever $q_i\le b_k(\bq)$ and $B_{\bq}\subseteq A_i$ whenever $q_i> b_k(\bq)$.

\item Saturated case. If $b_k(\mathbf{p})=1$, then $N_A\ge k$.
\end{enumerate}

\end{definition}

By definition, for a $k$-layered $\mathbf{p}$-arrangement $A_1, \dots, A_n$, the probability that at least $k$ of $A_1, \dots, A_n$ occur is $b_k(\mathbf{p})$, which is the sharp upper bound in \eqref{eq:event-count-bounds}. In the following proposition, we further show that all arrangements that achieve the sharp upper bound in \eqref{eq:event-count-bounds} are the $k$-layered $\mathbf{p}$-arrangements.

\begin{proposition}\label{prop:event-equality}
Let $\mathbf{p}\in [0,1]^n$, $A_1,\ldots,A_n\in\mA$ satisfy $\mathbb P(A_i)=p_i$, and set $N_A=\sum_{i=1}^n\id_{A_i}$. Then
\begin{equation}\label{eq:event-bound}
    \mathbb P(N_A\ge k)\le b_k(\mathbf{p}).
\end{equation}
Moreover, equality holds in \eqref{eq:event-bound} if and only if $A_1,\ldots,A_n$ form a $k$-layered $\mathbf{p}$-arrangement.
\end{proposition}

\begin{proof}
Lemma~\ref{lem:bk-closed} and \eqref{eq:event-count-bounds} together imply \eqref{eq:event-bound}. 
By the definition of  $k$-layered $\mathbf{p}$-arrangement, it is clear that $\p(N_A\ge k)=b_k(\mathbf{p})$ if $A_1,\ldots,A_n$ form a $k$-layered $\mathbf{p}$-arrangement. We will prove the ``only if" part. 

Conversely, suppose $\mathbb P(N_A\ge k)=b_k(\mathbf{p})$. If $b_k(\mathbf{p})=1$, then $N_A\ge k$ a.s., which is exactly the saturated condition. Assume now that $b_k(\mathbf{p})<1$, and put $b=b_k(\mathbf{p})$ and $B=\{N_A\ge k\}$. {If $b=0$, take $B_{\mathbf{p}}=\varnothing$. Otherwise, continuity and maximality in \eqref{eq:bk-def} give $\sum_{i=1}^n(p_i\wedge b)=kb$.} Hence,
\begin{equation}\label{eq:clipping-inequality}
    kb\le\mathbb E[N_A\id_B]=\sum_{i=1}^n\mathbb P(A_i\cap B)\le\sum_{i=1}^n(p_i\wedge b)=kb,
\end{equation}
which implies $\E[N_A\id_{B}]=kb$ and $\sum_{i=1}^n \p(A_i \cap B)=\sum_{i=1}^{n} p_i\wedge b$. As a result, we have $N_A=k$ on $B$, and $\mathbb P(A_i\cap B)=p_i\wedge b$ for each $i\in[n]$. If $p_i\le b$, then $\mathbb P(A_i\cap B)=p_i=\mathbb P(A_i)$, so $A_i\subseteq B$. If $p_i>b$, then $\mathbb P(A_i\cap B)=b=\mathbb P(B)$, so $B\subseteq A_i$.
{By definition of $B$, $N_A<k$ on $B^c$.} Thus $A_1,\dots,A_n$ form a $k$-layered $\mathbf{p}$-arrangement.
\end{proof}

We present a form of $k$-layered $\mathbf{p}$-arrangement $A_1, \dots, A_n$ in Remark~\ref{rem:k-layer}.

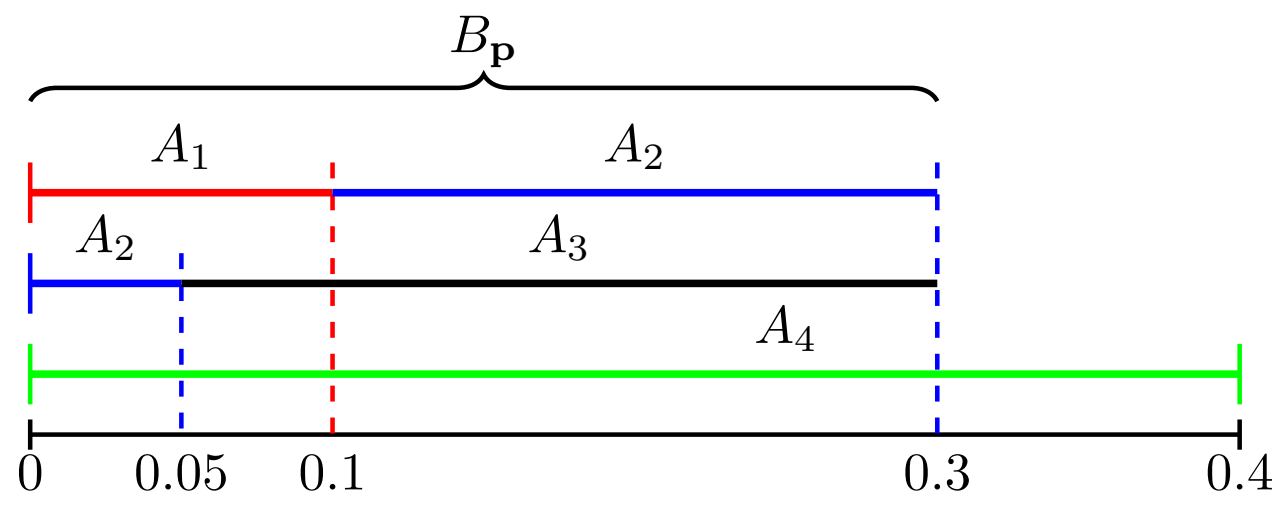
\begin{figure}
\centering
\begin{tikzpicture}
   \draw[line width=0.3mm] (0,-0.6)--(8,-0.6);
\draw[line width=0.3mm] (0,-0.5)--(0,-0.7);
\draw[line width=0.3mm] (8,-0.5)--(8,-0.7);
\node at (0,-0.85) {$0$};
\node at (8, -0.85) {$0.4$};

\draw [decorate,  decoration = {brace,    raise=3pt,    amplitude=5pt}, line width=0.3mm] (0,1.5) --  (6,1.5);
\node at (3,2) {$B_{\mathbf{p}}$};

\draw[line width=0.5mm,color=red] (0,1)--(2,1);
\draw[line width=0.3mm,color=red] (0,1.2)--(0,0.8);
\draw[line width=0.3mm,color=red,dashed] (2,1.2)--(2,-0.6);
\node at (2, -0.85) {$0.1$};
\node at (1, 1.3) {$A_1$};

\draw[line width=0.5mm,color=blue] (2,1)--(6,1);
\draw[line width=0.3mm,color=blue,dashed] (6,1.2)--(6,-0.6);
\node at (4, 1.3) {$A_2$};
\node at (6, -0.85) {0.3};

\draw[line width=0.5mm,color=blue] (0,0.4)--(1,0.4);
\draw[line width=0.3mm,color=blue] (0,0.2)--(0,0.6);
\draw[line width=0.3mm,color=blue,dashed] (1,0.6)--(1,-0.6);
\node at (0.5, 0.7) {$A_2$};

\draw[line width=0.5mm] (1,0.4)--(6,0.4);
\node at (1, -0.85){$0.05$};
\node at (3.5, 0.7) {$A_3$};

\draw[line width=0.5mm, color=green] (0,-0.2)--(8,-0.2);
\node at (5, 0.1) {{$A_4$}};
\draw[line width=0.3mm,color=green] (0,0)--(0,-0.4);
\draw[line width=0.3mm,color=green] (8,0)--(8,-0.4);
 \end{tikzpicture}
 \caption{$3$-layered $(0.1, 0.25, 0.25, {0.4})$-arrangement for $A_1$(red), $A_2$(blue), $A_3$(black), $A_4$(green) where $b_3(\mathbf{p})=0.3$}\label{fig:ex1}
 \end{figure}

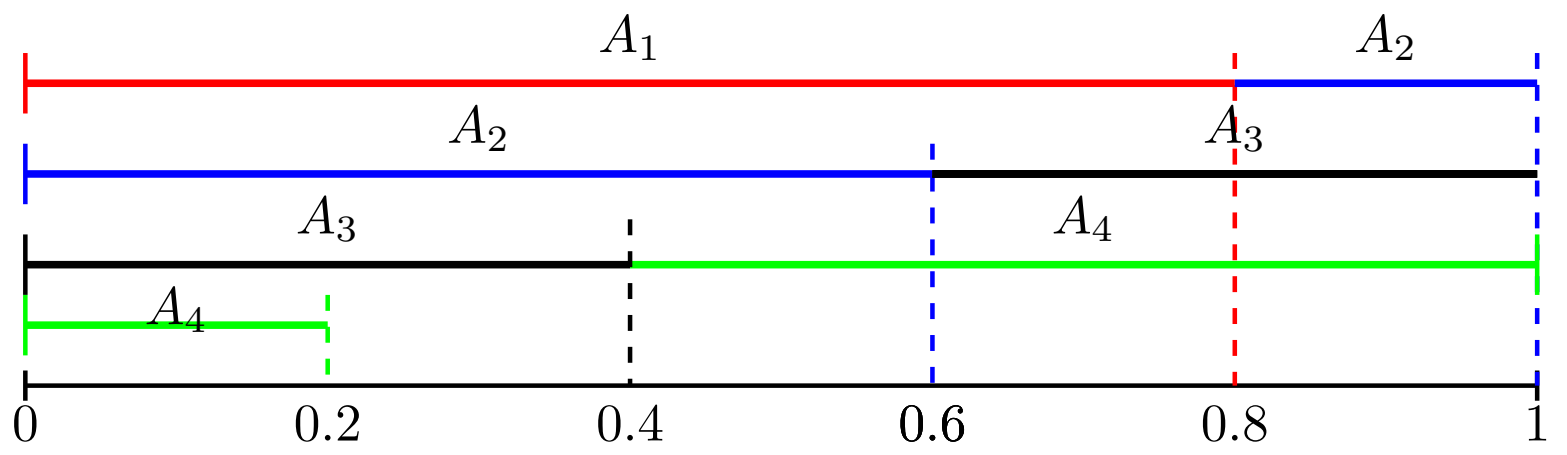
\begin{figure}
\centering
\begin{tikzpicture}
   \draw[line width=0.3mm] (0,-1)--(10,-1);
\draw[line width=0.3mm] (0,-0.9)--(0,-1.1);
\draw[line width=0.3mm] (10,-0.9)--(10,-1.1);
\node at (0,-1.25) {$0$};
\node at (10, -1.25) {$1$};

% \draw [decorate,  decoration = {brace,    raise=3pt,    amplitude=5pt}, line width=0.3mm] (0,1.5) --  (6,1.5);
% \node at (3,2) {$B_{\bp}$};

\draw[line width=0.5mm,color=red] (0,1)--(8,1);
\draw[line width=0.3mm,color=red] (0,1.2)--(0,0.8);
\draw[line width=0.3mm,color=red,dashed] (8,1.2)--(8,-1);
\node at (8, -1.25) {$0.8$};
\node at (4, 1.3) {$A_1$};

\draw[line width=0.5mm,color=blue] (8,1)--(10,1);
\draw[line width=0.3mm,color=blue,dashed] (10,1.2)--(10,-1);
\node at (9, 1.3) {$A_2$};

\draw[line width=0.5mm,color=blue] (0,0.4)--(6,0.4);
\draw[line width=0.3mm,color=blue] (0,0.2)--(0,0.6);
\draw[line width=0.3mm,color=blue,dashed] (6,0.6)--(6,-1);
\node at (3, 0.7) {$A_2$};

\draw[line width=0.5mm] (6,0.4)--(10,0.4);
\node at (6, -1.25){$0.6$};
\node at (8, 0.7) {$A_3$};

\draw[line width=0.5mm, color=black] (0,-0.2)--(4,-0.2);
\node at (2, 0.1) {{$A_3$}};
\draw[line width=0.3mm,color=black] (0,0)--(0,-0.4);
% \draw[line width=0.3mm,color=black] (4,0)--(4,-0.4);

\draw[line width=0.5mm, color=green] (4,-0.2)--(10,-0.2);
\node at (7, 0.1) {{$A_4$}};
% \draw[line width=0.3mm,color=green] (4,0)--(4,-0.4);
\draw[line width=0.3mm,color=green] (10,0)--(10,-0.4);

\draw[line width=0.3mm,color=black,dashed] (4,0.1)--(4,-1);
\node at (4, -1.25) {0.4};

\draw[line width=0.5mm] (6,0.4)--(10,0.4);
\node at (6, -1.25){$0.6$};

\draw[line width=0.5mm, color=green] (0,-0.6)--(2,-0.6);
\node at (1, -0.5) {{$A_4$}};
\draw[line width=0.3mm,color=green] (0,-0.4)--(0,-0.8);
% \draw[line width=0.3mm,color=green] (2,-0.4)--(2,-0.8);

\draw[line width=0.3mm,color=green,dashed] (2,-0.4)--(2,-1);
\node at (2, -1.25) {0.2};

\draw[line width=0.5mm] (6,0.4)--(10,0.4);
\node at (6, -1.25){$0.6$};

 \end{tikzpicture}
 \caption{$3$-layered $(0.8, 0.8, 0.8, 0.8)$-arrangement for $A_1$(red), $A_2$(blue), $A_3$(black), $A_4$(green) where $b_3(\mathbf{p})=1$}\label{fig:ex2}
 \end{figure}

\begin{remark}\label{rem:k-layer}
Without loss of generality, we assume $\p(A_1)\le \dots \le \p(A_n)$.
Let $a=b_k(\mathbf{p})$ and $r^*=\max\{i: p_i\le a\}$. Hence, we have $\p(A_i)\le a$ for $i \in [r^*]$ and $\p(A_i)>a$ for $i \in [n]\setminus[r^*]$. {For the folding operation below, place consecutive half-open intervals of the given lengths on the line and map them modulo the layer length onto the first layer.}
\begin{enumerate}[(i)]
\item Non-saturated case. {If $a=0$, fewer than $k$ of the $p_i$ are positive. Hence, every arrangement of $A_1, \dots, A_n$ satisfies that the probability of at least $k$ events occurring is $0$. If $0<a<1$, we have $\sum_{i=1}^{r^*}p_i=(k-(n-r^*))a>0$. Hence, $n-r^*<k$.} We fold the interval $[0, \sum_{i=1}^{r^*} p_i]$ into $(k-(n-r^*))$ layers, each of length $a$.
Since $p_i\le a$ for $i \in [r^*]$, this induces an overlap of the events $\{A_i\}_{i \in [r^*]}$ such that there exists an event $B_{\mathbf{p}}$ with $\p(B_{\mathbf{p}})=a$ and every point of $B_{\mathbf{p}}$ belongs to exactly $k-(n-r^*)$ of the  events $\{A_i\}_{i \in [r^*]}$ (see Figure~\ref{fig:ex1} for illustration). Next, let $B_{\mathbf{p}} \subseteq A_i$ for each $i \in [n]\setminus [r^*]$. As a result, on the set $B_{\mathbf{p}}$, there are exactly $k-(n-r^*)$ overlaps contributed by the first $r^*$ events and an additional $n-r^*<k$ overlaps by the remaining events. Outside {$B_{\mathbf{p}}$}, we have at most $n-r^*$ of the events $\{A_i\}_{i\in [n]}$ occurring. Hence, $A_1, \dots, A_n$ form a $k$-layered $\mathbf{p}$-arrangement.
\item Saturated case. If $b_k(\mathbf{p})=1$, by Lemma~\ref{lem:bk-closed}, we have $\sum_{i=1}^n p_i\ge k$. Hence, we can fold the interval $[0, \sum_{i=1}^n p_i]$ {into at least $k$ full layers of length one, with a possible final partial layer} (see Figure~\ref{fig:ex2} for illustration). As a result, we have $N_A\ge k$, and $A_1, \dots, A_n$ form a $k$-layered $\mathbf{p}$-arrangement.
\end{enumerate}

\end{remark}

\begin{example}\label{ex:1-n}
We present the following two special cases when $k=1$ and $k=n$. Without loss of generality, we assume $\mathbf{p}=(p_1, \dots, p_n)$ such that $p_1\le \dots \le p_n$.
\begin{enumerate}[(i)]
\item When $k=1$, we have $b_1(\mathbf{p})=(\sum_{i=1}^n p_i)\wedge 1$, which is the upper bound for the probability that at least one of the events $A_1, \dots, A_n$ occurs. In the case that $\sum_{i=1}^n p_i< 1$,  any mutually exclusive $\{A_i\}_{i \in [n]}$ is a $1$-layered $\mathbf{p}$-arrangement. When $\sum_{i=1}^n p_i\ge 1$, any $\{A_i\}_{i\in [n]}$ satisfying  $\bigcup_{i=1}^n A_i=\Omega$ is a  $1$-layered $\mathbf{p}$-arrangement.
\item When $k=n$, we have $b_{n}(\mathbf{p})=\min_{i \in [n]} p_i$, which is the upper bound for the probability that all events occur. {An $n$-layered $\mathbf{p}$-arrangement requires $A_1\subseteq A_i$ for all $i\in [n]$. One choice is the nested arrangement $A_1\subseteq\dots\subseteq A_n$.}
\end{enumerate}
\end{example}

% \begin{example}
% Suppose $F_1, F_2, F_3 \in \mathcal{M}$ and $t \in R$ such that  $t$ such that
% $$
% p_1=F_1(t)=0.50,\qquad p_2=F_2(t)=0.56,\qquad p_3=F_3(t)=0.60.
% $$ 
%  By Theorem \ref{thm:order}, the upper bound of $\p(X_{(2)}\le t)$ is 
% $$
% \min\Big\{1, \frac{p_1+p_2+p_3}{2},\ p_1+p_2\Big\}=\frac{p_1+p_2+p_3}{2}=0.83.
% $$
% By the proof of Theorem \ref{thm:order}, we organize the events $\{\overline{X^t_1}\le t\}$, $\{\overline{X^t_2}\le t\}$ and $\{\overline{X^t_3}\le t\}$ as Figure \ref{fig:ex1}. It is clear that exactly two of the events $\{\overline{X^t_1}\le t\}$, $\{\overline{X^t_2}\le t\}$ and $\{\overline{X^t_3}\le t\}$ occur at the same time. Hence, we have $\p(X_{(2)})\le \p(\overline{X^t_{(2)}})=0.83$ for all $(X_1, X_2, X_3) \in \mathcal{F}(F_1, F_2, F_3)$.
%\end{example}

The $k$-layered $\mathbf{p}$-tail-optimal random vector is defined as follows.

\begin{definition}
Let $\mathbf{p}\in [0,1]^n$ and $k \in [n]$. A random vector $\mathbf X$ is $k$-layered $\mathbf{p}$-tail-optimal if there exists a $k$-layered $\mathbf{p}$-arrangement events $A_1, \dots, A_n$ such that $A_i$ are $p_i$-tail events of $X_i$ for all $i\in [n]$.
\end{definition}

When $k=1$ and $\mathbf{p}\in\Delta_n^0$, layered tail-optimality
coincides with $\mathbf{p}$-tail counter-monotonicity.

\begin{proposition}\label{prop:klayer-to-tail}
For $\mathbf{p}\in \Delta_n^{0}$, $\mathbf{p}$-tail counter-monotonicity is equivalent to $1$-layered $\mathbf{p}$-tail-optimality.
\end{proposition}
\begin{proof}
Since $\mathbf{p} \in \Delta_n^{0}$, we have $b_1(\mathbf{p})=\sum_{i=1}^n p_i$.
If $\mathbf{X}$ is $\mathbf{p}$-tail counter-monotonic, there exist  mutually exclusive events $A_1, \dots, A_n$ such that 
 $A_i$ is a $p_i$-tail event of $X_i$ for all $i \in [n]$. Let $B_{\mathbf{p}}=\cup_{i=1}^n A_i$. It is clear that $N_A=1$ on $B_{\mathbf{p}}$ and $N_A={0}$ in $B_{\mathbf{p}}^c$. In the case that $\sum_{i=1}^n p_i=1$, we have $N_A=1$ a.s. Hence, $A_1, \dots, A_n$ form a $1$-layered $\mathbf{p}$-arrangement and $\mathbf{X}$ is $1$-layered $\mathbf{p}$-tail-optimal.

 Assume  $\mathbf{X}$ is $1$-layered $\mathbf{p}$-tail-optimal. By definition, there exists a $1$-layered $\mathbf{p}$-arrangement events $A_1, \dots, A_n$ such that $A_i$ are $p_i$-tail events of $X_i$ for all $i\in [n]$. By Proposition~\ref{prop:event-equality}, we have $\p\left(\bigcup_{i=1}^n A_i\right)=\p(\sum \id_{A_i}\ge 1)=\sum_{i=1}^n p_i$, which implies $A_1, \dots, A_n$ are mutually exclusive. Hence, $\mathbf{X}$ is $\mathbf{p}$-tail counter-monotonic.
\end{proof}

{\setlength{\emergencystretch}{1em}
Together with Theorem~\ref{thm:char}, we further show that $1$-layered $\mathbf{p}$-tail-optimality and $n$-layered $\mathbf{p}$-tail-optimality characterize counter-monotonicity and comonotonicity, respectively.\par}
\begin{theorem}\label{thm:k-layer}
The following statements hold.
\begin{enumerate}[(i)]
\item $\mathbf{X}$ is counter-monotonic if and only if $\mathbf{X}$ is $1$-layered $\mathbf{p}$-tail-optimal for all $\mathbf{p}\in \Delta_n^{0}$.
\item $\mathbf{X}$ is comonotonic if and only if $\mathbf{X}$ is $n$-layered $\mathbf{p}$-tail-optimal for all $\mathbf{p}=p\mathbf{1}_n$ where $p \in [0,1]$ and $\mathbf{1}_n=(1, \dots, 1)$.
\end{enumerate}
\end{theorem}
\begin{proof}
By Theorem~\ref{thm:char} and Proposition~\ref{prop:klayer-to-tail}, it is clear that (i) holds. 

Assume $\mathbf{X}$ is comonotonic. By \citet[Theorem 4]{bib:wang-zitikis-2021}, $\mathbf{X}$ shares a common $p$-tail event for all $p\in (0,1)$; that is, there exists $A$ such that $A$ is $p$-tail events of all $X_i$, $i\in [n]$.  It is clear that $(A, \dots, A)$ forms an
$n$-layered $p\mathbf{1}_n$ arrangement by Example~\ref{ex:1-n}. Hence, $\mathbf{X}$ is $n$-layered $p\mathbf{1}_n$-tail-optimal for all $p \in [0,1]$. Conversely, assume $\mathbf{X}$ is $n$-layered $p\mathbf{1}_n$-tail-optimal for all $p \in [0,1]$. By Example~\ref{ex:1-n}, we have $b_n(p\mathbf{1}_n)=p$. For a given $p\in [0,1]$, there exist $A_1, \dots, A_n$ such that $A_i$ is the $p$-tail event of $X_i$, $i\in [n]$, and   $A_1, \dots, A_n$ form a $n$-layered $p\mathbf{1}_n$ arrangement.  By definition, we have $\p(\sum_{i=1}^n \id_{A_i}=n)=\p(\bigcap_{i\in [n]} A_i)=p=\p(A_i)$, which implies $A_1=\dots=A_n$ a.s. Hence, $\mathbf{X}$ shares a common $p$-tail event for all $p\in [0,1]$.  By \citet[Theorem 4]{bib:wang-zitikis-2021}, we have $\mathbf{X}$ is comonotonic.
\end{proof}

Similar to Proposition~\ref{prop:increase}, we can also show that the $k$-layered $\mathbf{p}$-tail optimality is preserved under increasing marginal transformations.
\begin{proposition}\label{prop:layered-increasing}
Let $\mathbf{p} \in [0,1]^n$. If $\mathbf{X}$ is $k$-layered $\mathbf{p}$-tail optimal, then $(f_1(X_1), \dots, f_n(X_n))$ is  $k$-layered $\mathbf{p}$-tail optimal for all increasing functions $f_1, \dots, f_n$.
\end{proposition}
\begin{proof}
The proof is similar to the proof of Proposition~\ref{prop:increase}, since the $p_i$-tail event of $X_i$ is also the tail event of $f_i(X_i)$.
\end{proof}

Similar to $\mathbf{p}$-tail counter-monotonicity, the $k$-layered $\mathbf{p}$-tail optimality also exists in all Fr\'echet classes by Proposition~\ref{prop:layered-increasing}. We present a possible choice of copula for $k$-layered $\mathbf{p}$-tail optimality in the following remark.
\begin{remark}
Let $A_1,\ldots,A_n$ with $\mathbb P(A_i)=p_i$ be a $k$-layered $\mathbf{p}$-arrangement as we stated in Remark~\ref{rem:k-layer}. Let  $V_1,\ldots,V_n \sim \mathrm{U}[0,1]$ be independent of $A_1, \dots A_n$. For $i\in[n]$, let
\begin{equation*}
    U_i=(1-p_i+p_iV_i)\id_{A_i}+(1-p_i)V_i\id_{A_i^c}.
\end{equation*}
We have $U_i\sim U[0,1]$, and $A_i$ is a $p_i$-tail event of $U_i$. Set $X_i=F_i^{-1}(U_i)$, where $F_i^{-1}$ is the left quantile function. Since $F_i^{-1}$ is increasing, $A_i$ is also a $p_i$-tail event of $X_i$. Hence $\mathbf{X}$ is $k$-layered $\mathbf{p}$-tail optimal.
\end{remark}
% \begin{definition}\label{def:tail-domain}
% For a random vector $X=(X_1,\ldots,X_n)$, define its \emph{$k$-layered tail-optimality domain} by
% \begin{equation*}
% \begin{aligned}
% \mD_k(X)=\{\bp\in[0,1]^n:\ &\mbox{there exist }p_i\mbox{-tail events }A_i\mbox{ of }X_i,\ i\in[n],\\
% &\mbox{such that }A_1,\ldots,A_n\mbox{ form a }k\mbox{-layered }\bp\mbox{-arrangement}\}.
% \end{aligned}
% \end{equation*}
% \end{definition}

% Without loss of generality, assume that
% \[
% 0\le p_1\le \cdots \le p_n\le 1.
% \]
% For an $n$-copula $C$, let $\mathbb P_C(X_{(k)}\le t)$ denote the probability of $\{X_{(k)}\le t\}$ when $(X_1,\dots,X_n)\in\mathcal F(F_1,\dots,F_n)$ has the copula $C$. Equivalently, let $\mu_C$ denotes the probability measure on $[0,1]^n$ induced by $C$, then
% \[
% \mathbb P_C(X_{(k)}\le t)=\mu_C(A),
% \]
% where
% \[
% A:=\left\{(u_1,\dots,u_n)\in[0,1]^n:\sum_{i=1}^n \mathbf 1_{\{u_i\le p_i\}}\ge k\right\}.
% \]
% Hence the optimization problems
% \[
% \sup_{C\in\mathcal C_n}\mathbb P_C(X_{(k)}\le t)
% \qquad\text{and}\qquad
% \inf_{C\in\mathcal C_n}\mathbb P_C(X_{(k)}\le t)
% \]
% are equal to finding the largest and smallest possible mass on the set $A$ over all $n$-copulas.

\section{Bounds on the distribution of order statistics}\label{sec:distribution-bounds}
Section~\ref{sec:maximum} shows that mutually exclusive and concentrated tail events provide the upper and lower bounds on the maximum of a risk portfolio in the stochastic order respectively.
Let $X_{(k)}$ denote the $k$-th smallest order statistic of $\mathbf{X}$. In this section, we discuss bounds on risk measures for the order statistics of risks under dependence uncertainty using the layered arrangements in tail events that {need not be mutually exclusive or concentrated}.

Fix $t\in\mathbb R$. The event $\{X_{(k)}\le t\}$ occurs if and only if   at least $k$ of the events $\{X_1 \le t\}, \dots, \{X_n \le t\}$ occur. Hence, the distribution of $X_{(k)}$ is
\begin{equation}\label{eq:rank-cdf}\p(X_{(k)}\le t)=\p\left(\sum_{i=1}^n \id_{\{X_i\le t\}}\ge k\right).
\end{equation}
As we have shown in Section~\ref{sec:layered-arrangements}, the upper bound {on the probability of at least $k$ events occurring with probability vector $\mathbf{p} \in [0,1]^n$ is $b_k(\mathbf{p})$ and it is {achieved when those events form a $k$-layered $\mathbf{p}$-arrangement}. Therefore, we have the bounds on the distribution of the $k$-th smallest order statistic. 
The probability bounds are already provided by \cite{bib:ruger-1978}, and the attainability is established by 
\cite{bib:salako-2025}. The following proposition further 
uses the layered arrangements in Section~\ref{sec:layered-arrangements} to construct explicit
couplings with the given marginals that attain each bound at
the fixed threshold.

\begin{proposition}\label{prop:order}
Suppose $F_1, \dots, F_n \in \mathcal M$, $k \in [n]$ and fix $t \in \R$. Let $\mathbf{p}(t)=(F_1(t), \dots, F_n(t))$. There exist $(\overline{X_1^{t}}, \dots, \overline{X_n^{t}}) \in \mathcal{F}(F_1, \dots, F_n)$ and $(\underline{X_1^{t}}, \dots, \underline{X_n^{t}}) \in \mathcal{F}(F_1, \dots, F_n)$ such that $(-\overline{X_1^{t}}, \dots, -\overline{X_n^{t}})$ is $k$-layered $\mathbf{p}{(t)}$-tail optimal and $(\underline{X_1^{t}}, \dots, \underline{X_n^{t}})$ is $(n-k+1)$-layered $(\mathbf{1}_n-\mathbf{p}{(t)})$-tail optimal. Moreover, we have
$$1-b_{n-k+1}(\mathbf{1}_n-\mathbf{p}(t))
=\p\left(\underline{X^t_{(k)}}\le t\right)
\le \p(X_{(k)} \le t)
\le \p\left(\overline{X^t_{(k)}}\le t\right)=b_k(\mathbf{p}(t))$$
for all $\mathbf{X} \in \mathcal{F}(F_1, \dots, F_n)$ where $\overline{X^t_{(k)}}$ and $\underline{X^t_{(k)}}$ are the $k$-th order statistics of $(\overline{X_1^{t}}, \dots, \overline{X_n^{t}})$ and $(\underline{X_1^{t}}, \dots, \underline{X_n^{t}})$, respectively.
\end{proposition}

\begin{proof}
We show $(\overline{X_1^{t}}, \dots, \overline{X_n^{t}})$ and $(\underline{X_1^{t}}, \dots, \underline{X_n^{t}})$ attain the bounds by constructing them following the same logic in Proposition~\ref{prop:exist}. Write $p_i=F_i(t)$ and $\mathbf{p}=(p_1,\ldots,p_n)$.

Let $A_1, \dots A_n$ with $\p(A_i)=p_i$
be a $k$-layered $\mathbf{p}$-arrangement, $V_1, \dots V_n \sim \mathrm{U}[0,1]$ be independent and independent of $A_1, \dots, A_n$,  $$\overline{U_i}=\left(p_iV_i\right)\id_{A_i}+\left(p_i+(1-p_i)V_i\right)\id_{A_i^c}$$
and $\overline{X_i^t}=F_i^{-1}(\overline{U_i})$ for $i \in [n]$. By construction, it is clear that $\overline{X_i^t}\sim F_i$ and $A_i$ is a $p_i$-tail event of $-\overline{X_i^t}$ for $i \in [n]$. %{\color{blue}In fact, $\{\overline{X_i^t}\le t\}=\{\overline{U_i}\le F_i(t)\}=A_i$ up to null sets, including at atoms.} 
Hence, $(\overline{X_1^{t}}, \dots, \overline{X_n^{t}}) \in \mathcal{F}(F_1, \dots, F_n)$ and $(-\overline{X_1^{t}}, \dots, -\overline{X_n^{t}})$ is $k$-layered $\mathbf{p}$-tail optimal.
By Proposition~\ref{prop:event-equality}, Lemma~\ref{lem:bk-closed} and \eqref{eq:rank-cdf}, we further have $$\p\left(\overline{X^t_{(k)}}\le t\right)=\p\left(\sum_{i=1}^n \id_{A_i}\ge k\right)=b_k(\mathbf{p}).$$ 

For the lower bound, we have 
\begin{equation}\label{eq:rank-cdf-complement}\p\left(\sum_{i=1}^n \id_{\{X_i\le t\}}\ge k\right)=\p\left(\sum_{i=1}^n \id_{\{X_i> t\}}\le n- k\right)=1-\p\left(\sum_{i=1}^n \id_{\{X_i> t\}}\ge n- k+1\right).
\end{equation}

Let $A_1, \dots A_n$ with $\p(A_i)=1-p_i$
be an $(n-k+1)$-layered $(\mathbf{1}_n-\mathbf{p})$-arrangement, $V_1, \dots V_n \sim \mathrm{U}[0,1]$ be independent and independent of $A_1, \dots, A_n$,  $$\underline{U_i}=\left(p_iV_i\right)\id_{A^c_i}+\left(p_i+(1-p_i)V_i\right)\id_{A_i}$$
and $\underline{X_i^t}=F_i^{-1}({\underline{U_i}})$ for $i \in [n]$.
Hence, it is clear that $A_i$ is a $(1-p_i)$-tail event of $\underline{X_i^t}$, $i\in [n]$. Therefore, $(\underline{X_1^{t}}, \dots, \underline{X_n^{t}}) \in \mathcal{F}(F_1, \dots, F_n)$ and is $(n-k+1)$-layered $(\mathbf{1}_n-\mathbf{p})$-tail optimal.
By Proposition~\ref{prop:event-equality}, Lemma~\ref{lem:bk-closed} and \eqref{eq:rank-cdf}, we further have 
$$ \p\left(\sum_{i=1}^n \id_{\{\underline{X^t_i}> t\}}\ge n- k+1\right)= b_{n-k+1}(\mathbf{1}_n-\mathbf{p}).$$
Together with \eqref{eq:rank-cdf-complement}, we have the lower bound.
%{\color{blue}In each construction, conditional on the event pattern, the uniforms are independent and uniform on their corresponding intervals. Thus the joint law of the uniforms is a checkerboard copula with cutoffs $F_i(t)$ in coordinate $i$.}
% $$\p\left(\sum_{i=1}^n \id_{\{\underline{X^_i}\le t\}}\ge k\right)=\left(1-\min_{0\leq r\leq n-k}\frac{\sum_{j=r+1}^{n}1-p_{(j)}}{n-k+1-r}\right)_+,$$
% which completes the proof.
\end{proof}
% The example below illustrates how the tail events $\{X_i\le t\}$ overlap in the proof of Theorem \ref{thm:order}.

If we take the indicator cost function $c(z)=\id_{\{z\le t\}}$, Proposition~\ref{prop:order} gives optimizers for multi-marginal transport in Problem \eqref{eq:intro-cost}. On the other hand, the multi-marginal transport problem with indicator cost also provides another proof for the upper bound in Proposition~\ref{prop:order}.
For each $J\subseteq[n]$ with $|J|<k$, at least 
$k-\vert J\vert$ coordinates outside $J$ are at most $t$ whenever $X_{(k)}\le t$. Hence,
\[
 \id_{\{X_{(k)}\le t\}}
 \le \frac{1}{k-|J|}\sum_{i\notin J}\id_{\{X_i\le t\}}.
\]
By first taking expectations and then minimizing over all $J\subseteq[n]$ with $|J|<k$, we obtain the upper bound $b_k(\mathbf{p}(t))$ in Proposition~\ref{prop:order}.

Different from Theorem~\ref{thm:max} where we have a universal maximizer and minimizer for every $t \in \R$,
there are superscripts $t$ for the maximizer $(\overline{X_1^{t}}, \dots, \overline{X_n^{t}})$ and minimizer $(\underline{X_1^{t}}, \dots, \underline{X_n^{t}})$ as their constructions depend on the choice of threshold $t$. In the following example, a universal maximizer for all $t \in \R$ does not exist.

\begin{example}\label{ex:nounique}

Let $F_1,F_2,F_3\in \mathcal{M}$ satisfying
\[
\begin{aligned}
&F_1(0)=0,\qquad F_1(1)=0.01,\qquad F_1(2)=0.9,\qquad F_1(3)=1,\\
&F_2(0)=0,\qquad F_2(1)=0.05,\qquad F_2(2)=0.1,\qquad F_2(3)=1,\\
&F_3(0)=0,\qquad F_3(1)=0.05,\qquad F_3(2)=0.1,\qquad F_3(3)=1.
\end{aligned}
\]

For $t=1$, by Proposition~\ref{prop:order}, we have $\p(X_{(2)}\le 1)\le 0.055$ for all $(X_1, X_2, X_3) \in \mathcal{F}(F_1, F_2, F_3)$. For any $(\overline{X_1^1}, \overline{X_2^1}, \overline{X_3^1}) \in \mathcal{F}(F_1, F_2, F_3)$ that  achieves the upper bound $\p(\overline{X_{(2)}^1}\le 1)=0.055$, {write $A_i=\{\overline{X_i^1}\le1\}$.  We have  }
\[
\p(\overline{X_{(2)}^1}\leq 1)
= {\p(A_1\cap A_2)+\p(A_1\cap A_3)+\p(A_2\cap A_3)-2\p(A_1\cap A_2\cap A_3)}.
\]
Since {$\p(A_1\cap A_2\cap A_3)\ge0$ and $\p(A_1\cap A_i)\le\p(A_1)$ for $i=2,3$}, we have
\[
{\p(A_2\cap A_3)}\ge 2\times 0+0.055-0.01-0.01=0.035.
\]

For $t=2$, by Proposition~\ref{prop:order}, we have $\p(X_{(2)}\le {2})\le 0.2$ for all $(X_1, X_2, X_3) \in \mathcal{F}(F_1, F_2, F_3)$. For any $(\overline{X_1^2}, \overline{X_2^2}, \overline{X_3^2}) \in \mathcal{F}(F_1, F_2, F_3)$ that achieves the upper bound $\p(\overline{X_{(2)}^2}\le {2})=0.2$, write $B_i=\{\overline{X_i^2}\le 2\}$, and let $E$ be the event that at least two of $B_1,B_2,B_3$ occur. For any coupling that attains the bound at $t=2$, $\mathbb P(E)=0.2$ holds. And when $E$ occurs, at least one of $B_2,B_3$ must occur, that is,
\[
0.2=\mathbb P(E)
\le \mathbb P(B_2\cup B_3)
=0.1+0.1-\mathbb P(B_2\cap B_3).
\]
Thus, $\p(B_2\cap B_3)=0$, which conflicts with the probability at the upper-bound coupling for $t=1$.
Therefore, we cannot find $(\overline{X_1^1}, \overline{X_2^1}, \overline{X_3^1})= (\overline{X_1^2}, \overline{X_2^2}, \overline{X_3^2})$.
\end{example}
%While we may not be able to find a universal maximizer or minimizer when the marginals are different, we can find both a universal maximizer and a universal minimizer when the marginals are identical.

Example~\ref{ex:nounique} shows that we cannot find $(X_1, X_2, X_3) \in \mathcal{F}(F_1, F_2, F_3)$ such that $X_{(2)}\le_{\rm st} Y_{(2)}$ for {every} $(Y_1, Y_2, Y_3) \in \mathcal{F}(F_1, F_2, F_3)$.

\section{Stochastic dominance for order statistics}\label{sec:stochastic-order}
In this section, we provide necessary and sufficient conditions for the existence of a universal lower bound in stochastic order in Theorem~\ref{thm:universal-attainment}. We first present a theorem from \cite{bib:rychlik-1995}.
%For two distributions $F, G \in \mathcal{M}$, the Radon–Nikodym derivative $f$ of  the measure induced by $F$ with respect to the measure induced by $G$ is defined as
%$$f(x)=\lim_{h \downarrow 0}\frac{F(x+h)-F(x)}{G(x+h)-G(x)}~~~ \mbox{for}~ \d G\mbox{-almost every}~x.$$
Throughout this section, take the conventions $\sup\varnothing=-\infty$ and $\inf \varnothing=\infty$.

\begin{lemma}\label{lem:rychlik-attainment}
For $F_1, \dots, F_n \in \mathcal{M}$ and $k \in [n]$, let $G=\sum_{i=1}^n F_i$ and $F^{\uparrow}={1\wedge(G/k)}$.
There exists $\mathbf{X} \in \mathcal{F}(F_1,\dots, F_n)$ such that $X_{(k)} \sim {F}^{\uparrow}$ 
if and only if 
\begin{enumerate}[(a)]
    \item $\max_{i \in [n]} f_i\le 1/k$ {$\d G$-almost everywhere} on $(-\infty,\tau)$ and
    \item {if $\tau<\infty$,} $\sum_{i=1}^n \min\{P_i(\tau),{P}^{\uparrow}(\tau)\}\ge k{P}^{\uparrow}(\tau)$.
\end{enumerate}
hold, where
$\tau=\inf\left\{t\in\mathbb R: G(t)\ge k\right\}$,
{ $f_i$ is the Radon--Nikodym derivative of the measure induced by $F_i$ with respect to the measure induced by $G$, $P_i(t)=F_i(t)-F_i(t-)$, $i\in [n]$, and $P^\uparrow(t)=F^\uparrow(t)-F^\uparrow(t-)$.}
\end{lemma}

\begin{remark}\label{rem:con1}
{When $\tau$ is finite, we have $P_i(\tau)=\p(X_i=\tau)$ and $P^\uparrow(\tau)$ is the mass assigned to $\tau$ by $F^\uparrow$. Hence, if $F_1,\dots,F_n$ are continuous, condition (b) holds. In the case that $\tau=\infty$, since $F_1, \dots, F_n$ are cdfs on $\R$, we have $P_i(\infty)=0$ and $P^\uparrow(\infty)=0$. Hence, $\sum_{i=1}^n \min\{P_i(\tau),{P}^{\uparrow}(\tau)\}\ge k{P}^{\uparrow}(\tau)$ holds automatically.}
\end{remark}

\begin{theorem}\label{thm:universal-attainment}
Let $F_1,\ldots,F_n\in\mathcal M$,  $k\in [n]$,  $\mathbf{p}(t)=(F_1(t), \dots, F_n(t))$, $G(t)=b_k(\mathbf{p}(t))$ for $t \in \R$ and $\tau=\inf\{t \in \R: \sum_{i=1}^n F_i(t)\ge k\}$.
There exists $\mathbf{X}\in\mathcal F(F_1,\ldots,F_n)$ such that $X_{(k)}\le_{\rm st} Y_{(k)}$ for all $(Y_1, \dots, Y_n) \in \mathcal{F}(F_1, \dots, F_n)$
% \begin{equation}
%     \mathbb P(X_{(k)}\le t)=b_k(\boldsymbol p(t))
%     \quad\text{for every }t\in\mathbb R
%     \label{eq:universal-upper-cdf}
% \end{equation}
if and only if
both of the following conditions hold:
\begin{enumerate}
    \item[(i)] For every $i\in[n]$, the function $t\mapsto (G(t)-F_i(t))_+$ is increasing on $(-\infty,\tau)$.

    \item[(ii)] {If $\tau<\infty$,}
\begingroup
\[\sum_{i=1}^n \bigl[\bigl(F_i(\tau)-(F_i(\tau-)\wedge G(\tau-))\bigr)\wedge (1-G(\tau-))\bigr]\ge k(1-G(\tau-)).\]
\endgroup
\end{enumerate}
\end{theorem}

\begin{proof}
Lemma~\ref{lem:bk-closed} implies that  $G$ is a cdf. 
By Proposition~\ref{prop:order}, $G$ is the pointwise supremum of the attainable cdfs of the $k$-th order statistic. Thus a coupling with $X_{(k)}\le_{\rm st}Y_{(k)}$ for all admissible $Y$ exists if and only if there is an admissible coupling with $X_{(k)}\sim G$.

By definition of $b_k$, 
it holds that $G(t)<1$ for $t<\tau$, 
while $G(t)=1$ for $t\ge\tau$.
  Let $\widetilde F_i(t)=F_i(t)\wedge G(t)$ for all $t \in \R$ and  $i\in[n]$. 
Since both $F_i$ and $G$ are cdfs,  $\widetilde F_i\in \mathcal M$. 
By the definition of $b_k$, we have
$\sum_{i=1}^n\widetilde F_i(t)=kG(t)$  for all $t< \tau$. Since $G(t)=1$ for $t\ge \tau$, we have $\widetilde F_i=F_i$, $i \in [n]$, and  
$\sum_{i=1}^n \widetilde F_i(t)=\sum_{i=1}^n F_i(t)$ for all $t\ge \tau$. 
Moreover, by definition of $\tau$, we have $\sum_{i=1}^{n} F_i(t)\ge k$ for $t\ge\tau$. Therefore, we have
\begin{equation*}
1\wedge \frac{1}{k}\sum_{i=1}^n\widetilde F_i(t)=G(t) ~~~\mbox{for all}~~~
t\in\R.
\end{equation*}

 We next show that there exists $\mathbf{X} \in \mathcal{F}(F_1, \dots, F_n)$ such that 
$X_{(k)}\sim G$
if and only if there exists 
$(Z_1, \dots, Z_n) \in \mathcal{F}(\widetilde F_1, \dots, \widetilde F_n)$ such that $ Z_{(k)} \sim G$. 

\begin{enumerate}[(a)]
\item Suppose that there exists $\mathbf{X}\in \mathcal{F}(F_1, \dots, F_n)$ such that  $X_{(k)} \sim G$.
For $t<\tau$, at least $k$ of $\{X_i\le t\}_{i \in [n]}$ occur  on
$\{X_{(k)}\le t\}$, that is, $\sum_{i=1}^n\id_{\{X_i\le t, X_{(k)}\le t\}}\ge k\,\id_{\{X_{(k)}\le t\}}$. Note that $\p(X_i\le t, X_{(k)}\le t)\le F_i(t) \wedge G(t)=\widetilde F_i(t)$ for all $t \in \R$ and $i \in [n]$.
Hence, for $t< \tau$, we have
$$kG(t)=
k\p(X_{(k)}\le t)
\le
\sum_{i=1}^{n}\mathbb P(X_i\le t, X_{(k)}\le t)
\le
\sum_{i=1}^n \widetilde F_i(t)
=
kG(t),
$$
which implies $\p(X_i\le t, X_{(k)}\le t)=\widetilde F_i(t)$ for  all $t <\tau$ and $i \in [n]$.
For $t\ge\tau$, since $\mathbb P(X_{(k)}\le t)=G(t)=1$, we have $\p(X_i\le t, X_{(k)}\le t)=F_i(t)=\widetilde F_i(t)$, $i \in [n]$. Let $Z_i=X_i\vee X_{(k)}, i\in[n]$. We have $\p(Z_i\le t)=\p(X_i\le t, X_{(k)}\le t)$ for all $t \in \R$. Hence, 
we have $(Z_1, \dots, Z_n)\in {\mathcal{F}}(\widetilde F_1, \dots, \widetilde F_n)$. Moreover,  $Z_{(k)}=X_{(k)}\sim G$, which gives the ``only if" part.

\item  Suppose that there exists $(Z_1, \dots, Z_n) \in \mathcal{F}(\widetilde F_1, \dots, \widetilde F_n)$ such that $Z_{(k)} \sim G$.
Let  $U_i \sim \mathrm{U}[0,1]$ such that $Z_i=\widetilde{F}^{-1}_i(U_i)$, $i \in [n]$. 
Take $X_i=F_i^{-1}(U_i)$,  $i \in [n]$. It is clear that $\mathbf{X} \in \mathcal{F}(F_1, \dots, F_n)$.
Since $\widetilde F_i\le F_i$,
we have $X_i\le Z_i$, and hence $X_{(k)}\le Z_{(k)}$.
Therefore,
$$
G(t)=\mathbb P(Z_{(k)}\le t)\le\mathbb P(X_{(k)}\le t)\le G(t)
$$
where the last inequality follows from Proposition~\ref{prop:order}.
Hence, $X_{(k)} \sim G$, which completes the ``if" part.
\end{enumerate}
Note that $G=\widetilde F^\uparrow$ as defined in Lemma~\ref{lem:rychlik-attainment}. Hence, there exists 
$(Z_1, \dots, Z_n) \in \mathcal{F}(\widetilde F_1, \dots, \widetilde F_n)$ such that $ Z_{(k)} \sim G$ if and only if (a) and (b) in Lemma~\ref{lem:rychlik-attainment} hold for $\widetilde F_1, \dots, \widetilde F_n$.

Since $\widetilde F_i\le F_i$, we have $\tau \le \inf\{t \in \R: \sum_{i=1}^n {\widetilde F_i(t)}\ge k\}$. As $\sum_{i=1}^n \widetilde F_i(t)=\sum_{i=1}^n F_i(t) \ge k$ for  $t \ge \tau$, we have $\inf\{t \in \R: \sum_{i=1}^n \widetilde F_i{(t)}\ge k\}\le \tau$. Hence, $\tau=\inf\{t \in \R: \sum_{i=1}^n \widetilde F_i(t)\ge k\}$.

Note that
$G(t)-\widetilde F_i(t)=(G(t)-F_i(t))_+$.
Condition (i)  is equivalent to  $t \mapsto G(t)-\widetilde F_i(t)$
being increasing on  $(-\infty,\tau)$ for every $i\in[n]
$. {Since} $kG(t)=\sum_{i=1}^n \widetilde F_i(t)$ for $t <\tau$, condition (i) is equivalent to  
$$
    \widetilde F_i(t) - \widetilde F_i(s)
    \le G(t)-G(s)= \frac1k\sum_{j=1}^n (\widetilde F_j(t) - \widetilde F_j(s))
    \quad\text{for every pair of } s<t<\tau \text{ and every } i\in[n].
$$
That is, condition (i) holds if and only if condition (a)  in Lemma~\ref{lem:rychlik-attainment} holds for $\widetilde F_1, \dots, \widetilde F_n$.

\begingroup
If $\tau<\infty$, $\widetilde F_i(\tau)=F_i(\tau)$,
$\widetilde F_i(\tau-)=F_i(\tau-)\wedge G(\tau-)$ and
$G(\tau-)=b_k(\mathbf{p}(\tau-))$ by continuity of $b_k$. Therefore (ii) is
condition (b) of Lemma~\ref{lem:rychlik-attainment} for $\widetilde F_1,\dots,\widetilde F_n$.
 This completes the proof.
\endgroup
\end{proof}

Similarly, we provide the necessary and sufficient conditions for the upper bound in stochastic order.

\begin{proposition}\label{prop:upper}
Let $F_1,\ldots,F_n\in\mathcal M${,} $k\in[n]$,  $\mathbf{p}(t)=(F_1(t), \dots, F_n(t))$, $H(t)=1-b_{n-k+1}(\mathbf{1}_n-\mathbf{p}(t))$ for $t \in \R$ and $\sigma=\sup\{t \in \R: \sum_{i=1}^n F_i(t)\le k-1\}$. 
There exists $\mathbf{X}\in\mathcal F(F_1,\ldots,F_n)$ such that $Y_{(k)}\le_{\rm st} X_{(k)}$ for all $(Y_1, \dots, Y_n) \in \mathcal{F}(F_1, \dots, F_n)$
if and only if
both of the following conditions hold:
\begin{enumerate}
    \item[(iii)] For every $i\in[n]$, the function $t\mapsto (F_i(t)-H(t))_+$ is decreasing on $(\sigma, +\infty)$.

    \item[(iv)] {If $\sigma>-\infty$,}
\begingroup
\[\sum_{i=1}^n\bigl[\bigl((F_i(\sigma)\vee H(\sigma))-F_i(\sigma-)\bigr)\wedge H(\sigma)\bigr]\ge (n-k+1)H(\sigma).\]
\endgroup
\end{enumerate}
\end{proposition}
\begin{proof}
Let $\widehat F_i(t)=1-F_i((-t)-)$ for all $t \in {\R}$.
It is clear that  $(Y_1, \dots, Y_n)\in \mathcal{F}(F_1, \dots, F_n)$ if and only if $(-Y_1, \dots, -Y_n) \in \mathcal{F}(\widehat F_1, \dots, \widehat F_n)$ and $Y_{(k)}=-(-\mathbf{Y})_{(n-k+1)}$. Hence, there exists $\mathbf{X}\in\mathcal F(F_1,\ldots,F_n)$ such that $Y_{(k)}\le_{\rm st} X_{(k)}$ for all $(Y_1, \dots, Y_n) \in \mathcal{F}(F_1, \dots, F_n)$ if and only if there exists {$(Z_1,\ldots,Z_n)\in\mathcal F(\widehat F_1,\ldots,\widehat F_n)$ such that $Z_{(n-k+1)}\le_{\rm st} Y_{(n-k+1)}$} for all $(Y_1, \dots, Y_n) \in \mathcal{F}(\widehat  F_1,\ldots, \widehat F_n)$. 

Let 
$\widehat \tau= \inf\left\{t \in \R: \sum_{i=1}^n \widehat F_i(t)\ge n-k+1\right\}$.
We have
$\widehat \tau=\inf\left\{t \in \R: \sum_{i=1}^n F_i((-t)-)\le k-1\right\}=-\sup\left\{t \in \R: \sum_{i=1}^n F_i(t-)\le k-1\right\}$.
Since $F_i(t-)\le F_i(t)$, it is clear we have $\sigma\le-\widehat \tau$. For any $t$ such that $\sum_{i=1}^n F_i(t-)\le k-1$, we have $\sum_{i=1}^n F_i(s)\le \sum_{i=1}^n F_i(t-)\le k-1$ for all $s<t$. Hence, $\sigma \ge -\widehat{\tau}$. Therefore, $-\sigma=\widehat \tau$.

Let $\widehat{\mathbf{p}}(t)=(\widehat F_1(t), \dots, \widehat F_n(t))=\mathbf{1}_n-(F_1((-t)-), \dots, F_n((-t)-))$. Then, we have  $$(b_{n-k+1}(\widehat{\mathbf{p}}(t))-\widehat F_i(t))_+=(F_i((-t)-)-H((-t)-))_+.$$ 
It is clear that $t \mapsto (F_i((-t)-)-H((-t)-))_+$ is increasing on $(-\infty, -\sigma)$ if and only if $t \mapsto (F_i(t-)-H(t-))_+$ is decreasing on $(\sigma, \infty)$. {Since $F_i$ and $H$ are cdfs, $t\mapsto (F_i(t-)-H(t-))_+$ is decreasing on $(\sigma,\infty)$ if and only if $t\mapsto (F_i(t)-H(t))_+$ is decreasing on $(\sigma,\infty)$. This gives (iii).}

If $\sigma>-\infty$, $\widehat \tau<+\infty$. In this case,  we have $\widehat F_i(\widehat{\tau})=\widehat F_i(-\sigma)=1-F_i(\sigma-)$, $\widehat F_i(\widehat{\tau}-)=\widehat F_i((-\sigma)-)=1-F_i(\sigma)$, $b_{n-k+1}(\widehat{\mathbf{p}}(\widehat \tau-))=b_{n-k+1}(\widehat{\mathbf{p}}((-\sigma)-))=b_{n-k+1}(\mathbf{1}_n-\mathbf{p}(\sigma))=1-H(\sigma)$. Substituting these identities in (ii) of Theorem~\ref{thm:universal-attainment} proves (iv).
\end{proof}

\begin{remark}\label{rem:con2}
In the case that $F_1, \dots, F_n$ are continuous, we have $t\mapsto b_k(\mathbf{p}(t))$ is continuous, and hence $\widetilde F_1, \dots, \widetilde F_n$ and $\widehat F_1, \dots, \widehat F_n$ in the proofs of Theorem~\ref{thm:universal-attainment} and Proposition~\ref{prop:upper} are continuous. Therefore, conditions (ii) and (iv)  hold by Remark~\ref{rem:con1}.
\end{remark}

%  \begin{example}
% When $k=1$ and $k=n$, conditions (i) to (\textcolor{blue}{iv}) hold for all $F_1, \dots, F_n \in \mathcal{M}$.
% \begin{enumerate}[(1)]
% \item When $k=1$, by Example \ref{ex:1-n}, we have $b_1(\bp(t))=\sum_{i=1}^n F_i(t)$ for $t<\tau$. Hence, $(\textcolor{blue}{b_1(\bp(t))-F_i(t)})_+=\sum_{j\neq i} F_j(t)$, which is increasing with $t$. Hence, condition (i) holds.
% For condition (iii), we have $H(t)=\max_{i\in [n]}F_i(t)$ and $(F_i(t)-H(t))_+=0$. Hence, condition (iii) holds.
% By Remark \ref{rem:min}, the upper bound is achieved by co-monotonicity,  and the lower bound is achieved by $\bp$-tail counter-monotonicity for some $\bp\in \Delta_n$.
% \item When $k=n$, by Example \ref{ex:1-n}, we have $b_n(\bp(t))=\min_{i\in[n]} F_i(t)$. Hence, $(b_n(\bp(t))-F_i(t))_+=0$, which gives condition (i).
% For condition (iii), we have $H(t)=(\sum_{i=1}^n F_i(t)-(n-1))_+$. Hence, \textcolor{blue}{for $t>\sigma$, }$(F_i(t)-H(t))_+=n-1-\sum_{j\neq i}F_j(t)$, which is decreasing in $t$.
% By Theorem \ref{thm:max}, the upper bound is achieved by $\bp$-tail counter-monotonicity for some $\bp\in\Delta_n$, and the lower bound is achieved by co-monotonicity.
% \textcolor{blue}{In both cases the atomic conditions follow from the existence of these couplings.}
% \end{enumerate}
% \end{example}

 \begin{example}
When $k=1$ and $k=n$, conditions (i)--(iv) hold for all $F_1, \dots, F_n \in \mathcal{M}$.
\begin{enumerate}[(1)]
\item 
When $k=1$, by Example~\ref{ex:1-n}, we have $b_1(\mathbf{p}(t))=\sum_{i=1}^n F_i(t)$ 
for $t < \tau$. Hence, $(b_1(\mathbf{p}(t))-F_i(t))_+=\sum_{j\neq i} F_j(t)$, which is increasing with $t$. Hence, condition (i) holds. If there exists $F_i(\tau)-F_i(\tau-)\ge 1-b_k(\mathbf{p}(\tau-))$, it is clear that (ii) holds. If $F_i(\tau)-F_i(\tau-)< 1-b_k(\mathbf{p}(\tau-))$ for all $i \in [n]$, we have $\sum_{i=1}^n F_i(\tau)-F_i(\tau-)\ge 1-\sum_{i=1}^n F_i(\tau-)=1-b_k(\mathbf{p}(\tau-))$. Hence, (ii) holds.
For condition (iii), we have $H(t)=\max_{i\in [n]}F_i(t)$ and $(F_i(t)-H(t))_+=0$. Hence, condition (iii)  holds.
Since
$\sigma=\sup\{t \in \R: \sum_{i=1}^n F_i(t)\le 0\}$, $F_i(\sigma-)=0$. Therefore, $\sum_{i=1}^n \left[\left(F_i(\sigma)\vee H(\sigma)-F_i(\sigma-)\right) \wedge H(\sigma)\right]=nH(\sigma)$, and  (iv)  holds.
By Remark~\ref{rem:min}, the upper bound is achieved by comonotonicity, and the lower bound is achieved by $\mathbf{p}$-tail counter-monotonicity for some $\mathbf{p}\in \Delta_n$.

\item When $k=n$, by Example~\ref{ex:1-n},
we have $b_n(\mathbf{p}(t))=\min_{i \in [n]} F_i(t)$. Hence, $(b_n(\mathbf{p}(t))-F_i(t))_+=0$, which gives condition (i). 
    Since $\sum_{i=1}^n F_i(\tau)\ge n$.  $F_i(\tau)=1$ for all $i\in [n]$. Therefore,  
    $$\left(F_i(\tau)-\bigl(F_i(\tau-)\wedge b_k(\mathbf{p}(\tau-))\bigr)\right) \wedge\left(1-b_k(\mathbf{p}(\tau-))\right)=1-b_k(\mathbf{p}(\tau-),$$
    which gives condition (ii). For condition (iii), we have 
  $H(t)=(\sum_{i=1}^n F_i(t)-(n-1))_+$. Hence, for $t >\sigma$, $(F_i(t)-H(t))_+=n-1-\sum_{j\neq i}F_j(t)$,  which is decreasing in $t$.
  Moreover, $\sigma=\sup\{t \in \R: \sum_{i=1}^n F_i(t)\le n-1\}$.  Hence, $H(\sigma)=\sum_{i=1}^n F_i(\sigma)-(n-1)\le F_i(\sigma)$ for all $i \in [n]$. If there exists $F_i(\sigma)-F_i(\sigma-)\ge H(\sigma)$, it is clear that condition (iv) holds. If $F_i(\sigma)-F_i(\sigma-)< H(\sigma)$ for all $i \in [n]$, we have $\sum_{i=1}^n F_i(\sigma)-F_i(\sigma-)\ge \sum_{i=1}^n F_i(\sigma)-(n-1)=H(\sigma)$. Hence, condition (iv)  holds. By Theorem  \ref{thm:max}, the upper bound is achieved by $\mathbf{p}$-tail counter-monotonicity for some $\mathbf{p}\in \Delta_n$, and the lower bound is achieved by comonotonicity. 
\end{enumerate}
\end{example}

In the following proposition, we provide some Fr\'echet classes such that conditions (i)--(iv) hold.

\begin{proposition}\label{prop:marginal}
Let $F_1, \dots, F_n \in \mathcal{M}$.
\begin{enumerate}[(a)]
\item  If $F_1=\dots=F_n$, then conditions (i)--(iv) hold for all $k \in [n]$.
\item  If $F_1, \dots, F_n$ have density functions $f_1, \dots, f_n$ such that $\left(k \vee(n-k+1)\right)f_i\le \sum_{i=1}^n f_i$ for all $i\in [n]$, then conditions (i)--(iv) hold. 
\end{enumerate}
\end{proposition}

\begin{proof}
\begin{enumerate}[(a)]
\item Assume $F_1=\dots=F_n=F$.  We have $\tau=\inf\{t\in\mathbb R:F(t)\ge k/n\}$, and
$
    b_k(\mathbf{p}(t))=1\wedge({nF(t)}/{k}).
    $
For $t<\tau$, we have $F(t)<k/n$, and hence
$
    \bigl(b_k(\mathbf{p}(t))-F(t)\bigr)_+
    ={(n-k)F(t)}/{k},
$
which is increasing in $t$. Thus, condition (i) holds.
If $\tau$ is finite, then $F(\tau)\ge k/n$ and let $m=1-nF(\tau-)/k$. Thus $F(\tau)-F(\tau-)\ge (k/n)m$.
The left-hand side of (ii) is $n((F(\tau)-F(\tau-))\wedge m)$ and is at least $km$.
Apply the same argument to $\widehat F_i(t)=1-F_i((-t)-)$ and rank $n-k+1$ to obtain (iii) and (iv).

\item Since $F_1, \dots, F_n$ are continuous, conditions (ii) and (iv) hold by Remark~\ref{rem:con2}. We only need to check conditions (i) and (iii). 

We first prove that if $kf_i\le \sum_{i=1}^n f_i$ for all $i \in [n]$, then condition (i) holds. Let $s<t<\tau$ and $i \in [n]$. We need to show that $(G(t)-F_i(t))_+\ge (G(s)-F_i(s))_+$ for all $s<t<\tau$. If $G(s)\le F_i(s)$, then $(G(t)-F_i(t))_+\ge 0= (G(s)-F_i(s))_+$. Without loss of generality, assume $G(s)>F_i(s)$.

Let $J=\{j \in [n]: F_j(t)-F_j(s)\ge F_i(t)-F_i(s)\}$ and $r=\vert J\vert$. 
If $r\ge k$, it is clear that $\sum_{j=1}^n \left(F_j(t)-F_j(s)\right)\wedge \left(F_i(t)-F_i(s)\right)\ge k (F_i(t)-F_i(s))$. Now, assume $r<k$. Let $\Delta=\sum_{j=1}^n F_j(t)-F_j(s)$.
Since  $kf_j\le \sum_{l=1}^n f_l$ for all $j\in [n]$, we have $F_j(t)-F_j(s)\le \Delta/k$ for all $j \in [n]$. Hence,
$\sum_{j \in J} F_j(t)-F_j(s)\le r \Delta/k$. As a result, $$\sum_{j \notin J}F_j(t)-F_j(s)=\Delta-\sum_{j \in J} \left(F_j(t)-F_j(s)\right)\ge \frac{k-r}{k}\Delta\ge (k-r) (F_i(t)-F_i(s)).$$
Therefore, $$\sum_{j=1}^n \left(F_j(t)-F_j(s)\right)\wedge \left(F_i(t)-F_i(s)\right)\ge 
r (F_i(t)-F_i(s))+\sum_{j \notin J}F_j(t)-F_j(s)\ge k(F_i(t)-F_i(s)).$$
In conclusion, we have $\sum_{j=1}^n \left(F_j(t)-F_j(s)\right)\wedge \left(F_i(t)-F_i(s)\right)\ge k(F_i(t)-F_i(s))$.

 For $t<\tau$, by definition of $b_k$, we have $\sum_{j=1}^n F_j(t)\wedge G(t)=kG(t)$. Let $b=G(s)+F_i(t)-F_i(s)$. We have
\begin{align*}
\sum_{j=1}^n F_j(t)\wedge b &=\sum_{j=1}^n (F_j(s)+F_j(t)-F_j(s))\wedge \left(G(s)+F_i(t)-F_i(s)\right)\\ &\ge \sum_{j=1}^n (F_j(s)\wedge G(s))+\sum_{j=1}^n \left((F_j(t)-F_j(s))\wedge (F_i(t)-F_i(s))\right)\\
&\ge kG(s)+k(F_i(t)-F_i(s))=kb.
\end{align*}
Hence, by definition of $b_k(\mathbf{p}(t))$, we have $G(t)\ge b$.
Therefore, $$G(t)-F_i(t)\ge b-F_i(s)+(F_i(s)-F_i(t))=G(s)+(F_i(t)-F_i(s))-F_i(s)+(F_i(s)-F_i(t))=G(s)-F_i(s)>0.$$
In conclusion, we have $(G(t)-F_i(t))_+\ge (G(s)-F_i(s))_+$, which proves (i).

Let $\widehat f_i(t)=f_i(-t)$ for all $t \in \R$; that is, $\widehat f_i$ is the density of $\widehat F_i$ in the proof of Proposition~\ref{prop:upper}. By $(n-k+1)f_i\le \sum_{i=1}^n f_i$, we have $(n-k+1)\widehat f_i\le \sum_{i=1}^n \widehat f_i$. Hence, condition (i) in Theorem~\ref{thm:universal-attainment} holds for $\widehat F_1, \dots, \widehat F_n$ for the $n-k+1$ order. By $-X_{(k)}=(-\mathbf{X})_{(n-k+1)}$, condition (iii) holds.\qedhere
\end{enumerate}
\end{proof}

In the identical distributions case, Proposition~\ref{prop:identical} provides a {stochastic representation} that achieves the upper bound and lower bound in the stochastic order.

\begin{proposition}\label{prop:identical}
Suppose $F \in \mathcal M$ and $t \in \R$. 
Let $J$  be a random set that is uniformly distributed over all $\{K: K \subseteq [n]: \vert K\vert=k\}$,  {and let $V,W\sim \mathrm{U}[0,1]$ be independent of $J$.  Define} 
\begin{equation}\label{eq:identical}
    \underline{U_i} =\left\{\begin{aligned}
&\frac{k}{n} V, & i\in J,\\
&\frac{k}{n}+\left(1-\frac{k}{n}\right)W, & i\notin J.
\end{aligned}
\right., ~~~i\in [n]
\end{equation} 
and $(\underline{X_1}, \dots, \underline{X_n})=(F^{-1}( \underline{U_1}), \dots, F^{-1}( \underline{U_n}))$. Let $J'$ be a random set that is uniformly distributed over all $\{K: K \subseteq [n]: \vert K\vert=n-k+1\}$ and be independent of $V$ and $W$. Define
\begin{equation}\label{eq:identical-reflected}
    \overline{U_i} =\left\{\begin{aligned}
&\frac{n-k+1}{n} V, & i\in {J^\prime},\\
&\frac{n-k+1}{n}+\left(1-\frac{n-k+1}{n}\right)W, & i\notin {J^\prime}.
\end{aligned}
\right., ~~~i\in [n]
\end{equation} 
and $(\overline{X_1},\dots,\overline{X_n})=(-G^{-1}(\overline{U_1}),\dots,-G^{-1}(\overline{U_n}))$ where  $G$ is the distribution of $-X$ for $X\sim F$. Then, we have $(\underline{X_1}, \dots, \underline{X_n}) \in \mathcal{F}(F, \dots, F)$, $(\overline{X_1}, \dots, \overline{X_n}) \in \mathcal{F}(F, \dots, F)$,
$$\underline{X}_{(k)}\le_{\rm st} X_{(k)}\le_{\rm st} \overline{X}_{(k)}$$
for all $\mathbf{X} \in \mathcal{F}(F, \dots, F)$ and $t \in \R$. Moreover, 
$$\p\left(\overline{X}_{(k)}\le t\right)=\left(\frac{nF(t)-(k-1)}{n-k+1}\right)_+ ~~~\mbox{and}~~~ \p\left(\underline{X}_{(k)}\le t\right)=1 \wedge \frac{nF(t)}{k}.$$
\end{proposition}

\begin{proof}
% Let $J$  be a random set that is uniformly distributed over all $\{K: K \subset [n]: \vert K\vert=k\}$ and  $V,W \sim \mathrm{U}[0,1]$. 
% For $i\in [n]$, define
% \begin{equation}\label{eq:indentical}
%     U_i =\left\{\begin{aligned}
% &\frac{k}{n} V, & i\in J,\\
% &\frac{k}{n}+\left(1-\frac{k}{n}\right)W, & i\notin J.
% \end{aligned}
% \right.
% \end{equation}
For $\underline{U_1}, \dots, \underline{U_n}$ in \eqref{eq:identical}, we have $\underline{U_i}\le k/n$ for $i \in J$ and $\underline{U_i}> k/n$ for $i \notin J$ a.s. and $\p(i \in J)=k/n$. %\textcolor{blue}{It follows by mixing the two uniform distributions with weights $k/n$ and $1-k/n$ that each $\underline U_i$ is uniform on $[0,1]$.}
%Let $(\overline{X_1}, \dots, \overline{X_n})=(F^{-1}(U_1), \dots, F^{-1}(U_n))$.
Hence,
\begin{align*}
\left\{{\underline{X}}_{(k)}\le t\right\}
=\left\{\sum_{i=1}^n \id_{\{{\underline{X_i}}\le t\}}\ge k\right\}
=\left\{\sum_{i=1}^n \id_{\{\underline{U_i}\le {F(t)}\}}\ge k\right\}
=\left\{\sum_{i\in J} \id_{\{\underline{U_i}\le F(t)\}}+\sum_{i\notin J} \id_{\{\underline{U_i}\le F(t)\}}\ge k\right\}.
\end{align*}

For $t\in \R$ such that $F(t)> k/n$, we have $\sum_{i\in J} \id_{\{\underline{U_i}\le F(t)\}}=k$. Hence, $\p\left({\underline{X}}_{(k)}\le t\right)=1$.
For $t\in \R$ such that $F(t)\le k/n$, we have $\id_{\{\underline{U_i}\le F(t)\}}=0$ a.s. for all $i \notin J$. Hence, 
$$\left\{\sum_{i\in J} \id_{\{\underline{U_i}\le F(t)\}}+\sum_{i\notin J} \id_{\{\underline{U_i}\le F(t)\}}\ge k\right\}=\left\{\sum_{i\in J} \id_{\left\{V\le \frac{n}{k}F(t)\right\}}\ge k\right\}.$$ 
It is clear that  $\sum_{i\in J} \id_{\{V\le {nF(t)}/{k}\}}= k$ if $V\le nF(t)/k$ and $\sum_{i\in J} \id_{\{V\le{nF(t)}/{k}\}}= 0$ if $V> nF(t)/k$. Hence, $\p\left({\underline{X}}_{(k)}\le t\right)=\p(V\le nF(t)/k)=nF(t)/k$.
By taking $p_1=\dots=p_n=F(t)$ in Proposition~\ref{prop:order}, we have $$\p\left(\underline{X}_{(k)}\le t\right)=1 \wedge \frac{nF(t)}{k}\ge \p(X_{(k)}\le t)$$
for all $t\in \R$  and $\mathbf{X}\in \mathcal{F}(F, \dots, F)$, which implies $ \underline{X}_{(k)}\le_{\rm st}X_{(k)}$.

By the same argument, $(-\overline{\mathbf{X}})_{(n-k+1)}\le_{\rm st}{(-\mathbf{X})_{(n-k+1)}}$ follows for all $-\mathbf{X}\in\mathcal{F}({G,\dots,G})$.
By $X_{(k)}=-(-\mathbf{X})_{(n-k+1)}$, we have the upper bound.
%\textcolor{blue}{Its cdf is $1-\min\{1,n(1-F(t))/(n-k+1)\}$ by reflection and left limits.}
% By the same logic, we can find $(\underline{X_1}, \dots, \underline{X_n})$ such that 
% $$\p\left(\sum_{i=1}^n \id_{\{\underline{X_i}> t\}}\ge n- k+1\right)=1\wedge \frac{n(1-F(t))}{n-k+1}.$$
% Hence, we have
% $$\p(\underline{X_{(k)}}\le t)=\p\left(\sum_{i=1}^n \id_{\{\underline{X_i}\le t\}}\ge k\right)= 1-\p\left(\sum_{i=1}^n \id_{\{\underline{X_i}> t\}}\ge n-k+1\right)=\left(\frac{nF(t)-k+1)}{n-k+1}\right)_+.
% $$
% By taking $p_1=\dots=p_n=F(t)$ in Theorem \ref{thm:order}, we have 
% $\p\left(\underline{X_{(k)}}\le t\right)\le \p(X_{(k)}\le t)$, for all $t\in \R$  and $(X_1, \dots, X_n )\in \mathcal{F}(F, \dots, F)$, which implies $X_{(k)}\le_{\rm st} \underline{X_{(k)}}$.
\end{proof}
By taking $k=n$ in \eqref{eq:identical}, we have $(\underline{U_1}, \dots, \underline{U_n})=(V, \dots, V)$ and $(\underline{X_1}, \dots, \underline{X_n})$ is comonotonic. {Proposition} \ref{prop:identical} implies that the maximum of $X_1, \dots, X_n$ achieves the lower bound in the stochastic order when $\mathbf{X}$ is comonotonic, as shown in Theorem~\ref{thm:max}. {For $k=1$, $J$ is a random singleton uniformly distributed over $\{\{1\}, \dots, \{n\}\}$. Hence,  $(-X_1, \dots, -X_n)$ is $(1/n,\dots,1/n)$-tail counter-monotonic.} {Proposition} \ref{prop:identical} implies that the minimum of $X_1, \dots, X_n$ achieves the lower bound in stochastic {order under this coupling}, as shown in Remark~\ref{rem:min}. 

\begin{example}
We provide some examples with non-identical marginals that satisfy (b) in Proposition~\ref{prop:marginal}.
\begin{enumerate}[(1)]
\item Let $k=2$, $X_1, X_2\sim \mathrm{N}(0,1)$ and $X_3\sim \mathrm{N}(0,(3/4)^2)$. It can be verified that $2f_i\le \sum_{j=1}^n f_j$ for all $i \in [3]$.  
\item Let $k=2$, $X_1, X_2\sim \mathrm{U}(-1,1)$ and $X_3\sim \mathrm{U}(-3/4,3/4)$. It can be verified that $2f_i\le \sum_{j=1}^n f_j$ for all $i \in [3]$. 
\end{enumerate}
\end{example}

   Based on Theorem~\ref{thm:universal-attainment} and Proposition~\ref{prop:upper}, we have the bounds for law-invariant and monotone risk measures on the $k$-th order statistics of dependent risks. {The proof is straightforward and we omit it.}
\begin{corollary}\label{cor:order}
   Suppose that $\rho$ is law-invariant and monotone. If conditions (i) and (ii) hold, there exists $(\underline{X}_1, \dots, \underline{X}_n) \in \mathcal{F}(F_1, \dots, F_n)$ such that 
   $$\inf_{\mathbf{X}\in \mathcal F({F_1,\dots,F_n})} \rho(X_{(k)})=\rho(\underline{X}_{(k)}).$$
   If conditions (iii) and (iv) hold, there exists $(\overline{X}_1, \dots, \overline{X}_n) \in \mathcal{F}(F_1, \dots, F_n)$ such that 
    $$\sup_{\mathbf{X}\in \mathcal F({F_1,\dots,F_n})} \rho(X_{(k)})=\rho(\overline{X}_{(k)}).$$
If $F_1=\dots=F_n=F$, $(\overline{X}_1, \dots, \overline{X}_n)$ and $(\underline{X}_1, \dots, \underline{X}_n)$ are constructed in Proposition~\ref{prop:identical}.
\end{corollary}
%\begin{proof}
%The proof follows from law-invariance and monotonicity of $\rho$,  there exists $\underline{X}_{(k)}\le_{\rm st} X_{(k)}$ if condition (i) and (ii) hold and there exists $X_{(k)}\le_{\rm st}  \overline{X}_{(k)}$ if condition (iii) and (iv) hold. 
%\end{proof}

\begingroup

\section{Risk measure bounds for order statistics}\label{sec:risk-measure-bounds}

Given marginal distribution functions $F_1,\ldots,F_n$,
Proposition~\ref{prop:order} provides the bounds at a fixed threshold:
\[
H(t)\le \p(X_{(k)}\le t)\le G(t),\qquad t\in\R,\quad \mathbf{X}\in\mathcal F(F_1,\ldots,F_n),
\]
where $G(t)=b_k(\mathbf{p}(t))$, $H(t)=1-b_{n-k+1}(\mathbf{1}_n-\mathbf{p}(t))$ and $\mathbf{p}(t)=(F_1(t),\ldots,F_n(t))$.
Let $Y_G$ and $Y_H$ be random variables with distribution functions $G$
and $H$, respectively.
Suppose that the cost $c$ is increasing.
If conditions (i) and (ii) of Theorem~\ref{thm:universal-attainment} hold,
then a single coupling has $X_{(k)}\sim G$. By Corollary~\ref{cor:order},
the infimum in \eqref{eq:intro-rm} is $\rho(c(Y_G))$ for every
law-invariant and monotone risk measure $\rho$ for which the problem is
well defined. Similarly, conditions (iii) and (iv) of Proposition~\ref{prop:upper}
solve the supremum $\rho(c(Y_H))$. 

For identical marginals, 
both conditions (i) and (ii) of Theorem~\ref{thm:universal-attainment}
and conditions (iii) and (iv) of Proposition~\ref{prop:upper} 
follow from Proposition~\ref{prop:marginal}, 
which yields the following explicit bounds.

\begin{corollary}\label{cor:identical-risk-bounds}
Let $F\in\mathcal M$ and $X\sim F$. For each $\alpha\in(0,1)$, we have
\[
\inf_{\mathbf{X}\in\mathcal F(F,\ldots,F)}\VaR_\alpha(X_{(k)})
=\VaR_{1-k(1-\alpha)/n}(X),\qquad
\sup_{\mathbf{X}\in\mathcal F(F,\ldots,F)}\VaR_\alpha(X_{(k)})
=\VaR_{(n-k+1)\alpha/n}(X).
\]
Moreover, if $X\in L^1$, we have
\[
\inf_{\mathbf{X}\in\mathcal F(F,\ldots,F)}\ES_\alpha(X_{(k)})
=\RVaR_{1-k/n,\,1-k(1-\alpha)/n}(X),\qquad
\sup_{\mathbf{X}\in\mathcal F(F,\ldots,F)}\ES_\alpha(X_{(k)})
=\ES_{(n-k+1)\alpha/n}(X),
\]
\[
\inf_{\mathbf{X}\in\mathcal F(F,\ldots,F)}\E[X_{(k)}]
=\RVaR_{1-k/n,\,1}(X),
\quad\text{and}\quad
\sup_{\mathbf{X}\in\mathcal F(F,\ldots,F)}\E[X_{(k)}]
=\ES_{(n-k+1)/n}(X).
\]
All the stated extrema are attained.
\end{corollary}

\begin{proof}
By Proposition~\ref{prop:identical}, the universal infimum and supremum are attained by the order statistics
$\underline X_{(k)}$ and $\overline X_{(k)}$ with distribution functions
\[
\p(\underline X_{(k)}\le t)=1\wedge\frac{nF(t)}{k}
\quad\text{and}\quad
\p(\overline X_{(k)}\le t)=\left(\frac{nF(t)-(k-1)}{n-k+1}\right)_+,
\]
respectively. By definition of VaR, we obtain
\begin{align*}
\VaR_\alpha(\underline X_{(k)})
&=\inf\left\{t\in\R:F(t)\ge\frac{k(1-\alpha)}{n}\right\}
=\VaR_{1-k(1-\alpha)/n}(X),\\
\VaR_\alpha(\overline X_{(k)})
&=\inf\left\{t\in\R:F(t)\ge 1-\frac{n-k+1}{n}\alpha\right\}
=\VaR_{(n-k+1)\alpha/n}(X).
\end{align*}
Then integrating the above formulas yields
\begin{align*}
\ES_\alpha(\underline X_{(k)})
=\frac1\alpha\int_0^\alpha\VaR_{1-k(1-\beta)/n}(X)\,\d\beta
=\frac{n}{k\alpha}\int_{1-k/n}^{1-k(1-\alpha)/n}\VaR_\beta(X)\,\d\beta
=\RVaR_{1-k/n,\,1-k(1-\alpha)/n}(X),
\end{align*}
and 
\begin{align*}
\ES_\alpha(\overline X_{(k)})
=\frac1\alpha\int_0^\alpha\VaR_{(n-k+1)\beta/n}(X)\,\d\beta
=\frac{n}{(n-k+1)\alpha}\int_0^{(n-k+1)\alpha/n}\VaR_\beta(X)\,\d\beta
=\ES_{(n-k+1)\alpha/n}(X).
\end{align*}
Taking $\alpha=1$ in the above formulas yields the mean results.
\end{proof}

When any of conditions (i)--(iv) fails, the corresponding
bound is unattainable as a distribution function of $X_{(k)}$. 
Therefore, it is meaningful to study whether the bounds on VaR, ES
and the mean are attained.

\subsection{Attainment in distorted optimal transport}

A distortion function is an increasing map $g:[0,1]\to[0,1]$ with
$g(0)=0$ and $g(1)=1$. For a random variable $Y$, its
distorted expectation under $g$ is
\[
\rho_g(Y)=\int_0^\infty g(\p(Y>t))\,\d t-\int_{-\infty}^0\bigl(1-g(\p(Y>t))\bigr)\,\d t
\]
whenever both integrals are finite. For a Borel cost function
$c:\R^n\to\R$, the distorted optimal transport problems of
\citet{bib:liu-et-al-2026} with given marginals have the form
\begin{equation}\label{eq:distorted-rank-transport}
\inf_{\mathbf{X}\in\mathcal F(F_1,\ldots,F_n)}\rho_g(c(\mathbf{X}))
\quad\text{and}\quad
\sup_{\mathbf{X}\in\mathcal F(F_1,\ldots,F_n)}\rho_g(c(\mathbf{X})).
\end{equation}
For $c(\mathbf{x})=x_{(k)}$, these are the risk measure problems in
\eqref{eq:intro-rm} with $\rho=\rho_g$. Different choices of the distortion function $g$ 
yields different risk measures. For instance, taking $g(u)=u$, $\rho_g$ is the mean; take $g(u)=1\wedge(u/\alpha)$, $\rho_g$ is $\ES_\alpha$; take $g(u)=\id_{\{u>\alpha\}}$ and $g(u)=\id_{\{u\ge\alpha\}}$, 
$\rho_g$ is $\VaR_\alpha$ and $\VaR^+_\alpha$, respectively.

\begin{lemma}\label{lem:var-attainment-cost}
Let $F_1,\ldots,F_n\in\mathcal M$, let $g$ be a distortion function,
and let $c:\R^n\to\R$ be Borel measurable. Suppose that either of the
following conditions holds.
\begin{enumerate}[(a)]
\item $c(\mathbf{X})\in L^1$ for every $\mathbf{X}\in\mathcal F(F_1,\ldots,F_n)$, and
there is a finite constant $K$ such that $g(u)\le Ku$ and
$1-g(1-u)\le Ku$ for every $u\in[0,1]$.
\item $g$ is zero on a neighborhood of $0$ and one on a neighborhood of $1$.
\end{enumerate}
If $c$ is lower semicontinuous and $g$ is left-continuous on $(0,1]$,
the infimum in \eqref{eq:distorted-rank-transport} is attained.
If $c$ is upper semicontinuous and $g$ is right-continuous on $[0,1)$,
the supremum in \eqref{eq:distorted-rank-transport} is attained.
\end{lemma}

\begin{proof}
Write $\Pi$ for the weakly compact set of joint laws in
$\mathcal F(F_1,\ldots,F_n)$. We only need to prove attainment of the infimum,
since the proof of the supremum is symmetric.
For a joint law
$\pi\in\Pi$, if $\pi_j$ converges weakly to $\pi$, then 
$\pi(\{\mathbf{x}:c(\mathbf{x})>t\})\le\liminf_j \pi_j(\{\mathbf{x}:c(\mathbf{x})>t\})$ holds. As $g$ is
increasing and left-continuous on $(0,1]$, it holds that
\begin{equation}\label{eq:bounds-of-distort}
    g(\pi(\{\mathbf{x}:c(\mathbf{x})>t\}))\le\liminf_{j\to\infty}g(\pi_j(\{\mathbf{x}:c(\mathbf{x})>t\})),\qquad t\in\R.
\end{equation}
The following proof is divided into three steps.

(1) Under condition (a), we first show that $\sup_{\pi\in\Pi}\E_\pi[|c(\mathbf{X})|]<\infty$.
Suppose not. We could choose $\pi_j\in\Pi$ with
$\E_{\pi_j}[|c(\mathbf{X})|]\ge 2^j$ for every $j\ge1$. Then
$\sum_{j=1}^\infty 2^{-j}\pi_j$ belongs to $\Pi$,
which yields a joint law that violates $c(\mathbf{X})\in L^1$.
Thus, $\sup_{\pi\in\Pi}\E_\pi[|c(\mathbf{X})|]<\infty$ necessarily holds.

(2) For $M>0$, we show 
\begin{equation}\label{eq:distortion-negative-tail}
\lim_{M\to\infty}\sup_{\pi\in\Pi}\int_{-\infty}^{-M}\bigl(1-g(\pi(\{\mathbf{x}:c(\mathbf{x})>t\}))\bigr)\,\d t=0.
\end{equation}

Under (a), from step (1) and Theorem~2.3 of \citet{bib:ruschendorf-2013}, 
there exist functions $f_i$ with $\int f_i\,\d F_i<\infty$ such that
$|c(\mathbf{X})|\le\sum_i f_i(X_i)$ $\pi$-a.s. for every $\pi\in\Pi$.
The integral in \eqref{eq:distortion-negative-tail} is bounded by
\[
K\E_\pi[(-c(\mathbf{X})-M)_+]
\le K\sum_{i=1}^n\int(f_i(x)-M/n)_+\,\d F_i(x).
\]
The right-hand side tends to zero as $M\to\infty$.

Under (b), choose $\delta\in(0,1)$ such that $g(u)=1$ for
$u\in[1-\delta,1]$.
By tightness there is a compact set $C\subset\R^n$ with
$\pi(C)\ge1-\delta$ for every $\pi\in\Pi$. By lower
semicontinuity of $c$ and compactness of $C$, 
$\min_{\mathbf{x}\in C}c(\mathbf{x})$ exists and is finite. 
For every $t<\min_{\mathbf{x}\in C}c(\mathbf{x})$ and $\pi\in\Pi$, 
$C\subseteq\{\mathbf{x}\in\R^n:c(\mathbf{x})>t\}$ and $1-\delta\le \pi(C)\le \pi(\{\mathbf{x}:c(\mathbf{x})>t\})\le 1$ hold.
Hence $g(\pi(\{\mathbf{x}:c(\mathbf{x})>t\}))=1$ for every $t<\min_{\mathbf{x}\in C}c(\mathbf{x})$ and $\pi\in\Pi$. For
$M>\max\{0,-\min_{\mathbf{x}\in C}c(\mathbf{x})\}$, we have $-M<\min_{\mathbf{x}\in C}c(\mathbf{x})$ and thus
\[
\sup_{\pi\in\Pi}\int_{-\infty}^{-M}\bigl(1-g(\pi(\{\mathbf{x}:c(\mathbf{x})>t\})))\bigr)\,\d t=0.
\]

(3) Finally, we show the attainment of the infimum.

Under (a), the two integrals defining $\rho_g(c(\mathbf{X}))$
are bounded by $K\E_\pi[(c(\mathbf{X}))_+]$ and
$K\E_\pi[(-c(\mathbf{X}))_+]$, respectively. 
Under (b), both integrands
vanish sufficiently far because $g$ is constant near $0$ and $1$.
Thus $\rho_g(c(\mathbf{X}))$ is finite for every joint law $\pi\in\Pi$ of $\mathbf{X}$.
For $M>0$, consider
\[
J_M(\pi)=\int_{-M}^{\infty}
g\bigl(\pi(\{\mathbf{x}:c(\mathbf{x})>t\})\bigr)\,\d t-M.
\]
By \eqref{eq:bounds-of-distort}, $J_M(\pi)\le\liminf_{j\to\infty}J_M(\pi_j)$ holds whenever
$\pi_j$ converges weakly to $\pi$. Thus $J_M$ is lower semicontinuous.
For $\mathbf{X}$ with joint law $\pi$, we have
$J_M(\pi)-\rho_g(c(\mathbf{X}))=\int_{-\infty}^{-M}\bigl(1-g(\pi(\{\mathbf{x}:c(\mathbf{x})>t\}))\bigr)\,\d t$,
which tends to zero uniformly over $\Pi$ by \eqref{eq:distortion-negative-tail}. 
Therefore, $\rho_g(c(\mathbf{X}))$ is lower semicontinuous as a function of the joint law $\pi$,
and attains its minimum on the weakly compact set $\Pi$.
\end{proof}

\begin{remark}\label{rem:attain}
The function $\mathbf{x}\mapsto x_{(k)}$ is continuous.
Then Lemma~\ref{lem:var-attainment-cost}(b) applies and 
gives attainment of the infimum of $\VaR_\alpha(X_{(k)})$ and the supremum of
$\VaR^+_\alpha(X_{(k)})$ for arbitrary marginals. If all marginals have
finite first moments, then 
Lemma~\ref{lem:var-attainment-cost}(a) applies and 
gives attainment of both extrema for $\ES_\alpha(X_{(k)})$ and for the mean.
\end{remark}

\subsection{Value-at-Risk}\label{sec:value-at-risk}

The distribution bounds at a given threshold determine the VaR extrema.
Remark~\ref{rem:attain} already states that the infimum of $\VaR_\alpha(X_{(k)})$ is attained.
The following theorem mainly studies the attainment of the upper bounds.

\begin{theorem}\label{thm:var-attain}
For $0<\alpha<1$, we have
\begin{align}
\inf_{\mathbf{X}\in\mathcal F(F_1,\ldots,F_n)}\VaR_\alpha(X_{(k)})
&=G^{-1}(1-\alpha)
=\inf\{t\in\R:b_k(\mathbf{p}(t))\ge1-\alpha\}, \notag\\ 
\sup_{\mathbf{X}\in\mathcal F(F_1,\ldots,F_n)}\VaR_\alpha(X_{(k)})
&=H^{-1}(1-\alpha)
=\inf\{t\in\R:b_{n-k+1}(\mathbf{1}_n-\mathbf{p}(t))\le\alpha\}.
\label{eq:maximum-var}
\end{align}
The attainment of the supremum is determined as follows.
\begin{enumerate}[(i)]
\item When $H\bigl(H^{-1}(1-\alpha)-\bigr)<1-\alpha$,
the supremum is attained.

\item When $H\bigl(H^{-1}(1-\alpha)-\bigr)=1-\alpha$,
the supremum is attained if and only if
\begin{equation}\label{eq:cardinality-condition}
\bigl|\{i:\text{either }F_i\bigl(H^{-1}(1-\alpha)-\bigr)<1-\alpha
\text{ or }F_i\bigl(H^{-1}(1-\alpha)-\bigr)>F_i(t),\
\forall t<H^{-1}(1-\alpha)\}\bigr|\ge n-k+1.
\end{equation}
\end{enumerate}
\end{theorem}

\begin{proof}
By Proposition~\ref{prop:order}, we know
$G^{-1}(1-\alpha)\le \VaR_\alpha(X_{(k)})\le H^{-1}(1-\alpha)$ for every $\mathbf{X}\in\mathcal F(F_1,\ldots,F_n)$.
Taking $t=G^{-1}(1-\alpha)$ in Proposition~\ref{prop:order}, the corresponding
coupling $(\overline{X_1^t},\ldots,\overline{X_n^t})$ satisfies
$\p(\overline{X^t_{(k)}}\le t)=G(t)\ge1-\alpha$.
Thus, it attains the infimum of VaR.
Similarly, for each $t<H^{-1}(1-\alpha)$, there exists a coupling with
$\p(X_{(k)}\le t)=H(t)<1-\alpha$, and hence $\VaR_\alpha(X_{(k)})>t$.
Taking the supremum over these $t$ proves \eqref{eq:maximum-var}.
In the following, we verify the two assertions for attainment of the supremum.

(i) Applying Proposition~\ref{prop:order} to the reflected marginals at rank
$n-k+1$ and threshold $-H^{-1}(1-\alpha)$, there exists a coupling
$\mathbf{X}$ such that
\[
\p(X_{(k)}\ge H^{-1}(1-\alpha))
=1-H\bigl(H^{-1}(1-\alpha)-\bigr)>\alpha,
\]
where the last inequality follows from
$H\bigl(H^{-1}(1-\alpha)-\bigr)<1-\alpha$.
Hence $\p(X_{(k)}>t)>\alpha$ for every $t<H^{-1}(1-\alpha)$.
Together with \eqref{eq:maximum-var}, the upper VaR bound is attained.

(ii) \underline{The ``only if'' direction}. Suppose that a coupling
$\mathbf{X}\in\mathcal F(F_1,\ldots,F_n)$ attains the upper VaR bound.
We prove \eqref{eq:cardinality-condition} by contradiction.
Suppose that the set in \eqref{eq:cardinality-condition} contains fewer than
$n-k+1$ indices. The proof is divided into two steps.

(a) We first show that there exists $t_0<H^{-1}(1-\alpha)$ such that
every $X_i$ with index $i$ lying outside the set in
\eqref{eq:cardinality-condition} satisfies $X_i\le t_0$ almost surely
on $\{X_{(k)}<H^{-1}(1-\alpha)\}$.
Since $\mathbf{X}$ attains the upper VaR bound,
$\p(X_{(k)}\le t)<1-\alpha$ for every $t<H^{-1}(1-\alpha)$.
Letting $t$ increase to $H^{-1}(1-\alpha)$ and using
$H\bigl(H^{-1}(1-\alpha)-\bigr)=1-\alpha$, we obtain
$\p(X_{(k)}\ge H^{-1}(1-\alpha))=\alpha$.
Since $i$ lies outside the set in \eqref{eq:cardinality-condition}, we have
\begin{itemize}
\item[(i)] $F_i\bigl(H^{-1}(1-\alpha)-\bigr)\ge1-\alpha$, or equivalently
$\p(X_i\ge H^{-1}(1-\alpha))\le\alpha$.
By the proof of Proposition~\ref{prop:event-equality},
\[
\p\bigl(X_i\ge H^{-1}(1-\alpha),\ X_{(k)}\ge H^{-1}(1-\alpha)\bigr)
=\min\bigl\{\p(X_i\ge H^{-1}(1-\alpha)),\,\alpha\bigr\}
=\p(X_i\ge H^{-1}(1-\alpha)),
\]
which yields
$\p\bigl(X_i\ge H^{-1}(1-\alpha),\ X_{(k)}<H^{-1}(1-\alpha)\bigr)=0$.

\item[(ii)] There exists $t<H^{-1}(1-\alpha)$ such that
$F_i(t)=F_i\bigl(H^{-1}(1-\alpha)-\bigr)$.
Thus $\p(t<X_i<H^{-1}(1-\alpha))=0$.
We can choose some $t_0<H^{-1}(1-\alpha)$ with
$\p(t_0<X_i<H^{-1}(1-\alpha))=0$ for every such index $i$.
\end{itemize}
Together, (i) and (ii) imply $X_i\le t_0$ a.s. on
$\{X_{(k)}<H^{-1}(1-\alpha)\}$.

(b) Next, we show that $\p(X_{(k)}\le t_0)=1-\alpha$.
At least $k$ indices lie outside the set in \eqref{eq:cardinality-condition}.
By step (a), the corresponding $X_i$ all satisfy $X_i\le t_0$ a.s.
on $\{X_{(k)}<H^{-1}(1-\alpha)\}$. So $X_{(k)}\le t_0$ holds on this event.
On $\{X_{(k)}\ge H^{-1}(1-\alpha)\}$, we have $X_{(k)}>t_0$.
Therefore,
\[
\p(X_{(k)}\le t_0)
=\p\bigl(X_{(k)}<H^{-1}(1-\alpha)\bigr)
=1-\alpha.
\]
Thus, $\VaR_\alpha(X_{(k)})\le t_0<H^{-1}(1-\alpha)$ follows from the
definition of VaR, which contradicts the assumption that $\mathbf{X}$
attains the upper VaR bound.

\underline{The ``if'' direction}.
Suppose that \eqref{eq:cardinality-condition} holds.
We construct a coupling $\mathbf{X}\in\mathcal F(F_1,\ldots,F_n)$ that attains the upper
VaR bound. The proof is divided into two steps.

(i) We first construct a coupling $\mathbf{X}$ satisfying
$\p(X_{(k)}\ge H^{-1}(1-\alpha))=\alpha$.
By Remark~\ref{rem:k-layer}, we can construct an $(n-k+1)$-layered
$\bigl(\mathbf{1}_n-\mathbf{p}(H^{-1}(1-\alpha)-)\bigr)$-arrangement.
Let $B$ be the event on which at least $n-k+1$ events occur.
Then $\p(B)=1-H\bigl(H^{-1}(1-\alpha)-\bigr)=\alpha$.
In this construction, the part of the $i$th event in $B^c$ has
probability $\bigl(1-\alpha-F_i(H^{-1}(1-\alpha)-)\bigr)_+$.
At most $n-k$ of these probabilities are positive.
Choose the corresponding parts in $B^c$ in decreasing order
of probability, taking each to be a subset of the preceding one.

Conditional on the indicators of these events, generate
$U_1,\ldots,U_n$ independently satisfying
\[
U_i\sim
\begin{cases}
U\bigl(F_i(H^{-1}(1-\alpha)-),1\bigr),
& \text{if the $i$th event occurs},\\
U\bigl(0,F_i(H^{-1}(1-\alpha)-)\bigr),
& \text{otherwise}.
\end{cases}
\]
The $i$th event has probability $1-F_i(H^{-1}(1-\alpha)-)$,
so every $U_i$ is uniformly distributed on $[0,1]$.
Set $X_i=F_i^{-1}(U_i)$. Then $X_i$ has cdf $F_i$, and
$\{X_i\ge H^{-1}(1-\alpha)\}=\{U_i>F_i(H^{-1}(1-\alpha)-)\}$ a.s.
By the construction of $U_i$ and the definition of $B$,
$\{X_{(k)}\ge H^{-1}(1-\alpha)\}=B$ holds a.s.

(ii) For each fixed $t<H^{-1}(1-\alpha)$, we show that the coupling
$\mathbf{X}$ constructed in step (i) satisfies
$\p(t<X_{(k)}<H^{-1}(1-\alpha))>0$.
Divide the indices counted in \eqref{eq:cardinality-condition}
into the following two groups.
\begin{itemize}
\item[(G1)] Consider the indices satisfying
$F_i\bigl(H^{-1}(1-\alpha)-\bigr)<1-\alpha$.
By step (i), the events $B^c\cap\{X_i\ge H^{-1}(1-\alpha)\}$
for these indices are nested and have strictly positive probabilities.
Therefore,
\begin{equation}\label{eq:var-conditioning-event}
\p\!\left\{
B^c\cap\left(
\bigcap_{i\in[n],\,F_i(H^{-1}(1-\alpha)-)<1-\alpha}
\{X_i\ge H^{-1}(1-\alpha)\}
\right)\right\}>0.
\end{equation}
If there are no such indices, this event is $B^c$.
On the above intersection, $X_i\ge H^{-1}(1-\alpha)>t$ for every
$i\in[n]$ satisfying $F_i\bigl(H^{-1}(1-\alpha)-\bigr)<1-\alpha$.

\item[(G2)] Consider the indices satisfying
$F_i\bigl(H^{-1}(1-\alpha)-\bigr)\ge1-\alpha$ and
$F_i(s)<F_i\bigl(H^{-1}(1-\alpha)-\bigr)$ for every
$s<H^{-1}(1-\alpha)$. By step (i),
\[
\p\bigl(B^c\cap\{X_i\ge H^{-1}(1-\alpha)\}\bigr)
=\bigl(1-\alpha-F_i(H^{-1}(1-\alpha)-)\bigr)_+=0.
\]
Therefore, on the event in \eqref{eq:var-conditioning-event}, step (i) generates
$U_i$ from $U\bigl(0,F_i(H^{-1}(1-\alpha)-)\bigr)$, which implies
that $\{t<X_i<H^{-1}(1-\alpha)\}$ has strictly positive conditional
probability.
\end{itemize}
On the event in \eqref{eq:var-conditioning-event}, the event indicators in step (i)
are fixed a.s., so the $U_i$ are conditionally independent.
Moreover, the two groups together contain at least $n-k+1$ indices
by \eqref{eq:cardinality-condition}.
Thus, we obtain $\p(t<X_{(k)}<H^{-1}(1-\alpha))>0$.

Finally, combining steps (i) and (ii),
$\p(X_{(k)}>t)=\alpha+\p(t<X_{(k)}<H^{-1}(1-\alpha))>\alpha$
holds for every $t<H^{-1}(1-\alpha)$.
By the definition of VaR, $\VaR_\alpha(X_{(k)})\ge H^{-1}(1-\alpha)$.
Together with \eqref{eq:maximum-var}, the coupling $\mathbf{X}$
attains the upper VaR bound.
\end{proof}

The upper bound in \eqref{eq:maximum-var} is attained for every
$\alpha\in(0,1)$ if at least $n-k+1$ marginal distribution functions
are strictly increasing on $\R$. It is also attained for every $\alpha$
if all marginals have the same support. When $H\bigl(H^{-1}(1-\alpha)-\bigr)=1-\alpha$,
the following example provides a case where the supremum is unattained.

\begin{example}
For $n=3$, $k=2$, and $\alpha=1/2$, let $X_1$ have mass $1/2$
at each of $-1$ and $1$. Let $X_3$ have mass $3/4$ at $-1$
and $1/4$ at $1$. Let $X_2$ have mass $1/4$ at $1$, with its
remaining mass uniformly distributed on $(-1,0)$.
Note that $H(t)=0$ for $t<-1$, and for $-1\le t\le 0$,
\[
H(t)
=1-b_2\!\left(\frac12,\frac14-\frac{3t}{4},\frac14\right)
=\max\left\{\frac14,\frac12+\frac{3t}{8}\right\}.
\]
Then $H^{-1}(1/2)=0$ and $H(0-)=1/2$, and thus we have
$\{i:F_i(0-)<1/2\}=\varnothing$ and
$\{i:F_i(t)<F_i(0-)\ \text{for every }t<0\}=\{2\}$,
whereas $n-k+1=2$. By Theorem~\ref{thm:var-attain},
there is no coupling that attains the upper VaR bound of $0$.
\end{example}

\subsection{Mean and ES bounds}

Integrating the distribution bounds in Proposition~\ref{prop:order} over thresholds yields bounds on the mean
and ES for arbitrary marginals that are integrable. 
We first determine when the bounds for the mean are sharp.

\begin{theorem}\label{thm:mean-bounds}
Suppose that each $F_i$ has a finite first moment. Then
\begin{align}
\inf_{\mathbf{X}\in\mathcal F(F_1,\ldots,F_n)}\E[X_{(k)}]
&\ge
\int_0^\infty(1-G(t))\,\d t-\int_{-\infty}^0 G(t)\,\d t,
\label{eq:minimum-mean}\\
\sup_{\mathbf{X}\in\mathcal F(F_1,\ldots,F_n)}\E[X_{(k)}]
&\le
\int_0^\infty(1-H(t))\,\d t-\int_{-\infty}^0 H(t)\,\d t.
\label{eq:maximum-mean}
\end{align}
Both extrema are attained. Equality holds in
\eqref{eq:minimum-mean} if and only if conditions (i) and (ii) of
Theorem~\ref{thm:universal-attainment} hold, and in
\eqref{eq:maximum-mean} if and only if conditions (iii) and (iv) of
Proposition~\ref{prop:upper} hold.
\end{theorem}

\begin{proof}
By Proposition~\ref{prop:order},
$H(t)\le\p(X_{(k)}\le t)\le G(t)$ for every $t\in\R$ and
$\mathbf{X}\in\mathcal F(F_1,\ldots,F_n)$. Moreover, $G(t)\le\sum_{i=1}^n F_i(t)$ and
$1-H(t)\le\sum_{i=1}^n(1-F_i(t))$, so $G$ and $H$ have finite first
moments. The tail integral formula for the mean gives both inequalities.

Both extrema are attained by Lemma~\ref{lem:var-attainment-cost}
with $g(u)=u$ and $c(\mathbf{x})=x_{(k)}$.
For any $\mathbf{X}\in\mathcal F(F_1,\ldots,F_n)$, the difference between $\E[X_{(k)}]$
and the right-hand side of \eqref{eq:minimum-mean} is
$\int_{\R}\bigl(G(t)-\p(X_{(k)}\le t)\bigr)\,\d t$.
The integrand is nonnegative and right-continuous, so this difference
is zero if and only if $\p(X_{(k)}\le t)=G(t)$ for all $t\in\R$.
Theorem~\ref{thm:universal-attainment} gives the condition for equality.
For \eqref{eq:maximum-mean}, the corresponding integrand is
$\p(X_{(k)}\le t)-H(t)$, and Proposition~\ref{prop:upper} gives the
condition for equality.
\end{proof}

The next theorem gives the corresponding conditions under which the bounds for ES are sharp.

\begin{theorem}\label{thm:es-bounds}
Suppose that each $F_i$ has a finite first moment, and let
$\alpha\in(0,1)$ and $a=1-\alpha$. Then
\begin{align}
\inf_{\mathbf{X}\in\mathcal F(F_1,\ldots,F_n)}\ES_\alpha(X_{(k)})
&\ge \inf_{z\in\R}
\left\{z+\frac{1}{\alpha}\int_z^\infty(1-G(t))\,\d t\right\},
\label{eq:minimum-es}\\
\sup_{\mathbf{X}\in\mathcal F(F_1,\ldots,F_n)}\ES_\alpha(X_{(k)})
&\le \inf_{z\in\R}
\left\{z+\frac{1}{\alpha}\int_z^\infty(1-H(t))\,\d t\right\}.
\label{eq:maximum-es}
\end{align}
Both extrema are attained. Equality holds in
\eqref{eq:minimum-es} if and only if the following conditions hold.
\begin{enumerate}[(i)]
\item For every $i\in[n]$, the function $t\mapsto(G(t)-F_i(t))_+$ is
nondecreasing on $\{t\in\R:a<G(t)<1\}$.
\item At every $t\in\R$ such that $a<G(t-)<G(t)=1$,
\[
\sum_{i=1}^n\max\{0,G(t-)-F_i(t-),G(t)-F_i(t)\}\le n-k.
\]
\end{enumerate}
Equality holds in \eqref{eq:maximum-es} if and only if the following
conditions hold.
\begin{enumerate}
\item[(iii)] For every $i\in[n]$, the function $t\mapsto(F_i(t)-H(t))_+$
is nonincreasing on $\{t\in\R:a<H(t)<1\}$.
\item[(iv)] At every $t\in\R$ such that $H(t-)\le a<H(t)<1$,
\[
\sum_{i=1}^n\max\{0,F_i(t-)-a,F_i(t)-H(t)\}\le(k-1)\alpha.
\]
\end{enumerate}
\end{theorem}

\begin{proof}
We only prove the lower bound case. The upper bound case follows from
the fact that $X_{(k)}=-(-\mathbf{X})_{(n-k+1)}$, similar to the argument in
Proposition~\ref{prop:upper}.

By Proposition~\ref{prop:order},
$\p(X_{(k)}\le t)\le G(t)$ holds for every $\mathbf{X}\in\mathcal F(F_1,\ldots,F_n)$ and $t\in\R$.
Since $G(t)\le\sum_{i=1}^n F_i(t)$ and $1-G(t)\le\sum_{i=1}^n(1-F_i(t))$,
the distribution $G$ has a finite first moment.
Moreover, the cost $c(\mathbf{x})=x_{(k)}$ is continuous.
Thus, the ES infimum is attained by Lemma~\ref{lem:var-attainment-cost}.

Similar to the mean, equality holds in \eqref{eq:minimum-es} if and only if a coupling
$\mathbf{X}\in\mathcal F(F_1,\ldots,F_n)$ satisfies $\p(X_{(k)}\le t)=G(t)$ for every $t$
such that $G(t)>1-\alpha$.
Inequality \eqref{eq:minimum-es} follows from $\p(X_{(k)}\le t)\le G(t)$
for every $t\in\R$.
To prove that equality in \eqref{eq:minimum-es} is equivalent
to conditions (i) and (ii), we proceed in two steps.
Let $s=\inf\{t\in\R:G(t)>1-\alpha\}$.

(a) We first show that equality in \eqref{eq:minimum-es} holds if and
only if conditions (i) and (ii) of Theorem~\ref{thm:universal-attainment}
hold for the truncated distribution functions
$F_i(t)\id_{\{t\ge s\}}$, $i\in[n]$.
By the preceding argument, equality in \eqref{eq:minimum-es} holds
if and only if there exists $\mathbf{X}\in\mathcal F(F_1,\ldots,F_n)$ such that
$\p(X_{(k)}\le t)=G(t)$ for every $t\ge s$.
Note that $\p(X_i\vee s\le t)=F_i(t)\id_{\{t\ge s\}}$ and
$\p\bigl((X_1\vee s,\ldots,X_n\vee s)_{(k)}\le t\bigr)
=\p(X_{(k)}\le t)\id_{\{t\ge s\}}$.
Thus, the assertion follows from Theorem~\ref{thm:universal-attainment}.

(b) Next, we show that the two conditions in this theorem are equivalent
to conditions (i) and (ii) of Theorem~\ref{thm:universal-attainment},
respectively.
\begin{itemize}
\item[(b1)] Theorem~\ref{thm:universal-attainment} uses $t\mapsto(G(t)-F_i(t))_+\id_{\{t\ge s\}}$,
whose monotonicity is equivalent to condition (i).

\item[(b2)] If $G(s)=1$, the inequality in Theorem~\ref{thm:universal-attainment}
is $\sum_{i=1}^n F_i(s)\ge k$.
Since the case $G(t-)=G(t)=1$ is trivial, it suffices to consider
the case $1-\alpha<G(t-)<G(t)=1$.
At such a point $t>s$, condition (ii) of
Theorem~\ref{thm:universal-attainment} can be rewritten as
\begin{equation}\label{eq:condition-ii-last-step}
\sum_{i=1}^n\bigl(1-F_i(t)-(G(t-)-F_i(t-))_+\bigr)_+
\le(n-k)(1-G(t-)).
\end{equation}
From \eqref{eq:bk-def},
$\sum_{i=1}^n(G(t-)-F_i(t-))_+=(n-k)G(t-)$ holds.
Adding this identity to both sides of \eqref{eq:condition-ii-last-step},
\[
\sum_{i=1}^n\max\{0,\,G(t-)-F_i(t-),\,G(t)-F_i(t)\}\le n-k,
\]
which proves condition (ii).
\end{itemize}
Thus, equality holds in \eqref{eq:minimum-es} if and only if both
conditions (i) and (ii) hold.
\end{proof}

\par\endgroup

\begingroup
\section{Application to airport reserve capacity}\label{sec:airport-reserve-capacity}

We consider the capacity of a hypothetical reserve desk shared by eight
airports in an airline network. An airport makes one request on a day when
its disruption count exceeds a given trigger, and each slot at the desk
can serve one request that day. The decision makers seek the smallest number of
slots that covers all requests on at least $90\%$ of days. We use flight
records to construct the requests and examine how the capacity decision
changes under dependence uncertainty across airports.

\subsection{Data}

Our sample consists of scheduled domestic nonstop flights marketed by Delta Air Lines from the US Bureau of Transportation Statistics (BTS) Marketing Carrier On-Time Performance database.\footnote{\href{https://transtats.bts.gov/Fields.asp?gnoyr_VQ=FGK}{BTS TranStats, Marketing Carrier On-Time Performance}.}
The eight airports in the northeastern United States are Boston Logan
(BOS), John F.\ Kennedy (JFK), LaGuardia (LGA), Newark Liberty (EWR),
Philadelphia (PHL), Baltimore/Washington (BWI), Reagan National (DCA),
and Washington Dulles (IAD). A departure is disrupted if it is cancelled
or departs at least $60$ minutes late. Of the $925{,}822$ scheduled
departures, $24{,}726$ are cancelled and $68{,}827$ depart at least
$60$ minutes late. We aggregate by airport and scheduled flight date.
The panel contains a joint observation for each of the $1{,}461$ days
from January 2022 through December 2025. The first $1{,}096$ days are used
for calibration and the $365$ days in 2025 are used for evaluation.

We match the flight panel to airport station records from the Global
Historical Climatology Network Daily.\footnote{\href{https://www.ncei.noaa.gov/products/land-based-station/global-historical-climatology-network-daily}{NOAA National Centers for Environmental Information, Global Historical Climatology Network Daily}.}
A date is classified as severe if at least two of the eight stations each record a
fastest five-second wind of at least $18$ metres per second, snowfall
of at least $51$ millimetres, or thunder. A date is unclassified if
missing observations leave its status unresolved.
The calibration period contains $287$ severe-weather days and $808$
other days.\footnote{One date has no weather classification and is included only
in the pooled sample.} The evaluation period contains $130$ severe-weather
days and $235$ other days.

\subsection{Methodology}

Let $X_i$ denote the daily number of disrupted departures at airport $i$,
with distribution function $F_i$. Airport $i$ requests a slot if
$X_i>t_i$, where $t_i=\lfloor a v_i\rfloor$ and $v_i$ is its median
number of scheduled departures per day in the calibration period. We
consider the trigger proportions $a\in\{0.10,0.15,0.20,0.25\}$,
and use the same triggers throughout calibration and evaluation.

Let $Y_i=X_i-t_i$ have distribution function $G_i(y)=F_i(y+t_i)$.
Airport $i$ requests a slot when $Y_i>0$, with probability
$p_i=1-G_i(0)$. Write $\mathbf{p}=(p_1,\ldots,p_8)$ and let
$N=\sum_{i=1}^8\id_{\{Y_i>0\}}$ denote the total number of requests.
For $m<8$, the overflow event $\{N>m\}$ is exactly $\{Y_{(8-m)}>0\}$. 
By Proposition~\ref{prop:order}, we know
\begin{equation}\label{eq:airport-sharp}
    U(m):=\sup_{\mathbf{Y}\in\mathcal F(G_1,\ldots,G_8)}
    \mathbb{P}(N>m)=b_{m+1}(\mathbf{p}),
    \qquad m=0,\ldots,7.
\end{equation}
Set $U(8)=0$. For each fixed $m<8$, 
Proposition~\ref{prop:order} also gives a coupling that achieves $U(m)$.

Given a set $\mathcal D$ of calibration dates, write $N_d$ for the
total number of requests on date $d$ and $Y_{id}$ for the shifted
disruption count at airport $i$ on that date. Let
$\hat p_i=|\mathcal D|^{-1}\sum_{d\in\mathcal D}\id_{\{Y_{id}>0\}}$
be the empirical request probability at airport $i$. We compare four rules.
\begin{enumerate}[(i)]
    \item The \emph{Historical} rule uses
    $|\mathcal D|^{-1}\sum_{d\in\mathcal D}\id_{\{N_d>m\}}$,
    the empirical overflow frequency with requests aligned by date.
    \item The \emph{IND} rule uses the independent coupling. For $m<8$,
    the overflow probability is $\mathbb{P}(\sum_{i=1}^8 B_i>m)$, where
    $B_1,\ldots,B_8$ are independent Bernoulli variables with success
    probabilities $\hat p_1,\ldots,\hat p_8$, respectively.
    \item The \emph{CO} rule uses the comonotonic coupling. For $m<8$,
    the overflow probability is $\hat p_{(8-m)}$, the $(m+1)$-th largest
    empirical request probability.
    \item The \emph{Sharp} rule uses the sharp upper bound in
    \eqref{eq:airport-sharp} with $\mathbf{p}=(\hat p_1,\ldots,\hat p_8)$.
\end{enumerate}
The overflow probability is zero at $m=8$ under all four rules. Each
rule selects the smallest $m\in\{0,\ldots,8\}$ for which its calibration
overflow probability is at most $\varepsilon=10\%$. 
The sharp rule selects
\begin{equation}\label{eq:airport-capacity}
    m_{\mathrm{Sharp}}=\min\{m\in\{0,\ldots,8\}\colon U(m)\le\varepsilon\},
\end{equation}
where $U(m)$ is computed by using Lemma~\ref{lem:bk-closed} with
the ordered empirical probabilities $\hat p_{(1)}\le\cdots\le\hat p_{(8)}$.
For each selected capacity, we count days with $N_d>m$ in 2025.

\subsection{Results}

Figure~\ref{fig:airport-overflow} plots the four calibration overflow
curves for each trigger, where the first point 
below the dashed line ($10\%$ overflow probability) gives the selected capacity.
Table~\ref{tab:airport-overflow} reports these capacities and the
number of days they overflow on the common 2025 sample.

\begin{figure}[!ht]
    \centering
    \includegraphics[width=0.9\linewidth]{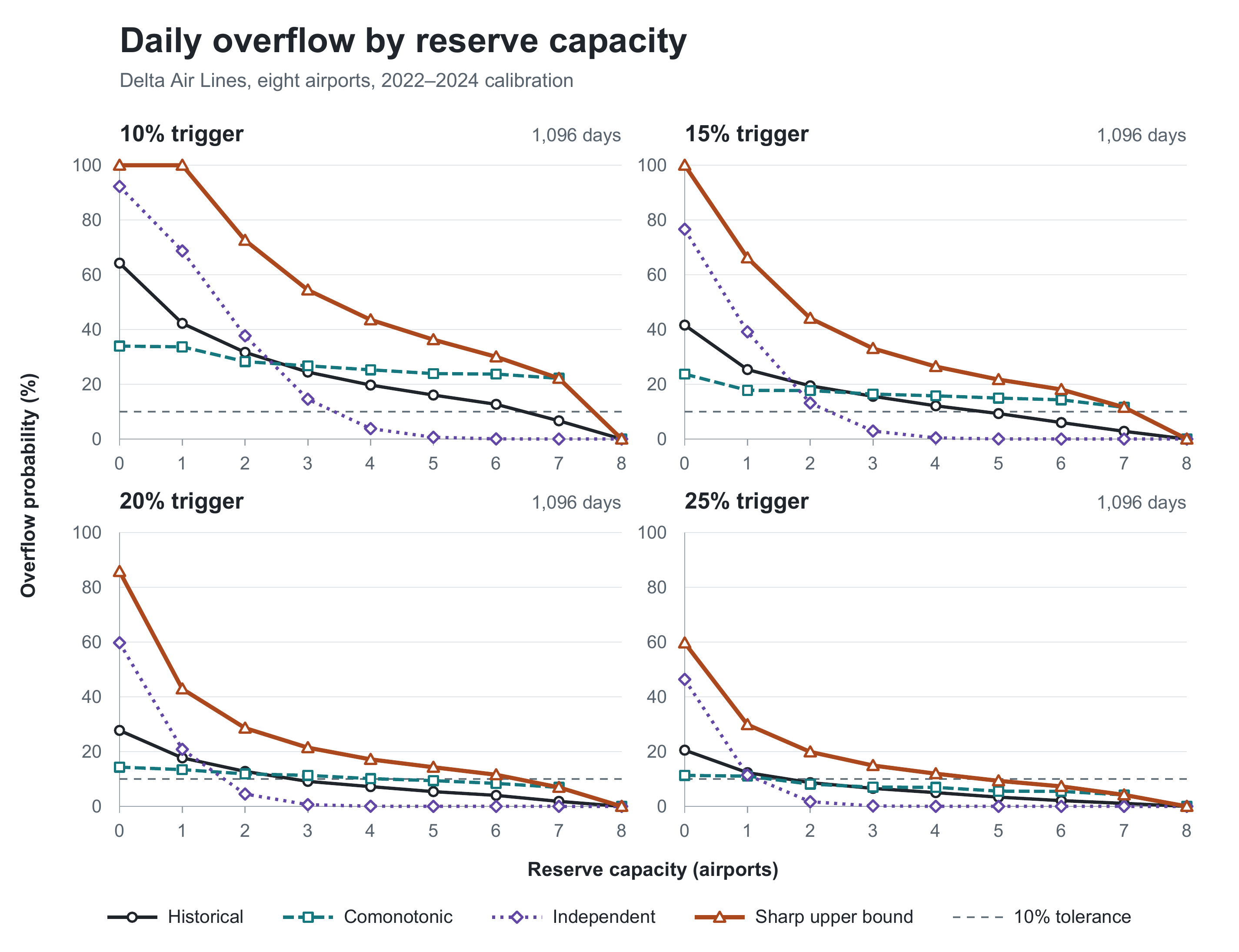}
    \caption{Calibration overflow probability based on the $1{,}096$ days in
    2022--2024.}
    \label{fig:airport-overflow}
\end{figure}

\begin{table}[!ht]
    \centering
    \caption{Reserve slots at a $90\%$ service target, with 2025 overflow days in parentheses.}
    \label{tab:airport-overflow}
    \begin{tabular}{@{}lcccc@{}}
        \toprule
        Trigger proportion & Historical & IND & CO & Sharp \\
        \midrule
        $10\%$ & $7\,(46)$ & $4\,(112)$ & $8\,(0)$  & $8\,(0)$ \\
        $15\%$ & $5\,(53)$ & $3\,(92)$  & $8\,(0)$  & $8\,(0)$ \\
        $20\%$ & $3\,(60)$ & $2\,(79)$  & $5\,(32)$ & $7\,(10)$ \\
        $25\%$ & $2\,(60)$ & $2\,(60)$  & $2\,(60)$ & $5\,(17)$ \\
        \bottomrule
    \end{tabular}
\end{table}

Historical and IND exceed the $10\%$ overflow probability in 2025 at
all four triggers. At the $10\%$ and $15\%$ triggers, CO and Sharp
both select eight slots. At the $20\%$ trigger, CO selects five slots
and Sharp selects seven, with sharp bounds of $14.29\%$ and $6.93\%$,
respectively. Both capacities meet the target in 2025, with $32$ and
$10$ overflow days, respectively. At the $25\%$ trigger, Historical,
IND, and CO all select two slots, with calibration overflow probabilities
of $8.67\%$, $1.69\%$, and $8.12\%$, respectively. The sharp bound at
two slots is $19.89\%$. It falls from $11.93\%$ at four slots to $9.33\%$
at five, so \eqref{eq:airport-capacity} recommends five slots. In the
2025 evaluation, five slots cover all requests on $43$ additional days,
reducing overflow from $60$ days ($16.44\%$) to $17$ days ($4.66\%$)
and meeting the $90\%$ service target.

Table~\ref{tab:airport-weather} reports the weather analysis at the
$25\%$ trigger. The first two rows use the capacities in
Table~\ref{tab:airport-overflow}, selected using all calibration days.
They report overflow separately on severe-weather and other days in
2025. The last two rows use capacities selected from the calibration
days in each weather group.

\begin{table}[!ht]
    \centering
    \caption{Reserve slots and 2025 overflow days in parentheses by observed weather at the $25\%$ trigger.}
    \label{tab:airport-weather}
    \begin{tabular}{@{}llrcccc@{}}
        \toprule
        Calibration data & 2025 weather & Days & Historical & IND & CO & Sharp \\
        \midrule
        All calibration days    & Severe & $130$ & $2\,(43)$ & $2\,(43)$ & $2\,(43)$ & $5\,(16)$ \\
        All calibration days    & Other  & $235$ & $2\,(17)$ & $2\,(17)$ & $2\,(17)$ & $5\,(1)$ \\
        Severe weather & Severe & $130$ & $6\,(12)$ & $3\,(34)$ & $8\,(0)$  & $8\,(0)$ \\
        Other days     & Other  & $235$ & $1\,(24)$ & $1\,(24)$ & $0\,(63)$ & $2\,(17)$ \\
        \bottomrule
    \end{tabular}
\end{table}

The five slots selected by Sharp using all calibration days lead to
overflow on $16$ of the $130$ severe-weather days in 2025 ($12.31\%$),
above the $10\%$ tolerance. If the planner can set capacity by weather
group and requires $90\%$ service in each group, both CO and Sharp
select eight slots for severe weather. For other days, CO selects zero
slots and Sharp selects two. On the same $235$ evaluation days, two
slots reduce overflow from $63$ days ($26.81\%$) to $17$ ($7.23\%$),
below the $10\%$ tolerance. 

When decision makers account for how demands coincide, the sharp bound gives the exact worst-case protection based on the given marginal probabilities. Therefore, the sharp bound provides the least shared capacity that meets a prescribed service target under dependence uncertainty.

\par\endgroup

% \section{Conclusion}\label{sec:con}

% Tail counter-monotonicity gives a coupling that maximizes the maximum for every threshold. The layered arrangements extend this construction to threshold probabilities for any order statistic. Working with one threshold at a time gives the sharp service guarantee for a capacity decision. The pointwise bounds also give exact VaR bounds even when the stochastic envelope is unattainable. The pairwise characterization describes all attainable rank laws for general marginals. On a finite grid, its linear formulation gives exact ES bounds and optimizes expected costs over these laws.

% In the airport illustration, the historical sample and an independence model both select one reserve assignment at the chosen trigger and service level. Two are needed to meet that level for all couplings of the pooled empirical marginals.  A further question is how to incorporate other information available to a planner, such as bounds on joint exceedances or uncertainty about the individual workload distributions.

\section{Conclusion}\label{sec:conclusion}

We study OT and dependence uncertainty problems for order statistics by proposing a new dependence structure called $\mathbf{p}$-tail counter-monotonicity. The $\mathbf{p}$-tail counter-monotonic coupling gives the largest maximum in stochastic order, and the comonotonic coupling gives the smallest (Theorem~\ref{thm:max}).  For a general order statistic, based on the results in \cite{bib:ruger-1978} and \cite{bib:salako-2025}, we construct explicit couplings that attain the sharp bounds on the distribution function of $X_{(k)}$ at each fixed threshold $t$ (Proposition~\ref{prop:order}) by the proposed $k$-layered $\mathbf{p}$-tail optimal dependence. Different from the maximum and minimum, for a general order statistic, a smallest or largest attainable law in stochastic order may fail to exist. Theorem~\ref{thm:universal-attainment} and Proposition~\ref{prop:upper} establish necessary and sufficient conditions on the marginals under which a single dependence structure attains the cdf bounds at all thresholds simultaneously. 
We also provide bounds for VaR, the mean, and ES and study their sharpness and attainment (Theorems~\ref{thm:var-attain}--\ref{thm:es-bounds}).
For marginals that fall outside the conditions, the methods for computing the exact bounds on expected costs and ES are given even when a stochastically extremal coupling does not exist (Appendix~\ref{app:attainable-distributions}).
Our results are illustrated by an application to the reserve capacity problem for eight US airports in Section~\ref{sec:airport-reserve-capacity}.

\begin{appendices}
\section{Attainable distributions of the order statistic}\label{app:attainable-distributions}

Note that Section~\ref{sec:risk-measure-bounds} only provides the sharp bounds for $\ES$ and the mean within a limited set of marginals. To compute the extrema for general marginals, the laws
that $X_{(k)}$ can have with any given marginals should be characterized. We first give a
linear programming formulation for finitely supported marginals and
then extend the characterization to arbitrary marginals. 

\subsection{Finitely supported marginals}

Let $z_1<\cdots<z_m$ contain the supports of all the marginals, and let $f_{i\ell}=\p(X_i=z_\ell)$ for $\mathbf{X}\in\mathcal F(F_1,\ldots,F_n)$. The distribution of the order statistic is determined by the probabilities $w_j=\p(X_{(k)}=z_j)$. For each $i,j,\ell$, denote the joint probability
\[
a_{ij\ell}=\p(X_{(k)}=z_j,\ X_i=z_\ell).
\]
The next proposition shows that the resulting linear constraints on these joint probabilities suffice to characterize the distribution of $X_{(k)}$.

\begin{proposition}\label{prop:finite}
Let $\mathbf{w}=(w_1,\ldots,w_m)\in\Delta_m$. There exists $\mathbf{X}\in\mathcal F(F_1,\ldots,F_n)$ with $\p(X_{(k)}=z_j)=w_j$ for every $j\in[m]$ if and only if there exist nonnegative $a_{ij\ell}$ satisfying
\begin{equation}\label{eq:finite-count}
\begin{aligned}
&\sum_{j=1}^{m} a_{ij\ell}=f_{i\ell} && i\in[n],\ \ell\in[m],\\
&\sum_{\ell=1}^{m} a_{ij\ell}=w_j && i\in[n],\ j\in[m],\\
&\sum_{i=1}^{n}\sum_{\ell<j} a_{ij\ell}\le (k-1)w_j && j\in[m],\\
&\sum_{i=1}^{n}\sum_{\ell>j} a_{ij\ell}\le (n-k)w_j && j\in[m].
\end{aligned}
\end{equation}
\end{proposition}

\begin{proof}
Since the ``only if'' direction is straightforward, it suffices to show the ``if'' direction.
Suppose that \eqref{eq:finite-count} holds. The construction proceeds in two steps.

(a) Take a partition $A_1,\ldots,A_m$ with $\mathbb{P}(A_j)=w_j$. We first construct
$\mathbf{X}$ on each $A_j$ with $w_j>0$ so that $X_{(k)}=z_j$. Fix $j$ with $w_j>0$.
In the matrix $(a_{ij\ell}/w_j)_{i,\ell}$,
take the rows and the single-entry sets as the first family.
Within this family, any two sets are disjoint or one contains the other.
The second family consists of the two disjoint sets of entries with $\ell<j$ and $\ell>j$.
By \eqref{eq:finite-count} and Theorem~1 of \citet{bib:budish-et-al-2013}, this
matrix is a convex combination of $0$--$1$ matrices with one $1$
in each row and at most $k-1$ and $n-k$ ones in the two blocks, respectively.
On $A_j$, draw one of these matrices with probability equal to its
coefficient in the convex combination, and set $X_i=z_\ell$ if row
$i$ has its $1$ in column $\ell$. Then on $A_j$, it holds that
$\sum_{i=1}^n \mathbf{1}_{\{X_i<z_j\}}\le k-1$ and
$\sum_{i=1}^n \mathbf{1}_{\{X_i>z_j\}}\le n-k$, which implies $X_{(k)}=z_j$.

(b) Next, we verify the marginal distributions. Fix $j$ with $w_j>0$.
By the convex combination in step (a), conditional on $A_j$, the probability of selecting a matrix with a $1$
in entry $(i,\ell)$ is $a_{ij\ell}/w_j$.
Therefore, $\mathbb{P}(X_i=z_\ell\mid A_j)=a_{ij\ell} / w_j$ holds.
By \eqref{eq:finite-count}, $a_{ij\ell}=0$ whenever $w_j=0$. Hence
\[
\mathbb{P}(X_i=z_\ell)
=\sum_{j:w_j>0}w_j\frac{a_{ij\ell}}{w_j}
=\sum_{j=1}^m a_{ij\ell}
=f_{i\ell}.
\]
Thus, this coupling $\mathbf{X}$ satisfies $w_j=\p(X_{(k)}=z_j)$ for every $j=1,\ldots,m$, and the probability vector $\mathbf{w}$ is attainable.
\end{proof}

The $a_{ij\ell}$ in \eqref{eq:finite-count} has $nm^2$ entries.
For each coordinate $i$ and each possible value $z_j$ of $X_{(k)}$,
define the joint probabilities
$b_{ij}=\p(X_{(k)}=z_j,\ X_i<z_j)=\sum_{\ell<j}a_{ij\ell}$,
$e_{ij}=\p(X_{(k)}=z_j,\ X_i=z_j)=a_{ijj}$, and
$h_{ij}=\p(X_{(k)}=z_j,\ X_i>z_j)=\sum_{\ell>j}a_{ij\ell}$.
These probabilities sum to $w_j=\p(X_{(k)}=z_j)$.

\begin{proposition}\label{prop:linear-size}
Let $\mathbf{w}\in\Delta_m$. There exists $\mathbf{X}\in\mathcal F(F_1,\ldots,F_n)$ with $\p(X_{(k)}=z_j)=w_j$ for every $j\in[m]$ if and only if there are nonnegative $b_{ij},e_{ij},h_{ij}$ with $e_{ij}\le f_{ij}$ for all $i\in[n]$ and $j\in[m]$ satisfying
\begin{equation}\label{eq:rank-prefix}
\begin{aligned}
&b_{ij}+e_{ij}+h_{ij}=w_j && i\in[n],\ j\in[m],\\
&\sum_i b_{ij}\le(k-1)w_j && j\in[m],\\
&\sum_i h_{ij}\le(n-k)w_j && j\in[m],\\
&\sum_{j\le r} b_{ij}\le \sum_{\ell<r}(f_{i\ell}-e_{i\ell}) && i\in[n],\ r\in[m],\\
&\sum_{j\ge r} h_{ij}\le \sum_{\ell>r}(f_{i\ell}-e_{i\ell}) && i\in[n],\ r\in[m].
\end{aligned}
\end{equation}
\end{proposition}

\begin{proof}
\underline{The ``only if'' direction.} By Proposition~\ref{prop:finite}, take nonnegative $a_{ij\ell}$ satisfying
\eqref{eq:finite-count}. Set $b_{ij}=\sum_{\ell<j}a_{ij\ell}$, $e_{ij}=a_{ijj}$ and $h_{ij}=\sum_{\ell>j}a_{ij\ell}$.
It suffices to verify the five constraints in \eqref{eq:rank-prefix}.
From \eqref{eq:finite-count}, we can easily obtain $e_{ij}\le f_{ij}$ and the first three constraints. For the fourth one,
\[
\sum_{j\le r} b_{ij}
=\sum_{\ell<r}\sum_{\ell<j\le r}a_{ij\ell}
\le \sum_{\ell<r}\sum_{j\ne\ell}a_{ij\ell}
=\sum_{\ell<r}(f_{i\ell}-e_{i\ell}),
\]
and the last one follows similarly.

\underline{The ``if'' direction.}
Suppose that there exist nonnegative $b_{ij},e_{ij},h_{ij}$ with $e_{ij}\le f_{ij}$
satisfying \eqref{eq:rank-prefix}. The proof is divided into two steps.

(a) We first construct $a_{ij\ell}$.
Since $\sum_j w_j=\sum_\ell f_{i\ell}=1$ and $b_{ij}+e_{ij}+h_{ij}=w_j$, we have
\[
\sum_j(b_{ij}+h_{ij})
=\sum_j(w_j-e_{ij})
=1-\sum_\ell e_{i\ell}
=\sum_\ell(f_{i\ell}-e_{i\ell}).
\]
Use $\big(\sum_{r<\ell}(f_{ir}-e_{ir}),\,
\sum_{r\le\ell}(f_{ir}-e_{ir})\big]$ for $f_{i\ell}-e_{i\ell}$,
$\big(\sum_{r<j}b_{ir},\,\sum_{r\le j}b_{ir}\big]$ for $b_{ij}$,
and $\big(\sum_\ell(f_{i\ell}-e_{i\ell})-\sum_{r\ge j}h_{ir},\,
\sum_\ell(f_{i\ell}-e_{i\ell})-\sum_{r>j}h_{ir}\big]$ for $h_{ij}$.
Set $a_{ijj}=e_{ij}$. For $\ell<j$, set
\[
a_{ij\ell}
=
\left(
\min\left\{\sum_{r\le\ell}(f_{ir}-e_{ir}),\,\sum_{r\le j}b_{ir}\right\}
-
\max\left\{\sum_{r<\ell}(f_{ir}-e_{ir}),\,\sum_{r<j}b_{ir}\right\}
\right)_+.
\]
For $\ell>j$, set
\[
a_{ij\ell}
=
\left(
\min\left\{\sum_{r\ge\ell}(f_{ir}-e_{ir}),\,\sum_{r\ge j}h_{ir}\right\}
-
\max\left\{\sum_{r>\ell}(f_{ir}-e_{ir}),\,\sum_{r>j}h_{ir}\right\}
\right)_+.
\]

(b) Next, we verify \eqref{eq:finite-count}.
By the last two constraints in \eqref{eq:rank-prefix},
\[
\sum_{r\le j}b_{ir}
\le \sum_{\ell<j}(f_{i\ell}-e_{i\ell}),\qquad
\sum_\ell(f_{i\ell}-e_{i\ell})-\sum_{r\ge j}h_{ir}
\ge \sum_{\ell\le j}(f_{i\ell}-e_{i\ell}).
\]
Thus the interval for $b_{ij}$ overlaps only intervals indexed by
$\ell<j$ in positive length, and the interval for $h_{ij}$ only those
indexed by $\ell>j$.
Summing the intersection lengths in (a), we obtain
$\sum_{j\ne\ell}a_{ij\ell}=f_{i\ell}-e_{i\ell}$,
$\sum_{\ell<j}a_{ij\ell}=b_{ij}$, and $\sum_{\ell>j}a_{ij\ell}=h_{ij}$.
Since $a_{ijj}=e_{ij}$, the above equalities yield the two marginal constraints:
\[
\sum_j a_{ij\ell}=f_{i\ell}
\quad \text{and}\quad
\sum_\ell a_{ij\ell}=b_{ij}+e_{ij}+h_{ij}=w_j.
\]
The two order statistic constraints follow from
\[
\sum_i\sum_{\ell<j}a_{ij\ell}=\sum_i b_{ij}\le(k-1)w_j
\quad \text{and}\quad
\sum_i\sum_{\ell>j}a_{ij\ell}=\sum_i h_{ij}\le(n-k)w_j.
\]
Thus by Proposition~\ref{prop:finite}, the probability vector $\mathbf{w}$ is attainable.
\end{proof}

Let $\mathcal W_k$ be the set of probability vectors $\mathbf{w}$ that are attainable as distributions of $X_{(k)}$. By Proposition~\ref{prop:linear-size}, $\mathcal W_k$ is the projection of a polytope described by $O(nm)$ variables and constraints. For each cost function $c$ in \eqref{eq:intro-cost}, the expected cost is $\sum_j c(z_j)w_j$, so its lower and upper bounds are obtained by linear programming over $\mathcal W_k$. In particular, taking $c(x)=x$ gives
\[
\inf_{\mathbf{X}\in\mathcal F(F_1,\ldots,F_n)}\E[X_{(k)}]
=\min_{\mathbf{w}\in\mathcal W_k}\sum_j z_jw_j
\quad\text{and}\quad
\sup_{\mathbf{X}\in\mathcal F(F_1,\ldots,F_n)}\E[X_{(k)}]
=\max_{\mathbf{w}\in\mathcal W_k}\sum_j z_jw_j.
\]
%For a general distortion risk measure with distortion function $g$, the objective in \eqref{eq:intro-rm} is
%\[
%z_1+\sum_{j=1}^{m-1}(z_{j+1}-z_j)g\left(\sum_{\ell>j}w_\ell\right).
%\]
For the ES bounds at level $\alpha\in(0,1)$, we have
\begin{subequations}\label{eq:finite-es}
\begin{align}
\sup_{\mathbf{X}\in\mathcal F(F_1,\ldots,F_n)}\ES_\alpha(X_{(k)})
&=\frac{1}{\alpha}\max\left\{\sum_j z_jv_j:\mathbf{w}\in\mathcal W_k,\ 0\le v_j\le w_j,\ \sum_jv_j=\alpha\right\},\label{eq:finite-es-maximum}\\
\inf_{\mathbf{X}\in\mathcal F(F_1,\ldots,F_n)}\ES_\alpha(X_{(k)})
&=\min_{s\in[m]}\left\{z_s+\frac{1}{\alpha}\min_{\mathbf{w}\in\mathcal W_k}\sum_j(z_j-z_s)_+w_j\right\}.\label{eq:finite-es-minimum}
\end{align}
\end{subequations}

The reason is simple. For each $\mathbf{w}\in\mathcal W_k$ and each event $A$ with
$\p(A)=\alpha$, set $v_j=\p(A\cap\{X_{(k)}=z_j\})$. By atomlessness,
we can write
\[
\ES_\alpha(X_{(k)})
=\frac{1}{\alpha}\max_{A:\,\p(A)=\alpha}
\E[X_{(k)}\mathbf{1}_A]
=\frac{1}{\alpha}\max\left\{\sum_j z_jv_j:
0\le v_j\le w_j,\ \sum_jv_j=\alpha\right\}.
\]
Maximizing over $\mathbf{w}\in\mathcal W_k$ gives the upper ES bound.
Note that
$\ES_\alpha(X_{(k)})=\min_{t\in\mathbb R}
\{t+\frac{1}{\alpha}\sum_j(z_j-t)_+w_j\}$.
For each $\mathbf{w}\in\mathcal W_k$, the objective is strictly
decreasing up to $t=\VaR_\alpha(X_{(k)})$ and increasing thereafter.
Thus, it follows that
\[
\begin{aligned}
&\inf_{\mathbf{X}\in\mathcal F(F_1,\ldots,F_n)}\ES_\alpha(X_{(k)})
=\min_{\mathbf{w}\in\mathcal W_k}\min_{t\in\mathbb R}
\left\{t+\frac{1}{\alpha}\sum_j(z_j-t)_+w_j\right\}\\
&=\min_{\mathbf{w}\in\mathcal W_k}\min_{s\in[m]}
\left\{z_s+\frac{1}{\alpha}\sum_j(z_j-z_s)_+w_j\right\}
=\min_{s\in[m]}\left\{z_s+\frac{1}{\alpha}
\min_{\mathbf{w}\in\mathcal W_k}\sum_j(z_j-z_s)_+w_j\right\}.
\end{aligned}
\]

\subsection{General marginals}\label{sec:general-marginals}

We now allow arbitrary marginal distribution functions $F_1,\ldots,F_n$. Let $\mu_i$ be the probability measure with distribution function $F_i$, and let $\mathcal K$ be the set of probability laws of $X_{(k)}$ over $\mathcal F(F_1,\ldots,F_n)$. 

\begin{proposition}\label{prop:pairwise}
A probability measure $\nu$ belongs to $\mathcal K$ if and only if, for each $i\in[n]$, there is a bivariate probability measure $\gamma_i$ with first marginal $\nu$ and second marginal $\mu_i$ satisfying for every Borel set $D\subseteq\R$,
\begin{align*}
\sum_{i=1}^n\gamma_i\bigl(\{(y,x):y\in D,\ x<y\}\bigr)&\le (k-1)\nu(D),\\
\sum_{i=1}^n\gamma_i\bigl(\{(y,x):y\in D,\ x>y\}\bigr)&\le (n-k)\nu(D).
\end{align*}
\end{proposition}

\begin{proof}
Since the ``only if'' direction is straightforward, it suffices to show the ``if'' direction.
Suppose that $\gamma_1,\ldots,\gamma_n$ satisfy the given conditions.
We show that $\nu$ belongs to $\mathcal K$.
By the standard method, we only need to construct a coupling on each of the finite grids.
Then by taking a weak limit, we can prove the same argument holds for a general $\nu$.

For $r=1,2,\ldots$, let $T_r$ be the increasing map
$$
T_r(x)=(-r)\vee\left(r\wedge\frac{\lfloor rx\rfloor}{r}\right).
$$
For fixed $r$, write $z_1<\cdots<z_m$ for the values of $T_r$ and set
$a_{ij\ell}=\gamma_i\bigl(\{(y,x):T_r(y)=z_j,\ T_r(x)=z_\ell\}\bigr)$,
$w_j=(T_r)_\#\nu(\{z_j\})$ and $f_{i\ell}=(T_r)_\#\mu_i(\{z_\ell\})$.
Then the marginal constraints in \eqref{eq:finite-count} hold.
Since $T_r(x)<T_r(y)$ implies $x<y$, we have
$$
\begin{aligned}
\sum_i\sum_{\ell<j}a_{ij\ell}
&=\sum_i\gamma_i\bigl(\{(y,x):T_r(y)=z_j,\ T_r(x)<T_r(y)\}\bigr)\\
&\le \sum_i\gamma_i\bigl(\{(y,x):T_r(y)=z_j,\ x<y\}\bigr)
\le (k-1)w_j.
\end{aligned}
$$
The bound for $\ell>j$ follows similarly. 
By Proposition~\ref{prop:finite}, we can construct a coupling with coordinate
laws $(T_r)_\#\mu_i$ and rank law $(T_r)_\#\nu$.
This completes the proof that $\nu$ belongs to $\mathcal K$.
\end{proof}

Therefore, for a law-invariant risk measure $\rho$, the two bounds in \eqref{eq:intro-rm} can be taken over random variables with laws in $\mathcal K$.

\end{appendices}

\end{document}